\documentclass[letterpaper, 10 pt, conference]{ieeeconf}  

\IEEEoverridecommandlockouts                              
\usepackage{amsmath,amssymb,bm,bbm}
\usepackage{color, subfig}
\usepackage{graphicx}
\usepackage{epstopdf}
\usepackage{epsfig}
\usepackage{graphics} 
\usepackage{epstopdf}
\usepackage{enumerate}
\usepackage{caption}
\usepackage{amsmath} 
\usepackage{amssymb, dsfont}  
\usepackage{subfig,pgfplots}
\pgfplotsset{ 
  compat=newest, 
  legend style =
  {font=\small \sffamily},
  label style = {font=\small\sffamily},
every tick label/.append style={font=\small}}
\usepackage{tikz}

\usetikzlibrary {positioning}

\def\beq{\begin{equation}}
\def\eeq{\end{equation}}
\newtheorem{theorem}{Theorem}

\newtheorem{lemma}{Lemma}

\newtheorem{example}{Example}
\newtheorem{remark}{Remark}
\newcommand{\abs}[1]{\lvert#1\rvert}
\newcommand{\ds}{\displaystyle}

\newcommand{\ba}{\begin{array}}
\newcommand{\ea}{\end{array}}

\newcommand{\be}{\begin{equation}}
\newcommand{\ee}{\end{equation}}
\newcommand{\eps}{\varepsilon}

\newcommand{\mc}{\mathcal}

\newcommand{\1}{\mathbbm{1}}

\newcommand{\R}{\mathbb{R}}
\newcommand{\N}{\mathbb{N}}

\renewcommand{\P}{\mathbb{P}}

\renewcommand{\span}{\text{span}}
\DeclareMathOperator{\supp}{supp}

\def\1{\mathds{1}}

\def\N{\mathbb{N}}

\def\R{\mathbb{R}}

\def\P{\mathbb{P}}

 \def \l {3.3}
\def \h {2.5}

\title{\LARGE \bf
On stochastic imitation dynamics in large-scale networks
}

\author{Lorenzo Zino, Giacomo Como, and Fabio Fagnani
\thanks{The authors are with the  Department of Mathematical Sciences ``G.L.~Lagrange'', 
        Politecnico di Torino, corso Duca degli Abruzzi 24, 10129 Torino, Italy
        {\tt\small \{lorenzo.zino, giacomo.como, fabio.fagnani\}@polito.it}}%
\thanks{L.~Zino is also with the Department of Mathematics, Universit{\`a} di Torino.}%
        \thanks{G.~Como is also with the Department of Automatic Control, Lund University, where is a member of the excellence centres LCCC and ELLIIT.  His work was partially supported by the Swedish Research Council through Project Research Grant 2015-04066 and by the Compagnia di San Paolo.}%
}

\begin{document}

\maketitle
\thispagestyle{empty}
\pagestyle{empty}

\begin{abstract}

We consider a broad class of stochastic imitation dynamics over networks, encompassing several well known learning models such as the replicator dynamics. In the considered models, players have no global information about the game structure: they only know their own current utility and the one of neighbor players contacted through pairwise interactions in a network. In response to this information, players update their state according to some stochastic rules. For potential population games and complete interaction networks, we prove convergence and long-lasting permanence close to the evolutionary stable strategies of the game. These results refine and extend the ones known for deterministic imitation dynamics as they account for new emerging behaviors including meta-stability of the equilibria. Finally, we discuss extensions of our results beyond the fully mixed case, studying imitation dynamics where agents interact on complex communication networks.

\end{abstract}

\section{Introduction}

This work is concerned with the analysis of stochastic imitation dynamics for potential population games in large-scale networks. Imitation dynamics belong to the broader class of learning processes and are used to model the evolution of behaviors and strategies in social and biological systems \cite{Weibull1995, Bjornerstedt1996, Hofbauer2003}. In contrast to most of the other learning dynamics such as best response and its noisy versions \cite{Ellison1993,Fudenberg1998} or logit learning \cite{Blume1993}, imitation dynamics require very little information about the game and its structure to be known by the players: each player only has to know his/her own current utility and to be able to observe the one of some of her/his fellow players. The strength of imitation dynamics models relies, thus, in their very limited assumptions, which allow for applying them to many real world situations, in which players may not even be aware of all the possible actions that can be played.

Imitation dynamics are based on interactions between individuals in the population. By contacting another player, an individual gets aware of a new action (the one played by the contacted player) and of its utility. Then the player may revise his/her own action using this information, in a stochastic fashion, in order to increase his/her utility. In this work, we provide analytical results for fully mixed populations, where all the possible pairs of individuals are allowed to interact. Numerical simulations suggest the possibility to extend our results to more general cases.

Stochastic learning dynamics have been analyzed in the literature from different points of view. On the one hand, for models in which more information of the game structure is assumed  (e.g., noisy best response, logit learning), the evolution of the strategies in the population and its asymptotic behavior have been analyzed \cite{Montanari2010, Marden2012,Barreiro-Gomez2016, Tatarenko2017}, where conditions for convergence close to Nash equilibria (NE) and estimates of the convergence time have been determined. On the other hand, models assuming little information and based on pairwise contacts have been mostly tackled by considering their deterministic counterparts.  Available result in this area, dealing with local stability, can be found in \cite{Nachbar1990, Hofbauer2000, Sandholm2001, Sandholm2010,Panageas2016}. Recently, we proposed an analysis of the global stability for the relevant class of potential population games in \cite{cdc2017}. Considerations of the behavior of stochastic dynamics can be found in \cite{Ellison1993,Fudenberg2006}, however, the results in the literature are mostly limited to some specific dynamics, and to their applications for some specific classes of games, e.g., $2$-action games with linear reward \cite{Fudenberg2006}.

This work contributes to extend the understanding of imitation dynamics. For the special and relevant class of potential games, we prove convergence and long-lasting permanence close to the set of evolutionary stable strategies (ESS) of the (continuous) sub-game restricted to the support of the initial population configuration. This result refines and extends what is known for deterministic imitation dynamics \cite{cdc2017}, for which convergence close to the NE have been proven. Hence, the study of the stochastic process deepens the knowledge of the learning dynamics, allowing for the analysis of some interesting phenomena that are not captured by deterministic imitation dynamics models.

The paper is organized as follows. In Section \ref{sec:potential}, we introduce potential population games. In Section \ref{sec:imitation-dynamics} we present the family of the stochastic imitation dynamics studied in this paper, we briefly recall the main results known in the deterministic approach to this dynamics, and we propose an example showing a relevant phenomenon that is not caught by the deterministic approach. The main contributions of this paper are the results on the long-run behavior of the stochastic imitation dynamics, which are presented in Section \ref{sec:analysis}. Then, in Section \ref{sec:simulations} we discuss extensions of our results beyond the fully mixed case presenting some numerical simulations in cases where the players communicate on a complex networks of interactions. Finally, Section \ref{sec:conclusion} concludes the work and presents future works.

\section{Potential population games}\label{sec:potential}

In this work, we deal with population games whereby the reward for playing a given action depends only on the empirical frequency of the actions played in the population. In order to formally define this class of games, we introduce the following notions.

Let $\mc A$ be a finite set of actions with $\abs{\mc A}=m$ and $\mc V=\{1,\dots,n\}$ a finite set of players (agents). Let us define the \emph{configuration} of the system as a vector $y\in\mc A^{n}$, where $y_v=i$ means that node $v\in\mc V$ plays action $i\in\mc A$. For large scale systems, i.e., as $n$ grows large, the size of the configuration space grows exponentially in $n$ making its analysis almost unfeasible. This issue is addressed in the following, by defining a lower dimensional observable on the system: the type.

We let $\mc X$ be the unitary simplex of continuous probability vectors over $\mc A$, i.e.,
\be
\mc X=\left\{x\in\R_+^{\mc A}:\sum_{i\in\mc A}x_i=1\right\}_,
\ee
and, denoting the set $S_n:=\{0,1/n,\dots, 1\}$, we let
 $\mc X^{(n)}=\mc X\cap S_n^{\mc A}$
be its discrete counterpart. Vertices of the simplex are denoted as $\delta^{(i)}$, $i\in\mc A$.

The \emph{type} of the system is the $m$-dimensional vector $x\in\mc X^{(n)}$, whose $i$-th component $x_i$ counts the fraction of individuals playing action $i$ in the configuration of the system $y$.
Given a type of the system $x$, its \emph{support} $\supp_{x}$ is the subset of actions played by at least one individual:
\be\label{eq:support}
\supp_{x}:=\{i\in\mc A: x_i>0\}.
\ee
We say that $x$ is an interior point of its support if $\exists\, \eps>0$ (non-depending on the population size) such that $x_i\neq 0\implies x_i>\eps$. The definition of interior points becomes non-trivial when $n\to \infty$. In fact, types with a negligible fraction of players using an action are not interior of their support.

We remark that the size of the space of types $\mc X^{(n)}$ grows only polynomially in $n$, making the analysis on the space of type more tractable, as we will see in the following of this work. However, in general the type of the system $x$ is not a sufficient observable to represent the state of the system and to study its evolution, since it does not capture the topological structure of the network of interaction.

We can now introduce population games. For this class of games, the reward of any player playing action $i\in\mc A$ depends only on the type of the system. We denote it as $r_i(x)$. Let
$
r:\mc X\to\R^{\mc A}
$ 
be a Lipschitz-continuous reward vector function, whose entry $r_i(x)$ represents the reward received by any player playing action $i\in\mc A$ when the type of the system is $x$. We can naturally define the set of NE of the (continuous) population game as \be  \mc N:=\left\{x\in\mc X: x_i>0\implies r_i(x)\geq r_j(x),\,\forall j\in\mc A\right\}, \ee
and its set of ESS as
\be\ba{l}  \mc S:=\{x\in\mc X: \exists\,\eps>0, \forall\,y \in\mc X\smallsetminus\{x\}: \\[4pt] \ds\qquad\qquad||x-y||<\eps \implies(x-y)^Tr(y)<0\}. \ea\ee
In plain words, NE are configurations where no one of the players could have an  increase of the reward by flipping his/her action (while the other players do not change their action). ESS are configurations such that, if a small fraction of ``mutant'' players changes their action, the reward of the mutants is less than the one of the other players. From the literature, $\mc S\subseteq \mc N$ \cite{Sandholm2010a}, which comes straightforward from our insights on the two definitions. 

Given a subset of actions $\mc B\subseteq\mc A$, these two definitions can be naturally adapted to the game restricted to $\mc B$ by considering configurations with $\text{supp}_{x}=\text{supp}_{y}=\mc B$ and changing each occurrence of $\mc A$ to $\mc B$.

In this work, we will stick our analysis to the case of potential games \cite{Monderer1996}, i.e., when it exists a potential function $\Phi:\mc X\to\R$, that is continuous on the simplex $\mc X$, continuously differentiable in its interior, with gradient extendable by continuity to its boundary $\partial\mc X$, such that $\forall\,i,j\in\mc A$ and $\forall\,x\in\mc X$
\be\label{eq:potential}
r_j(x)-r_i(x)=\frac{\partial}{\partial x_j}\Phi(x)-\frac{\partial}{\partial x_i}\Phi(x).
\ee

If a game is potential, then NE are critical points of $\Phi$  \cite{Scutari2006}, while isolated local maxima of $\Phi$ form the set $\mc S$ \cite{Sandholm2010}. The same two relationships hold between critical points of $\Phi$ restricted to $\mc B$ and the equilibria of the restricted game.\medskip

\begin{example}[Congestion games]\label{ex:congestion}
An important class of potential games are congestion games \cite{Monderer1996,Rosenthal1973}. In their simplest formulation, we suppose that the reward of agents playing action $i\in \mc A$ only depends on $x_i$. Therefore, the reward vector functions have the form $r_i(x)=r_i(x_i)$, and \be\Phi(x)=\sum_{i\in\mc A}\Psi_i(x_i),\ee
is a potential of the game, where $\Psi_i$ is an anti-derivative of $r_i$. If the functions $r_i$, $i\in\mc A$, are monotone decreasing (e.g., negative rewards representing costs for the use of resources), the potential $\Phi(x)$ is concave and has a global maximum $\bar x$, that is the only NE of the game and it is also ESS.
\end{example}\medskip

\section{Stochastic Imitation Dynamics}\label{sec:imitation-dynamics}

Having defined the game, we introduce here the family of learning dynamics the players adopt in order to improve their reward. In this paper, we consider imitation dynamics, arising when individuals modify their actions in response to pairwise interactions  \cite{Hofbauer2003,cdc2017}. When two players meet, they can discuss about the action they are playing and the corresponding rewards. Then, depending on the difference between the two rewards and possibly other factors, a player either keeps playing the same action, or updates his/her action to the one of the contacted individual.

Each player is identified by a node on a graph $\mc G=(\mc V,\mc E)$, where the set of links $\mc E$ represents the network of interactions between players, in the sense that the presence of the link $(u,v)$ has to be interpreted as the possibility of a contact between player $u$ and player $v$. Given a player $u\in \mc V$, we define the set of player with whom $u$ can interact as the neighbors of $u$, 
that are the players that can influence the action played by $u$.

The update mechanism acts as follows. Each node is equipped with an independent Poisson clock with rate $\lambda$. When the clock associated with player $u$ clicks, the node gets activated, contacts a randomly chosen neighbor $v$, and updates his/her action according to the following probabilistic law. When a player $u$ that plays action $i$ contacts a player $v$ that uses action $j$, then $u$ updates his/her action to $j$ with a probability that depends only on the two actions $i$ and $j$ and on their difference in reward. In formula, we call this probability $f_{ij}(x)$, where, $\forall\,i,j\in\mc A$, the function $f_{ij}:\R^{\mc A}\to(0,1)$ is Lipschitz-continuous on $\mc X$ such that
\be\label{eq:sign}
\text{sgn}\left(f_{ij}(x)-f_{ji}(x)\right)=\text{sgn}\left(r_j(x)-r_i(x)\right),\,\forall\,x\in\mc X.
\ee

In plain words, if the reward for playing action $j$ is greater than the one for playing $i$, then the probability that a player updates his/her action from $i$ to $j$ is greater than the probability that a player flips its state from $j$ to $i$. Moreover, the probability that a player changes its action and copies the action of the contacted player is always greater than $0$. We present here two relevant examples of imitation dynamics.\medskip

\begin{example}[Replicator equation]\label{ex:replicator}
If we let $f_{ij}(x)\propto r_j(x),$ $\forall \,i,j\in\mc A\,,$
the imitation dynamics reduces to the replicator equation \cite{Hofbauer1998,Taylor1978, Schuster1983}. Hence, imitation dynamics encompass and generalize the replicator equation.
\end{example}\smallskip
\begin{example}\label{ex:atan}
A very general family of imitation dynamics can be defined using the following probabilities
\be\label{eq:atan} f_{ij}(x)=\frac{1}{2}+\frac{1}{\pi}\arctan(K_{ij}(r_j(x)-r_i(x)))\,,\ee
for $i,j\in\mc A$, where $K_{i,j}>0$. 
\end{example}\medskip

In this work we will consider the fully mixed case, i.e., when the graph $\mc G$ is a complete graph (with self-loops). Under this hypothesis, the neighbors of each player coincides with the whole set of nodes. In this case, the probability that a generic player $u$ contacts someone that plays action $i\in\mc A$ is proportional to the fraction of players playing action $i$ all over the population, i.e., $x_i$. Therefore, under this assumption, the whole state of the system is represented through the type of the system, instead of its configuration.

Let us consider the $m$-dimensional process $X(t)$ on $\mc X^{(n)}$, with $t\in\R^+$, following the time evolution of the type of the system. This stochastic process induced by the imitation dynamics is a Markov jump process, since the state updates are governed by independent Poisson clocks and the copying probabilities depend only on the type. The admissible transitions of the process are those in which one component $j$ of the type is increased by $1/n$ and another one $i$ is decreased by the same quantity, as an effect of an agent that changes its action from $i$ to $j$. We denote by $\lambda^{(n)}_{ij}$ the rate associated with this transition. This rate can be computed as follows. Since a fraction $x_i$ of players play action $i$, the activation rate of a player using action $i$ is $\lambda\cdot n x_i$. We multiply this rate by the probability that the contacted player uses action $j$, that is $x_j$, and, then, by the copying probability $f_{ij}(x)$. Therefore, we finally obtain, $\forall\,i,j\in\mc A$,
\be\label{eq:rate}
\lambda^{(n)}_{ij}(x)=n\lambda x_i x_jf_{ij}(x).
\ee

\begin{remark}\label{rem:deterministic} 
As already mentioned, many works \cite{Nachbar1990, Hofbauer2000, Sandholm2001, Sandholm2010,Panageas2016} study learning dynamics through deterministic processes $x(t)$, which approximates the evolution of the stochastic process $X(t)$ arbitrarily well, for large-scale populations and bounded times \cite{Kurtz1970,Kurtz1971}. This deterministic process is  the  solutions of the $m$-dimensional Cauchy problem
\be\label{eq:ode}\left\{\ba{l}
\dot x=\lambda\,\text{diag}(x)(F^T(x)-F(x))x\\[7pt]
x(0)=X(0),
\ea\right.\ee
where $F$ is a matrix function defined, entry-wise, as
$
F(x)_{ij}=f_{ij}(x),$ $\forall\,i,j\in\mc A$.

In \cite{cdc2017} we analyzed such deterministic imitation dynamics, proving that the solutions of \eqref{eq:ode} converge to NE of the game restricted to the support of the initial condition. Being the potential a global Lyapunov function \cite{Sandholm2010}, isolated local maxima of $\Phi$ are asymptotically stable equilibria of \eqref{eq:ode}. On the contrary, NE that are isolated local minima and saddle points are unstable. 
\end{remark}\smallskip

However, the deterministic approach in Remark \ref{rem:deterministic} fails in grasping many interesting aspects of the imitation dynamics, specifically concerning with their long-run behavior.

A first behavior that is not caught by the deterministic models is the asymptotic behavior of the system. If fact, $X(t)$ admits a set of absorbing states, coinciding with the pure configurations $\delta^{(i)}$, and, being all the functions $f_{ij}$ are always non null, the system is eventually absorbed in a pure configuration. However, as we will prove in the following, this absorbing event takes an enormous time to occur.

On the contrary, a more subtle and relevant aspect of the imitation dynamics that the deterministic approach fails to grasp is the different nature of the NE \eqref{eq:ode} converges to. The following example gives us an intuition of this issue, which will be deeply discussed and analyzed in the next section.\medskip

\begin{example}\label{ex:example}
\begin{figure}
\centering
\subfloat[Reward functions]{\definecolor{mycolor1}{rgb}{0.00000,0.44700,0.74100}%
\definecolor{mycolor2}{rgb}{0.85000,0.32500,0.09800}%
\begin{tikzpicture}

\begin{axis}[%
 axis lines=middle,
 x   axis line style={-},
y   axis line style={->},
ytick style={draw=none},
ymajorticks=false,
width=\l cm,
height=\h cm,
at={(0cm,0cm)},
scale only axis,
xmin=0,
xmax=1,
xlabel={$x_i$},
ymin=0,
ymax=1,
ylabel={$r_i(x_i)$},
axis background/.style={fill=white},
every axis x label/.style={
    at={(.78,.09)},
    anchor=west,
},
every axis y label/.style={
    at={(0,.8)},
    anchor=west,
},
]
\addplot [color=mycolor1,solid,forget plot,thick]
  table[row sep=crcr]{%
0	0.66666875\\
0.0050251256281407	0.658963571031371\\
0.0100502512562814	0.651480148987564\\
0.0150753768844221	0.644215776764717\\
0.0201005025125628	0.637167747258965\\
0.0251256281407035	0.630333353366445\\
0.0301507537688442	0.623709887983293\\
0.0351758793969849	0.617294644005646\\
0.0402010050251256	0.61108491432964\\
0.0452261306532663	0.605077991851413\\
0.050251256281407	0.599271169467099\\
0.0552763819095477	0.593661740072836\\
0.0603015075376884	0.58824699656476\\
0.0653266331658292	0.583024231839007\\
0.0703517587939698	0.577990738791715\\
0.0753768844221105	0.573143810319019\\
0.0804020100502513	0.568480739317056\\
0.085427135678392	0.563998818681962\\
0.0904522613065327	0.559695341309874\\
0.0954773869346734	0.555567600096928\\
0.100502512562814	0.551612887939261\\
0.105527638190955	0.547828497733009\\
0.110552763819095	0.544211722374308\\
0.115577889447236	0.540759854759296\\
0.120603015075377	0.537470187784107\\
0.125628140703518	0.53434001434488\\
0.130653266331658	0.531366627337751\\
0.135678391959799	0.528547319658855\\
0.14070351758794	0.525879384204329\\
0.14572864321608	0.52336011387031\\
0.150753768844221	0.520986801552934\\
0.155778894472362	0.518756740148338\\
0.160804020100503	0.516667222552658\\
0.165829145728643	0.51471554166203\\
0.170854271356784	0.512898990372591\\
0.175879396984925	0.511214861580478\\
0.180904522613065	0.509660448181826\\
0.185929648241206	0.508233043072773\\
0.190954773869347	0.506929939149454\\
0.195979899497487	0.505748429308007\\
0.201005025125628	0.504685806444567\\
0.206030150753769	0.503739363455272\\
0.21105527638191	0.502906393236257\\
0.21608040201005	0.502184188683658\\
0.221105527638191	0.501570042693614\\
0.226130653266332	0.501061248162259\\
0.231155778894472	0.50065509798573\\
0.236180904522613	0.500348885060165\\
0.241206030150754	0.500139902281698\\
0.246231155778894	0.500025442546468\\
0.251256281407035	0.500002798750609\\
0.256281407035176	0.500069263790259\\
0.261306532663317	0.500222130561554\\
0.266331658291457	0.50045869196063\\
0.271356783919598	0.500776240883624\\
0.276381909547739	0.501172070226673\\
0.281407035175879	0.501643472885913\\
0.28643216080402	0.502187741757479\\
0.291457286432161	0.50280216973751\\
0.296482412060302	0.50348404972214\\
0.301507537688442	0.504230674607508\\
0.306532663316583	0.505039337289748\\
0.311557788944724	0.505907330664998\\
0.316582914572864	0.506831947629393\\
0.321608040201005	0.507810481079072\\
0.326633165829146	0.508840223910168\\
0.331658291457286	0.509918469018821\\
0.336683417085427	0.511042509301165\\
0.341708542713568	0.512209637653337\\
0.346733668341709	0.513417146971474\\
0.351758793969849	0.514662330151712\\
0.35678391959799	0.515942480090187\\
0.361809045226131	0.517254889683037\\
0.366834170854271	0.518596851826397\\
0.371859296482412	0.519965659416403\\
0.376884422110553	0.521358605349194\\
0.381909547738693	0.522772982520904\\
0.386934673366834	0.52420608382767\\
0.391959798994975	0.525655202165629\\
0.396984924623116	0.527117630430917\\
0.402010050251256	0.52859066151967\\
0.407035175879397	0.530071588328026\\
0.412060301507538	0.53155770375212\\
0.417085427135678	0.533046300688089\\
0.422110552763819	0.534534672032069\\
0.42713567839196	0.536020110680197\\
0.4321608040201	0.537499909528609\\
0.437185929648241	0.538971361473443\\
0.442211055276382	0.540431759410833\\
0.447236180904523	0.541878396236917\\
0.452261306532663	0.543308564847831\\
0.457286432160804	0.544719558139711\\
0.462311557788945	0.546108669008695\\
0.467336683417085	0.547473190350917\\
0.472361809045226	0.548810415062516\\
0.477386934673367	0.550117636039627\\
0.482412060301508	0.551392146178387\\
0.487437185929648	0.552631238374932\\
0.492462311557789	0.553832205525398\\
0.49748743718593	0.554992340525923\\
0.50251256281407	0.556108936272642\\
0.507537688442211	0.557179285661693\\
0.512562814070352	0.55820068158921\\
0.517587939698492	0.559170416951332\\
0.522613065326633	0.560085784644194\\
0.527638190954774	0.560944077563933\\
0.532663316582915	0.561742588606685\\
0.537688442211055	0.562478610668586\\
0.542713567839196	0.563149436645774\\
0.547738693467337	0.563752359434384\\
0.552763819095477	0.564284671930554\\
0.557788944723618	0.564743667030419\\
0.562814070351759	0.565126637630115\\
0.5678391959799	0.565430876625781\\
0.57286432160804	0.565653676913551\\
0.577889447236181	0.565792331389562\\
0.582914572864322	0.565844132949951\\
0.587939698492462	0.565806374490854\\
0.592964824120603	0.565676348908408\\
0.597989949748744	0.565451349098749\\
0.603015075376884	0.565128667958013\\
0.608040201005025	0.564705598382337\\
0.613065326633166	0.564179433267858\\
0.618090452261307	0.563547465510712\\
0.623115577889447	0.562806988007035\\
0.628140703517588	0.561955293652963\\
0.633165829145729	0.560989675344634\\
0.638190954773869	0.559907425978184\\
0.64321608040201	0.558705838449749\\
0.648241206030151	0.557382205655465\\
0.653266331658292	0.555933820491469\\
0.658291457286432	0.554357975853898\\
0.663316582914573	0.552651964638887\\
0.668341708542714	0.550813079742574\\
0.673366834170854	0.548838614061095\\
0.678391959798995	0.546725860490586\\
0.683417085427136	0.544472111927183\\
0.688442211055276	0.542074661267024\\
0.693467336683417	0.539530801406245\\
0.698492462311558	0.536837825240981\\
0.703517587939699	0.53399302566737\\
0.708542713567839	0.530993695581548\\
0.71356783919598	0.527837127879651\\
0.718592964824121	0.524520615457816\\
0.723618090452261	0.521041451212179\\
0.728643216080402	0.517396928038877\\
0.733668341708543	0.513584338834047\\
0.738693467336683	0.509600976493824\\
0.743718592964824	0.505444133914345\\
0.748743718592965	0.501111103991746\\
0.753768844221106	0.496599179622164\\
0.758793969849246	0.491905653701736\\
0.763819095477387	0.487027819126598\\
0.768844221105528	0.481962968792886\\
0.773869346733668	0.476708395596737\\
0.778894472361809	0.471261392434287\\
0.78391959798995	0.465619252201673\\
0.78894472361809	0.459779267795031\\
0.793969849246231	0.453738732110497\\
0.798994974874372	0.447494938044208\\
0.804020100502513	0.441045178492301\\
0.809045226130653	0.434386746350912\\
0.814070351758794	0.427516934516177\\
0.819095477386935	0.420433035884233\\
0.824120603015075	0.413132343351216\\
0.829145728643216	0.405612149813263\\
0.834170854271357	0.39786974816651\\
0.839195979899497	0.389902431307094\\
0.844221105527638	0.38170749213115\\
0.849246231155779	0.373282223534816\\
0.85427135678392	0.364623918414228\\
0.85929648241206	0.355729869665523\\
0.864321608040201	0.346597370184836\\
0.869346733668342	0.337223712868305\\
0.874371859296482	0.327606190612065\\
0.879396984924623	0.317742096312254\\
0.884422110552764	0.307628722865007\\
0.889447236180904	0.297263363166461\\
0.894472361809045	0.286643310112753\\
0.899497487437186	0.275765856600019\\
0.904522613065327	0.264628295524395\\
0.909547738693467	0.253227919782018\\
0.914572864321608	0.241562022269024\\
0.919597989949749	0.22962789588155\\
0.924623115577889	0.217422833515733\\
0.92964824120603	0.204944128067708\\
0.934673366834171	0.192189072433612\\
0.939698492462312	0.179154959509582\\
0.944723618090452	0.165839082191753\\
0.949748743718593	0.152238733376264\\
0.954773869346734	0.138351205959249\\
0.959798994974874	0.124173792836845\\
0.964824120603015	0.10970378690519\\
0.969849246231156	0.0949384810604182\\
0.974874371859296	0.0798751681986674\\
0.979899497487437	0.0645111412160738\\
0.984924623115578	0.0488436930087736\\
0.989949748743719	0.032870116472904\\
0.994974874371859	0.0165877045046005\\
1	0\\
};

\addplot [color=red,dashed,forget plot,thick]
  table[row sep=crcr]{%
0	0.5\\
1	0.5\\
};

\end{axis}
\end{tikzpicture}}\qquad\subfloat[Potential]{\definecolor{mycolor1}{rgb}{0.00000,0.44700,0.74100}%
\definecolor{mycolor2}{rgb}{0.85000,0.32500,0.09800}%
\begin{tikzpicture}

\begin{axis}[%
 axis lines=middle,
 x   axis line style={-},
y   axis line style={->},
ytick style={draw=none},
ymajorticks=false,
width=\l cm,
height=\h cm,
at={(0cm,0cm)},
scale only axis,
xmin=0,
xmax=1,
xlabel={$x_1$},
ymin=0,
ymax=1,
ylabel={$\Phi(x_1)$},
axis background/.style={fill=white},
every axis x label/.style={
    at={(.78,.09)},
    anchor=west,
},
every axis y label/.style={
    at={(0,.8)},
    anchor=west,
},
]
\addplot [color=mycolor1,solid,forget plot,thick]
  table[row sep=crcr]{%
0	0.333333333333333\\
0.0050251256281407	0.348058557426661\\
0.0100502512562814	0.362096888857624\\
0.0150753768844221	0.375468263360503\\
0.0201005025125628	0.388192371808974\\
0.0251256281407035	0.400288660216106\\
0.0301507537688442	0.411776329734365\\
0.0351758793969849	0.422674336655611\\
0.0402010050251256	0.433001392411097\\
0.0452261306532663	0.442775963571474\\
0.050251256281407	0.452016271846784\\
0.0552763819095477	0.460740294086467\\
0.0603015075376884	0.468965762279356\\
0.0653266331658292	0.476710163553678\\
0.0703517587939698	0.483990740177057\\
0.0753768844221105	0.490824489556509\\
0.0804020100502513	0.497228164238448\\
0.085427135678392	0.503218271908678\\
0.0904522613065327	0.508811075392403\\
0.0954773869346734	0.514022592654218\\
0.100502512562814	0.518868596798115\\
0.105527638190955	0.523364616067478\\
0.110552763819095	0.527525933845088\\
0.115577889447236	0.531367588653121\\
0.120603015075377	0.534904374153145\\
0.125628140703518	0.538150839146126\\
0.130653266331658	0.541121287572422\\
0.135678391959799	0.543829778511788\\
0.14070351758794	0.546290126183372\\
0.14572864321608	0.548515899945717\\
0.150753768844221	0.550520424296761\\
0.155778894472362	0.552316778873837\\
0.160804020100503	0.553917798453673\\
0.165829145728643	0.55533607295239\\
0.170854271356784	0.556583947425505\\
0.175879396984925	0.55767352206793\\
0.180904522613065	0.558616652213971\\
0.185929648241206	0.559424948337329\\
0.190954773869347	0.5601097760511\\
0.195979899497487	0.560682256107774\\
0.201005025125628	0.561153264399236\\
0.206030150753769	0.561533431956766\\
0.21105527638191	0.561833144951038\\
0.21608040201005	0.562062544692122\\
0.221105527638191	0.562231527629481\\
0.226130653266332	0.562349745351974\\
0.231155778894472	0.562426604587855\\
0.236180904522613	0.56247126720477\\
0.241206030150754	0.562492650209764\\
0.246231155778894	0.562499425749274\\
0.251256281407035	0.562500021109131\\
0.256281407035176	0.562502618714563\\
0.261306532663317	0.56251515613019\\
0.266331658291457	0.562545326060031\\
0.271356783919598	0.562600576347494\\
0.276381909547739	0.562688109975386\\
0.281407035175879	0.562814885065908\\
0.28643216080402	0.562987614880655\\
0.291457286432161	0.563212767820615\\
0.296482412060302	0.563496567426174\\
0.301507537688442	0.563844992377112\\
0.306532663316583	0.564263776492602\\
0.311557788944724	0.564758408731211\\
0.316582914572864	0.565334133190906\\
0.321608040201005	0.565995949109042\\
0.326633165829146	0.566748610862372\\
0.331658291457286	0.567596627967045\\
0.336683417085427	0.568544265078603\\
0.341708542713568	0.569595541991981\\
0.346733668341709	0.570754233641513\\
0.351758793969849	0.572023870100923\\
0.35678391959799	0.573407736583334\\
0.361809045226131	0.574908873441261\\
0.366834170854271	0.576530076166614\\
0.371859296482412	0.578273895390698\\
0.376884422110553	0.580142636884215\\
0.381909547738693	0.582138361557256\\
0.386934673366834	0.584262885459312\\
0.391959798994975	0.586517779779268\\
0.396984924623116	0.588904370845401\\
0.402010050251256	0.591423740125386\\
0.407035175879397	0.594076724226289\\
0.412060301507538	0.596863914894574\\
0.417085427135678	0.599785659016099\\
0.422110552763819	0.602842058616116\\
0.42713567839196	0.606032970859272\\
0.4321608040201	0.609358008049606\\
0.437185929648241	0.612816537630558\\
0.442211055276382	0.616407682184958\\
0.447236180904523	0.62013031943503\\
0.452261306532663	0.623983082242397\\
0.457286432160804	0.627964358608072\\
0.462311557788945	0.632072291672466\\
0.467336683417085	0.636304779715383\\
0.472361809045226	0.640659476156023\\
0.477386934673367	0.645133789552979\\
0.482412060301508	0.649724883604241\\
0.487437185929648	0.654429677147192\\
0.492462311557789	0.659244844158609\\
0.49748743718593	0.664166813754666\\
0.50251256281407	0.66919177019093\\
0.507537688442211	0.674315652862364\\
0.512562814070352	0.679534156303324\\
0.517587939698492	0.684842730187561\\
0.522613065326633	0.690236579328224\\
0.527638190954774	0.695710663677851\\
0.532663316582915	0.701259698328381\\
0.537688442211055	0.706878153511141\\
0.542713567839196	0.712560254596857\\
0.547738693467337	0.718299982095653\\
0.552763819095477	0.724091071657037\\
0.557788944723618	0.729927014069925\\
0.562814070351759	0.735801055262615\\
0.5678391959799	0.741706196302809\\
0.57286432160804	0.747635193397601\\
0.577889447236181	0.753580557893476\\
0.582914572864322	0.759534556276321\\
0.587939698492462	0.76548921017141\\
0.592964824120603	0.771436296343416\\
0.597989949748744	0.777367346696409\\
0.603015075376884	0.783273648273847\\
0.608040201005025	0.789146243258589\\
0.613065326633166	0.794975928972883\\
0.618090452261307	0.800753257878377\\
0.623115577889447	0.806468537576111\\
0.628140703517588	0.812111830806521\\
0.633165829145729	0.817672955449438\\
0.638190954773869	0.823141484524081\\
0.64321608040201	0.828506746189078\\
0.648241206030151	0.833757823742433\\
0.653266331658292	0.838883555621564\\
0.658291457286432	0.843872535403267\\
0.663316582914573	0.848713111803748\\
0.668341708542714	0.853393388678594\\
0.673366834170854	0.857901225022795\\
0.678391959798995	0.86222423497073\\
0.683417085427136	0.866349787796179\\
0.688442211055276	0.870265007912314\\
0.693467336683417	0.8739567748717\\
0.698492462311558	0.877411723366301\\
0.703517587939699	0.880616243227468\\
0.708542713567839	0.883556479425955\\
0.71356783919598	0.886218332071907\\
0.718592964824121	0.888587456414864\\
0.723618090452261	0.89064926284376\\
0.728643216080402	0.892388916886921\\
0.733668341708543	0.893791339212077\\
0.738693467336683	0.894841205626342\\
0.743718592964824	0.895522947076234\\
0.748743718592965	0.895820749647659\\
0.753768844221106	0.895718554565922\\
0.758793969849246	0.895200058195716\\
0.763819095477387	0.894248712041132\\
0.768844221105528	0.892847722745667\\
0.773869346733668	0.890980052092193\\
0.778894472361809	0.88862841700299\\
0.78391959798995	0.885775289539734\\
0.78894472361809	0.882402896903481\\
0.793969849246231	0.878493221434699\\
0.798994974874372	0.874028000613238\\
0.804020100502513	0.868988727058356\\
0.809045226130653	0.863356648528686\\
0.814070351758794	0.85711276792228\\
0.819095477386935	0.850237843276561\\
0.824120603015075	0.842712387768366\\
0.829145728643216	0.834516669713915\\
0.834170854271357	0.825630712568828\\
0.839195979899497	0.816034294928113\\
0.844221105527638	0.805706950526188\\
0.849246231155779	0.79462796823684\\
0.85427135678392	0.782776392073285\\
0.85929648241206	0.770131021188093\\
0.864321608040201	0.756670409873267\\
0.869346733668342	0.742372867560186\\
0.874371859296482	0.727216458819622\\
0.879396984924623	0.711179003361744\\
0.884422110552764	0.694238076036124\\
0.889447236180904	0.676371006831714\\
0.894472361809045	0.657554880876874\\
0.899497487437186	0.637766538439354\\
0.904522613065327	0.616982574926297\\
0.909547738693467	0.595179340884237\\
0.914572864321608	0.572332941999111\\
0.919597989949749	0.548419239096251\\
0.924623115577889	0.523413848140378\\
0.92964824120603	0.497292140235604\\
0.934673366834171	0.470029241625452\\
0.939698492462312	0.441600033692821\\
0.944723618090452	0.411979152960013\\
0.949748743718593	0.381140991088726\\
0.954773869346734	0.349059694880052\\
0.959798994974874	0.315709166274482\\
0.964824120603015	0.281063062351884\\
0.969849246231156	0.24509479533154\\
0.974874371859296	0.207777532572129\\
0.979899497487437	0.169084196571701\\
0.984924623115578	0.128987464967725\\
0.989949748743719	0.0874597705370468\\
0.994974874371859	0.0444733011959276\\
1	1.16573417585641e-15\\
};

\addplot[only marks,mark=o,mark options={scale=1.2},text mark as node=true,thick] coordinates {(0,0.3333333)};
\addplot[only marks,mark=x,mark options={scale=1.4},text mark as node=true,black,thick] coordinates {(0.25,0.5625)};
\addplot[only marks,mark=*,mark options={scale=1.2},text mark as node=true,black,thick] coordinates {(0.75,0.8958)};
\addplot[only marks,mark=o,mark options={scale=1.2},text mark as node=true,black,thick] coordinates {(1,0)};

\end{axis}
\end{tikzpicture}}
\caption{Reward functions $r_1$ (blue solid) and $r_2$ (red dashed) and potential $\Phi$ of the game in Example \ref{ex:example}. Empty circles correspond to local minima, filled cycles to local maxima (i.e., ESS), crosses to saddle points.}
\label{fig:potential}
\end{figure}
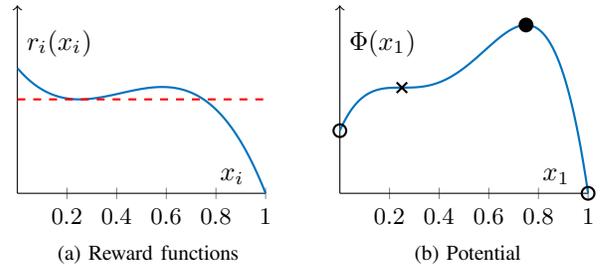

We consider a $2$-action congestion game (Example \ref{ex:congestion}) with reward functions
\be
r_1(x_1)=9-\left(4x_1-3\right)\left(4x_1-1\right)^2,\quad r_2(x_2)=9,
\ee
resulting into the potential
\be
\Phi(x)=-16x_1^4 + \frac{80}{3} x_1^3 - 14 x_1^2 + 3 x_1+9.
\ee
Figure \ref{fig:potential} depicts the two reward functions and the potential.

The system can be reduced to a one-dimensional system, since $x_2=1-x_1$, allowing for a complete analysis of the system. Critical points of the potential can be computed: they are the two local minima coinciding with the pure configurations $\delta^{(1)}$ and $\delta^{(2)}$, the saddle point in $A=(1/4,3/4)$, and the global maximum in $B=(3/4,1/4)$. Therefore, $\mc N=\{A,B\}$ and $\mc S=\{B\}$. Without any loss in generality, we consider the imitation dynamics presented in Example \ref{ex:atan} and we notice that the deterministic approximation obtained using \eqref{eq:ode} converges to $A$ if $x_1(0)<1/4$, or to $B$ otherwise. 

From the Monte Carlo simulations of the stochastic process in  Fig. \ref{fig:simulations} (with $\lambda=1$), we appreciate a different behavior of the system close to the two equilibria: if the system converges close to $A$, it spends there a (reasonably) short amount of time, then it converges close to $B$. Once the system is close to $B$, it spends there a very long time. In fact, the absorbing event is not seen in the time-horizons of our simulations.

\begin{figure}
\centering
\subfloat[$X_1(0)=0.001$, $n=2500$]{\begin{tikzpicture}
\definecolor{mycolor1}{rgb}{0.00000,0.44700,0.74100}%
\begin{axis}[
 axis lines=middle,
 x   axis line style={->},
y   axis line style={-},
    width=\l cm,
height=\h cm,
at={(0cm,0cm)},
scale only axis,
xmin=0,
xtick={0,2000,4000},
xmax=5500,
xlabel={$t$},
ymin=0,
ymax=1,
ylabel={$X_1(t)$},
axis background/.style={fill=white},
]

\addplot [color=mycolor1,solid,forget plot,thick]
  table[row sep=crcr]{%
0	0.004\\
60.7693828773781	0.0856\\
80.7044512752223	0.108\\
97.0346477858789	0.1248\\
111.731272534268	0.1528\\
125.867578122363	0.1472\\
139.35311977837	0.1768\\
152.469305161669	0.164\\
164.435871873536	0.196\\
175.332723304136	0.2088\\
187.28514017889	0.1968\\
198.80763453316	0.178\\
210.77933405002	0.1956\\
222.638975262792	0.1988\\
233.964901659787	0.2092\\
244.21408447747	0.2244\\
255.274049113214	0.2012\\
266.422863025913	0.1988\\
278.386411951336	0.1708\\
291.221683926357	0.1484\\
304.593378600007	0.1692\\
317.151529961675	0.1804\\
329.425294138566	0.166\\
341.807246196766	0.1724\\
354.368065037407	0.1828\\
365.640997275596	0.2012\\
376.700571678271	0.2236\\
386.556664275928	0.2372\\
396.551200413149	0.2404\\
406.521377153278	0.2508\\
415.947795457469	0.27\\
425.533835756194	0.2492\\
434.979952386162	0.2688\\
443.761441416298	0.2912\\
452.578360931963	0.2632\\
461.88324293186	0.2488\\
471.82540614848	0.2544\\
481.979137685488	0.2256\\
492.244912417452	0.2296\\
503.439537463204	0.184\\
515.16942025466	0.192\\
527.26583254379	0.1776\\
539.164912764238	0.2064\\
550.493664360016	0.204\\
561.679530914338	0.2008\\
572.55583573194	0.216\\
582.993576234165	0.2064\\
593.861114623631	0.1912\\
605.38225301461	0.1816\\
617.50215852902	0.208\\
628.046632039751	0.2296\\
637.987987368359	0.2448\\
647.922792891109	0.2492\\
657.805229400298	0.2356\\
667.890730067778	0.2356\\
677.749459824166	0.2444\\
687.786856295607	0.1924\\
698.802801167784	0.2172\\
709.441126212851	0.2028\\
720.287561819002	0.218\\
730.855501676792	0.2268\\
740.912354676792	0.2268\\
751.925778320293	0.2044\\
764.043658269254	0.1756\\
777.070069688865	0.1716\\
790.80954428985	0.1564\\
803.544803734018	0.1636\\
816.011080012552	0.1692\\
827.959455922747	0.1852\\
840.022888359522	0.1924\\
851.795391034474	0.1844\\
863.986860334233	0.182\\
876.06502245213	0.1652\\
888.245745551441	0.184\\
900.284116727974	0.1912\\
913.118526796775	0.1688\\
925.751380437906	0.1632\\
938.295599395758	0.196\\
949.547640819344	0.2112\\
960.05344266748	0.2408\\
970.516200980785	0.204\\
981.456972683519	0.2064\\
993.134368524105	0.1936\\
1004.02180020123	0.212\\
1015.3022401125	0.1896\\
1027.6091673882	0.164\\
1040.8967594572	0.148\\
1056.01719245066	0.1424\\
1070.66641850864	0.1608\\
1084.57866249792	0.1656\\
1096.66040134681	0.192\\
1108.44079597622	0.1744\\
1119.81022656882	0.196\\
1131.21980352879	0.1928\\
1143.44189423601	0.202\\
1153.99082161985	0.2188\\
1164.98932426625	0.206\\
1175.40092725334	0.2364\\
1185.84854194738	0.2204\\
1195.93230585726	0.2148\\
1206.66468912902	0.2076\\
1217.71497512748	0.1988\\
1229.06609537761	0.198\\
1239.81117359294	0.2196\\
1250.11859661865	0.2316\\
1260.19395326558	0.2412\\
1270.44036244796	0.2084\\
1280.55027202214	0.2428\\
1290.77967330103	0.2364\\
1300.68046840079	0.2492\\
1309.99780608118	0.238\\
1320.34440146553	0.234\\
1330.53702305015	0.222\\
1340.60956817852	0.2532\\
1349.7151669141	0.2868\\
1358.05960591453	0.2952\\
1366.61022988741	0.3104\\
1375.19395612399	0.2968\\
1383.60648005758	0.3312\\
1391.72773759641	0.324\\
1399.39030185863	0.3552\\
1407.27518535516	0.3464\\
1415.17805331315	0.3544\\
1422.95089569481	0.356\\
1430.99896401239	0.3464\\
1439.06647445683	0.356\\
1446.61134255475	0.3832\\
1454.29803856433	0.376\\
1461.93447964443	0.3832\\
1469.62599953906	0.384\\
1477.06421415245	0.4288\\
1484.41986281046	0.492\\
1491.83262043631	0.5016\\
1499.13954961204	0.5232\\
1506.43872509739	0.5656\\
1514.13307610226	0.6096\\
1521.6998938428	0.618\\
1529.48241101632	0.6668\\
1537.71248221159	0.6692\\
1545.41490225175	0.6612\\
1553.84686310677	0.706\\
1562.76826534744	0.7228\\
1571.58846912277	0.7236\\
1580.49448806672	0.7028\\
1589.46440681885	0.73\\
1598.90854375509	0.742\\
1608.88413408053	0.7788\\
1618.93602414486	0.7724\\
1628.98060114329	0.7372\\
1638.1809953153	0.726\\
1647.60827068087	0.7396\\
1657.2674417439	0.742\\
1666.28338112441	0.7412\\
1675.4101689652	0.7444\\
1685.00189866218	0.7308\\
1694.77598989434	0.7612\\
1704.43636190417	0.7544\\
1713.75048960821	0.7328\\
1723.19617015601	0.7448\\
1732.4794959413	0.7248\\
1741.70669889258	0.7368\\
1751.45370276735	0.7624\\
1761.47979953322	0.7528\\
1771.74205618538	0.7632\\
1781.74340334378	0.7664\\
1791.66876933626	0.756\\
1801.64354119825	0.7632\\
1812.00352695414	0.7576\\
1822.18746273455	0.7808\\
1833.33134456518	0.7736\\
1843.5252023833	0.7688\\
1853.38943656456	0.7656\\
1863.92334414913	0.7824\\
1874.05227575416	0.7688\\
1883.94011083958	0.7488\\
1893.44154565335	0.7496\\
1902.52893010707	0.6992\\
1910.85443756771	0.7124\\
1919.62230154574	0.7284\\
1929.03152391268	0.738\\
1938.8394221247	0.7524\\
1948.81520515966	0.7492\\
1957.89373417197	0.7404\\
1967.24722296203	0.7468\\
1977.02520784664	0.7732\\
1987.65007467933	0.77\\
1997.97220004445	0.7476\\
2007.64936860755	0.7484\\
2017.39225821074	0.758\\
2026.59855536766	0.7276\\
2035.56783163546	0.7316\\
2044.8289263651	0.7492\\
2054.19512125094	0.7276\\
2063.07093217709	0.7436\\
2072.92040877937	0.7508\\
2082.61793917178	0.742\\
2092.08241798933	0.7372\\
2101.46237760923	0.742\\
2110.73512738706	0.7432\\
2119.64008313988	0.72\\
2128.86948780438	0.7128\\
2137.25933354002	0.6848\\
2145.62422336299	0.712\\
2154.63906935654	0.7056\\
2163.708320266	0.736\\
2172.69732956627	0.7216\\
2181.64230958425	0.7216\\
2190.77339729255	0.736\\
2200.36445276602	0.7456\\
2209.86156393621	0.748\\
2219.46229792166	0.7224\\
2228.71085481837	0.7376\\
2237.76182672535	0.72\\
2246.84624563609	0.7176\\
2256.02887530515	0.7352\\
2265.11821545758	0.732\\
2274.55660252013	0.7296\\
2283.65755420431	0.7224\\
2292.5465146556	0.7376\\
2302.50427665354	0.7668\\
2312.19304035837	0.7588\\
2322.5589701948	0.7692\\
2332.51024376725	0.7396\\
2342.13982302079	0.7276\\
2351.54819513237	0.7356\\
2361.23344392887	0.7244\\
2370.15597011624	0.7364\\
2379.6802948172	0.758\\
2389.72100398357	0.7572\\
2399.5863032008	0.77\\
2409.98592940556	0.7684\\
2419.40164684714	0.7484\\
2428.82527377667	0.7508\\
2438.35549520807	0.7388\\
2447.6775801905	0.7308\\
2457.08659814156	0.7476\\
2466.46883705408	0.746\\
2476.22003237043	0.746\\
2485.22766612747	0.7148\\
2494.59789816955	0.7308\\
2504.13995681829	0.7344\\
2513.70629291139	0.7512\\
2523.72371182011	0.7512\\
2533.29365146611	0.7576\\
2543.41899822334	0.7488\\
2552.68196262008	0.7248\\
2562.06846426838	0.7336\\
2571.49844638121	0.7352\\
2580.14756927398	0.712\\
2588.94680745545	0.712\\
2597.92761387889	0.7392\\
2606.79792581094	0.7488\\
2616.46429999496	0.7368\\
2626.07186263291	0.7544\\
2635.54910697096	0.732\\
2644.31374981632	0.7152\\
2653.47143628191	0.7344\\
2662.96865881601	0.7456\\
2672.84638956662	0.7568\\
2682.60083023453	0.7712\\
2692.43962454286	0.7348\\
2702.05464988132	0.7428\\
2711.53487126663	0.7636\\
2721.55829787085	0.7516\\
2731.19126249522	0.7508\\
2741.23065254842	0.75\\
2750.78449940313	0.7524\\
2759.97056556925	0.726\\
2769.67673057604	0.7516\\
2779.13288589784	0.7476\\
2789.06126730799	0.754\\
2798.78464902152	0.7452\\
2808.22046361656	0.7348\\
2816.73622793094	0.6812\\
2824.88889704275	0.6924\\
2833.50923230977	0.7068\\
2842.61214486689	0.7268\\
2852.12006697353	0.742\\
2861.55349403667	0.7324\\
2871.14666746843	0.7476\\
2880.37970111892	0.7444\\
2890.82050237983	0.7888\\
2901.03846908839	0.7568\\
2910.65433156508	0.7656\\
2921.03639101862	0.7792\\
2930.79345326875	0.7632\\
2940.56872744022	0.78\\
2951.10231971647	0.7568\\
2960.76280046185	0.7408\\
2971.07751238539	0.7808\\
2981.32368874615	0.7488\\
2991.20132458997	0.7792\\
3001.90298032857	0.7808\\
3012.52703157925	0.7552\\
3022.11151682108	0.732\\
3031.45627371001	0.752\\
3041.09625058587	0.7488\\
3051.50580944121	0.7464\\
3060.81215782683	0.7328\\
3070.27839924882	0.752\\
3079.89916139836	0.732\\
3089.15525383553	0.7352\\
3097.95535924597	0.7164\\
3106.68215043579	0.7084\\
3115.5547641783	0.7116\\
3124.17751348586	0.718\\
3133.37232689281	0.7396\\
3143.22253623853	0.7388\\
3152.8943199799	0.7284\\
3162.24718218967	0.7548\\
3171.75656520119	0.7676\\
3181.0761319524	0.7708\\
3191.25927802721	0.75\\
3201.09886278768	0.7388\\
3210.52550284399	0.7212\\
3219.57683735924	0.7212\\
3228.5987543788	0.742\\
3238.36785145622	0.7676\\
3249.01195892414	0.77\\
3259.27909703235	0.754\\
3269.75176056451	0.7788\\
3279.55597766324	0.754\\
3289.63966969016	0.7436\\
3298.93474197684	0.7496\\
3308.87430301047	0.7616\\
3318.83900410489	0.7656\\
3329.08662487105	0.768\\
3339.36835446376	0.7712\\
3349.46886071323	0.7576\\
3359.32468825851	0.7712\\
3370.12346498579	0.792\\
3380.68308712209	0.772\\
3390.6939621611	0.7664\\
3400.32360122065	0.7536\\
3409.88417500555	0.7392\\
3419.41154203698	0.7696\\
3429.19949396029	0.7376\\
3438.63517942938	0.732\\
3448.00492060671	0.7624\\
3457.96289799439	0.7472\\
3467.59905998813	0.7272\\
3476.60492965822	0.7304\\
3485.54002405943	0.7072\\
3494.18366858572	0.72\\
3503.61515614654	0.742\\
3513.28811073367	0.7484\\
3522.60484284189	0.7588\\
3532.73249276703	0.7828\\
3543.20920503704	0.754\\
3552.53808422675	0.7332\\
3561.93833859272	0.7572\\
3571.66240057377	0.7532\\
3581.05642578931	0.7452\\
3590.81189534214	0.7452\\
3599.99939603994	0.7692\\
3609.95840787776	0.746\\
3619.31433535549	0.7444\\
3628.93301718023	0.7572\\
3638.80535351556	0.7468\\
3648.25224448726	0.7348\\
3658.00764515243	0.7676\\
3667.74865051905	0.7572\\
3677.42534914682	0.7508\\
3687.33216058834	0.7612\\
3697.51118960523	0.776\\
3707.41590023674	0.7464\\
3717.00719754426	0.7376\\
3726.51586320884	0.728\\
3735.92393465107	0.7432\\
3745.87776319964	0.7528\\
3755.94124439579	0.7776\\
3766.02871585642	0.7488\\
3775.10694992298	0.7216\\
3784.82090766572	0.7296\\
3794.02436637027	0.7424\\
3803.65872396058	0.7648\\
3813.70055229197	0.7616\\
3823.38410190756	0.7392\\
3832.76292605074	0.7504\\
3842.31105972765	0.732\\
3852.32216731687	0.7784\\
3862.18338078014	0.7584\\
3871.6484145449	0.7576\\
3880.77574000521	0.7136\\
3889.6524169056	0.736\\
3899.35378590849	0.7596\\
3909.68842132431	0.7604\\
3919.51001212902	0.7596\\
3928.96769039018	0.7364\\
3938.02506252607	0.734\\
3947.18750175054	0.7524\\
3956.51001716008	0.7172\\
3965.37847210732	0.7204\\
3974.08927727893	0.7188\\
3983.12606933872	0.698\\
3991.34974420969	0.6812\\
3999.86409383787	0.6868\\
4008.48462584597	0.706\\
4017.92662282175	0.754\\
4027.75670669686	0.7596\\
4037.88549015944	0.7676\\
4047.97879901652	0.7492\\
4057.73212807123	0.7628\\
4067.74960866027	0.7668\\
4077.84788873332	0.7708\\
4087.60362445321	0.7388\\
4096.96407478368	0.7456\\
4106.38171232504	0.744\\
4115.81300074781	0.7272\\
4125.1668477579	0.7384\\
4134.09094814762	0.7456\\
4143.04503252358	0.7248\\
4152.52255150481	0.7352\\
4161.89517695491	0.7576\\
4171.03696160906	0.7248\\
4180.42829520979	0.7472\\
4190.63756937387	0.7672\\
4201.37400065109	0.784\\
4211.45734922429	0.7712\\
4221.39280457053	0.7616\\
4230.67211562843	0.7344\\
4239.47263986817	0.7488\\
4248.68309682618	0.7216\\
4257.75495960079	0.7264\\
4266.19966151979	0.7248\\
4275.31602769483	0.7296\\
4285.19124412981	0.78\\
4295.0531310945	0.7588\\
4304.8691424428	0.754\\
4315.27818302631	0.7676\\
4326.05835191742	0.7876\\
4336.13578691749	0.7596\\
4345.68819521869	0.75\\
4355.08751082416	0.7428\\
4364.67261768646	0.7412\\
4374.04334378067	0.7428\\
4383.52869542177	0.7604\\
4393.27676759698	0.7532\\
4402.61770532166	0.7284\\
4411.92015393562	0.746\\
4422.14774519776	0.746\\
4431.36626103129	0.7332\\
4440.84423321579	0.7556\\
4450.54978800377	0.7516\\
4460.38168829596	0.758\\
4470.20783476433	0.7412\\
4480.05536606415	0.7468\\
4489.68870942358	0.7548\\
4499.32157502304	0.7424\\
4509.21890730036	0.756\\
4518.90706265556	0.7544\\
4528.74613663601	0.748\\
4538.20668205762	0.7552\\
4548.15539971427	0.7544\\
4558.07080289471	0.7608\\
4567.90301404155	0.7688\\
4578.07181311021	0.7664\\
4588.94555922689	0.7608\\
4598.75490797476	0.7696\\
4608.5254364786	0.72\\
4617.54242685955	0.7456\\
4626.85453758638	0.7464\\
4636.78398635253	0.7752\\
4647.03121840047	0.7496\\
4657.06637118803	0.76\\
4666.62448075141	0.7528\\
4676.61022208552	0.7464\\
4686.01414266577	0.7296\\
4695.50561224355	0.7348\\
4704.78380967957	0.738\\
4713.97616494877	0.7348\\
4723.33567213704	0.7548\\
4733.38446492724	0.7604\\
4743.37302727302	0.7428\\
4752.36705516743	0.7268\\
4761.08115255037	0.7068\\
4770.37246250802	0.7612\\
4780.35157634585	0.7756\\
4790.29709537827	0.7676\\
4800.07972441246	0.766\\
4810.44168503121	0.7564\\
4820.68164429494	0.7636\\
4831.18955533498	0.7812\\
4842.09786691209	0.7628\\
4851.4505720866	0.7348\\
4860.89349049351	0.7516\\
4870.97366945103	0.774\\
4881.16013748467	0.7796\\
4891.86328618582	0.7668\\
4902.27007013241	0.7624\\
4912.02593870069	0.7736\\
4922.33865945449	0.7768\\
4932.28389822523	0.744\\
4941.31145313536	0.7152\\
4950.28056490384	0.7312\\
4960.00827610651	0.7728\\
4970.08791260178	0.7664\\
4979.47743026769	0.7544\\
4989.50983262516	0.7576\\
5000.00065550816	0.7952\\
};
\end{axis}\end{tikzpicture}}\subfloat[$X_1(0)=0.3$, $n=2500$]{\begin{tikzpicture}
\definecolor{mycolor1}{rgb}{0.00000,0.44700,0.74100}%
\begin{axis}[
 axis lines=middle,
 x   axis line style={->},
y   axis line style={-},
    width=\l cm,
height=\h cm,
at={(0cm,0cm)},
scale only axis,
xmin=0,
xtick={0,2000,4000},
xmax=5500,
xlabel={$t$},
ymin=0,
ymax=1,
ylabel={$X_1(t)$},
axis background/.style={fill=white},
]

\addplot [color=mycolor1,solid,forget plot,thick]
  table[row sep=crcr]{%
0	0.3\\
9.45374151938731	0.2924\\
18.6315049256967	0.3044\\
27.5010530022459	0.3352\\
35.9740663439905	0.3552\\
44.0512276086423	0.3644\\
52.3061545720621	0.3656\\
60.4823340404989	0.3928\\
68.2784368070708	0.4548\\
76.0683922344986	0.4916\\
83.7370244087402	0.5296\\
91.634289932558	0.5992\\
99.7413194858048	0.666\\
108.387766956515	0.7184\\
117.860141591109	0.7392\\
127.76324694348	0.7188\\
137.192898751811	0.7036\\
146.169111483289	0.72\\
155.473941169166	0.7188\\
164.82129593535	0.7108\\
174.650656509049	0.7472\\
185.100098076869	0.7552\\
195.417381269848	0.774\\
206.23765750073	0.7384\\
216.189139772887	0.7488\\
226.588420273118	0.7396\\
236.471625218157	0.738\\
246.776377851778	0.7544\\
257.458058427236	0.7616\\
267.648241104311	0.7276\\
277.799300939845	0.732\\
287.179819742795	0.724\\
297.203305525075	0.7532\\
307.929199770535	0.7524\\
318.586245475676	0.7544\\
329.222449811906	0.766\\
339.33133653498	0.7772\\
350.153245092859	0.7744\\
360.217080817553	0.716\\
370.049271751415	0.7668\\
380.610795756657	0.762\\
390.2334032464	0.7368\\
400.269052199033	0.7332\\
410.536916294787	0.7604\\
420.459812517164	0.7184\\
429.464237440472	0.7216\\
438.949549145131	0.7356\\
448.823185691928	0.7248\\
458.57438701259	0.7136\\
468.04759521291	0.7428\\
478.393280318935	0.7412\\
488.035744473406	0.7512\\
497.771581999921	0.7508\\
508.207025932958	0.7668\\
518.489564891828	0.756\\
528.286112367844	0.7352\\
538.288592525587	0.7556\\
548.516786607851	0.7292\\
558.45198420932	0.7232\\
568.136409411499	0.7428\\
578.32375823825	0.7612\\
588.75822175577	0.7552\\
598.791052603446	0.752\\
609.721157326437	0.7692\\
620.432784091975	0.7456\\
630.183534409226	0.7432\\
640.30220221667	0.7548\\
650.36475286882	0.7196\\
659.655712142543	0.7192\\
669.284558305582	0.7308\\
679.12246272475	0.7556\\
689.225242612721	0.7448\\
698.658780964597	0.7216\\
708.270994238531	0.7292\\
718.509281941753	0.7428\\
728.675412760165	0.7416\\
738.737833898711	0.774\\
749.404217992966	0.7708\\
760.060438090141	0.7456\\
770.231859926305	0.744\\
780.621312008072	0.7524\\
791.401822800355	0.7864\\
802.168964828737	0.7568\\
812.372213897666	0.7316\\
821.807378454635	0.7108\\
831.254796823638	0.7464\\
841.176213468557	0.7408\\
850.939593371155	0.7564\\
861.378853514952	0.7488\\
871.546353347678	0.7488\\
881.677903075611	0.7244\\
891.419734835579	0.7212\\
901.180824734669	0.7496\\
911.107864054368	0.7348\\
921.531702469939	0.746\\
932.169434847921	0.756\\
942.242972910336	0.7416\\
952.383734869592	0.7324\\
961.885413428648	0.7208\\
971.38947496322	0.7248\\
980.926363228931	0.7204\\
990.373396207663	0.7116\\
999.943480636437	0.7144\\
1009.848589379	0.7424\\
1020.11431726413	0.7436\\
1029.8644366767	0.7328\\
1039.69483005934	0.768\\
1050.53244388167	0.7556\\
1060.45298231754	0.7444\\
1070.97151976428	0.756\\
1081.17275024744	0.7436\\
1091.31719707183	0.738\\
1101.67931688849	0.7608\\
1112.0597624184	0.7496\\
1122.07219464327	0.7556\\
1132.29010625577	0.7352\\
1142.62739684829	0.7496\\
1153.52766519416	0.7628\\
1164.1191600988	0.7444\\
1174.12774115789	0.7432\\
1184.0004764233	0.7288\\
1193.90024658051	0.7476\\
1204.13686363028	0.7584\\
1214.41319912314	0.7576\\
1224.29723146136	0.7236\\
1233.59412495061	0.7284\\
1243.15406829272	0.7232\\
1252.14672852276	0.7252\\
1262.31032170063	0.7436\\
1272.00404821746	0.7256\\
1281.60787542261	0.7248\\
1291.45240643414	0.726\\
1301.159325349	0.736\\
1311.74568192179	0.7392\\
1321.77051344125	0.7332\\
1331.3814259698	0.7308\\
1341.04936015195	0.7464\\
1351.79814059794	0.7456\\
1361.62727230933	0.7212\\
1371.50586461296	0.7408\\
1381.14500944905	0.7344\\
1391.10993220799	0.7228\\
1400.80183703964	0.7492\\
1411.46426676807	0.7504\\
1421.88539250823	0.7436\\
1432.82261034336	0.7596\\
1443.34280946239	0.7352\\
1453.83515413137	0.7416\\
1463.74403415062	0.7244\\
1473.51727096599	0.7628\\
1483.93040732675	0.744\\
1493.40563521427	0.7172\\
1503.040840745	0.7116\\
1512.14104168627	0.72\\
1522.10366374953	0.7248\\
1531.94797336724	0.722\\
1541.70752670653	0.7512\\
1552.47555634968	0.7688\\
1563.58642745113	0.7796\\
1574.40193524152	0.786\\
1585.40767990904	0.7744\\
1596.47839895736	0.7732\\
1607.67640798397	0.786\\
1618.27505447578	0.74\\
1628.36727273823	0.7424\\
1638.23030333316	0.7436\\
1648.80502205736	0.7356\\
1658.55972963091	0.728\\
1668.13224762259	0.7196\\
1677.64986434615	0.7388\\
1687.81699348418	0.752\\
1697.9323668643	0.7256\\
1707.86672894203	0.7524\\
1718.09516593808	0.7448\\
1727.7733649648	0.732\\
1737.29026009197	0.7444\\
1747.04731257979	0.742\\
1756.83798014681	0.7464\\
1767.02233057513	0.7492\\
1777.05113414714	0.7444\\
1787.21924295734	0.7552\\
1797.40107944021	0.7328\\
1807.42004418161	0.742\\
1817.41483319997	0.762\\
1827.64624202489	0.756\\
1838.36599942495	0.7668\\
1848.73001944607	0.7492\\
1858.39120250902	0.7256\\
1867.78990183895	0.7136\\
1877.43022252131	0.7292\\
1887.20505375006	0.7424\\
1896.70212294749	0.6968\\
1906.1032145247	0.7244\\
1915.82597177088	0.7196\\
1925.21232960205	0.7096\\
1934.15749263207	0.7064\\
1943.7699877261	0.7332\\
1953.22176739271	0.7336\\
1963.38514478426	0.7424\\
1973.30446677383	0.7388\\
1983.20480682363	0.7204\\
1992.91411980115	0.7304\\
2002.46817304532	0.7452\\
2012.56624889047	0.73\\
2022.51521924494	0.752\\
2032.72602451184	0.7544\\
2042.73427056615	0.7452\\
2052.60454124627	0.736\\
2062.56759178459	0.74\\
2072.47374454598	0.75\\
2082.37675985962	0.7364\\
2092.98012605145	0.7464\\
2103.07178979691	0.7584\\
2113.96734794958	0.786\\
2125.96025526582	0.7824\\
2136.701001056	0.7528\\
2146.29507302852	0.7364\\
2156.02122425939	0.726\\
2165.65114777427	0.7152\\
2175.32841845682	0.738\\
2185.3649891859	0.7556\\
2196.30730197151	0.764\\
2206.56744832414	0.756\\
2216.71040792877	0.7588\\
2226.88464227332	0.7408\\
2236.84596327191	0.7488\\
2247.53726329733	0.7748\\
2257.89238905916	0.746\\
2268.22602657367	0.7648\\
2278.68810866339	0.7568\\
2288.55649162675	0.754\\
2298.89776467378	0.7592\\
2309.32675015691	0.744\\
2319.14788315615	0.7292\\
2329.00884939011	0.7324\\
2338.89618170171	0.7336\\
2348.61518530525	0.7284\\
2358.03753227209	0.7548\\
2367.77909096003	0.7056\\
2376.91642193877	0.6952\\
2386.04275565308	0.7196\\
2396.32393123405	0.7456\\
2406.72018825984	0.7728\\
2417.42887239456	0.7596\\
2428.46020126495	0.7668\\
2439.63139676964	0.796\\
2450.25447460561	0.736\\
2460.33172651845	0.734\\
2470.5476104162	0.7376\\
2480.02762983735	0.7184\\
2489.72771582662	0.742\\
2499.9325917467	0.7644\\
2510.37007342742	0.7536\\
2520.77131661531	0.7644\\
2531.3198405491	0.7588\\
2541.84688895625	0.756\\
2551.87369098518	0.7176\\
2561.76477281206	0.7284\\
2571.40232710103	0.7364\\
2581.15211029739	0.748\\
2591.99378254995	0.7668\\
2602.31948882455	0.7316\\
2612.19928565089	0.7176\\
2621.28103627609	0.7064\\
2631.00189090651	0.7524\\
2641.22763788269	0.7488\\
2650.89653628633	0.736\\
2661.00416548337	0.7492\\
2670.60824181711	0.7292\\
2680.41692020685	0.7424\\
2690.57108947763	0.754\\
2700.9665458942	0.7572\\
2711.04038027631	0.764\\
2721.47453621475	0.7632\\
2731.52897387486	0.7612\\
2742.02304590613	0.7476\\
2752.29295326647	0.7664\\
2763.10365221724	0.77\\
2773.76674308987	0.7732\\
2784.46231183659	0.7456\\
2794.14396433928	0.7224\\
2804.25394555671	0.754\\
2814.86359251315	0.7416\\
2824.6160161751	0.7176\\
2834.08683525983	0.7388\\
2843.69346591583	0.73\\
2853.71990611933	0.7328\\
2863.32170815469	0.718\\
2872.76006995337	0.7148\\
2882.54671441632	0.7352\\
2892.96950742928	0.7368\\
2902.88490853471	0.7404\\
2912.82850286486	0.7428\\
2922.87152897816	0.7344\\
2932.96598899122	0.7524\\
2943.22607553474	0.7364\\
2953.24433726537	0.7032\\
2962.33267847024	0.696\\
2971.77820942337	0.714\\
2981.05836547492	0.7136\\
2990.74148511737	0.7448\\
3000.75499062883	0.7548\\
3011.07695768247	0.7572\\
3021.09028392444	0.7416\\
3031.25381379356	0.7564\\
3041.01653673136	0.7524\\
3051.07934883211	0.7456\\
3061.13248089794	0.76\\
3071.54147722488	0.762\\
3082.12429052393	0.7516\\
3091.93804330497	0.7312\\
3101.96487101442	0.7308\\
3111.76539395273	0.7308\\
3121.57946080629	0.7248\\
3131.43590132453	0.7376\\
3141.31181167531	0.7172\\
3151.11377298082	0.748\\
3161.32279655318	0.7576\\
3171.4840238907	0.7628\\
3182.21668410679	0.7612\\
3192.94571552053	0.7752\\
3204.13157824366	0.756\\
3214.05448025196	0.758\\
3224.69519366079	0.736\\
3234.43241704547	0.74\\
3244.9876044006	0.77\\
3255.89644205573	0.774\\
3266.45785120347	0.7576\\
3276.78337054673	0.7564\\
3286.63297699708	0.7476\\
3297.24390227281	0.7728\\
3308.1036573886	0.7456\\
3318.38699023552	0.7404\\
3328.43886210389	0.7472\\
3338.98227798584	0.7792\\
3350.08874931806	0.794\\
3361.01120350975	0.762\\
3371.50799503832	0.7416\\
3381.64835172775	0.7384\\
3391.75529161435	0.7348\\
3401.39683567915	0.7224\\
3410.81931581836	0.7232\\
3420.86889791538	0.7436\\
3431.57322617427	0.7636\\
3441.65378665011	0.7448\\
3451.24210925515	0.7476\\
3461.7337216342	0.7676\\
3472.11824455864	0.7432\\
3481.77742640073	0.72\\
3491.41209405852	0.6996\\
3500.78615084854	0.7336\\
3511.20680908208	0.7496\\
3521.17223160754	0.7556\\
3531.52476902161	0.7652\\
3541.77354511313	0.7752\\
3552.50064921241	0.752\\
3562.90835804129	0.7316\\
3572.45558680544	0.7216\\
3581.94902058541	0.732\\
3591.90299651714	0.7636\\
3602.37306771718	0.734\\
3612.58268831512	0.74\\
3622.46483849789	0.7364\\
3632.00860275917	0.7276\\
3641.88208500332	0.7448\\
3651.81890400833	0.7464\\
3662.26790507492	0.7692\\
3673.24554521972	0.7756\\
3683.59567481118	0.7456\\
3693.65636542317	0.742\\
3704.14523867916	0.7972\\
3715.56845331129	0.7864\\
3726.08032341505	0.7544\\
3736.73533871228	0.7348\\
3747.08059886988	0.7328\\
3756.9228965102	0.7312\\
3766.8027621454	0.7372\\
3776.79458928293	0.7356\\
3786.66837005674	0.7528\\
3796.81266521828	0.7572\\
3807.43089113581	0.7652\\
3818.14022810637	0.7568\\
3828.62240065892	0.7576\\
3839.22102865773	0.7572\\
3849.46229578556	0.766\\
3860.20492418809	0.7616\\
3870.57332379301	0.7604\\
3881.27027233606	0.7476\\
3891.08880869128	0.7504\\
3901.26662917116	0.7336\\
3911.16576380271	0.7292\\
3920.63146221296	0.7368\\
3930.61491921752	0.7584\\
3941.61864443614	0.766\\
3952.51525222932	0.7708\\
3963.48819382742	0.752\\
3974.10771305978	0.7564\\
3984.51827147917	0.7452\\
3994.22197839081	0.7184\\
4004.12207591716	0.7568\\
4014.25850459332	0.7468\\
4024.17497358431	0.7484\\
4034.165982702	0.7368\\
4044.0672201626	0.7468\\
4053.93117435568	0.75\\
4064.44913474468	0.7584\\
4074.74870913182	0.7608\\
4084.59628451195	0.746\\
4094.43539270999	0.7536\\
4104.14110705717	0.7472\\
4114.74851306817	0.7588\\
4125.02167864974	0.7428\\
4135.15371802449	0.7592\\
4145.13545290527	0.7468\\
4155.4312597305	0.7468\\
4165.28092407563	0.7\\
4174.5526240999	0.6992\\
4183.68521636238	0.7444\\
4193.28291513569	0.7436\\
4202.8205621462	0.7408\\
4213.1275884845	0.774\\
4224.10290793357	0.762\\
4234.69188301799	0.7496\\
4244.74227166761	0.7408\\
4254.90765404824	0.7412\\
4264.7413375147	0.7552\\
4275.15673298792	0.7368\\
4285.53696766935	0.734\\
4295.39944155055	0.7636\\
4305.64359311453	0.772\\
4316.73639363687	0.7768\\
4327.77004604929	0.7724\\
4339.15751569798	0.7672\\
4349.4721574888	0.7592\\
4359.34599847206	0.7148\\
4368.89450418392	0.7284\\
4379.48739126167	0.768\\
4390.40390301299	0.7564\\
4400.46745337236	0.746\\
4410.96925070843	0.7552\\
4420.99615666221	0.7592\\
4431.02868743456	0.7372\\
4441.61366999661	0.7464\\
4451.77254173902	0.7416\\
4461.43281919742	0.7012\\
4470.35413131444	0.6996\\
4479.50417358148	0.7016\\
4488.27685764302	0.6832\\
4496.94581450467	0.6956\\
4506.53079059937	0.7312\\
4515.9645978876	0.712\\
4525.4462336127	0.702\\
4534.88867125587	0.7284\\
4544.23963952759	0.6976\\
4554.25046986596	0.7508\\
4564.43995595388	0.7324\\
4574.285067372	0.7328\\
4583.85656274934	0.7288\\
4593.67486130317	0.7316\\
4603.73626203365	0.7432\\
4614.00979587343	0.7648\\
4624.76109066157	0.77\\
4634.79811654687	0.7468\\
4645.38834483911	0.7536\\
4655.5139194639	0.756\\
4666.17714864591	0.7524\\
4676.45822503432	0.7584\\
4686.39900454778	0.7424\\
4696.6648781935	0.7596\\
4707.14919492713	0.7452\\
4716.72733609444	0.7328\\
4726.70833245971	0.7236\\
4736.09759953487	0.7076\\
4745.51820973758	0.7096\\
4754.82207984596	0.7288\\
4764.44505241014	0.754\\
4774.55010399426	0.7388\\
4784.17248106531	0.7152\\
4793.32551906014	0.7276\\
4803.32697842423	0.7276\\
4813.19643726133	0.7296\\
4822.98506372039	0.76\\
4833.16734467079	0.7748\\
4843.8873418168	0.7704\\
4854.04703320801	0.76\\
4864.27203495296	0.75\\
4873.90482701361	0.718\\
4883.37261636757	0.7352\\
4893.24920737896	0.7316\\
4903.17735406805	0.7516\\
4913.87263439077	0.7576\\
4923.81980976714	0.7312\\
4933.46551106213	0.7188\\
4942.11744352326	0.69\\
4951.44379985725	0.7248\\
4960.87298057564	0.7428\\
4970.21440691186	0.7324\\
4979.97462390616	0.7472\\
4989.95278225305	0.7424\\
5000.00420173232	0.7452\\
};
\end{axis}\end{tikzpicture}}\\
\subfloat[$X_1(0)=0.001$, $n=25000$]{\begin{tikzpicture}
\definecolor{mycolor1}{rgb}{0.00000,0.44700,0.74100}%
\begin{axis}[
 axis lines=middle,
 x   axis line style={->},
y   axis line style={-},
xticklabel style={
        /pgf/number format/fixed,
        /pgf/number format/precision=5
},
scaled x ticks=false,
    width=\l cm,
height=\h cm,
at={(0cm,0cm)},
scale only axis,
xmin=0,
xtick={0,3000,6000,9000},
xmax=10500,
xlabel={$t$},
ymin=0,
ymax=1,
ylabel={$X_1(t)$},
axis background/.style={fill=white},
]

\addplot [color=red,solid,forget plot,thick]
  table[row sep=crcr]{%
0	0.001\\
94.5147605585202	0.0876\\
131.535698554084	0.1416\\
160.449075687867	0.1601\\
186.204268620177	0.1927\\
209.556010044686	0.2055\\
231.389860740334	0.2217\\
252.992839671474	0.2199\\
274.570735021929	0.2171\\
296.155839554218	0.218\\
317.987222825967	0.2254\\
338.891292969082	0.2388\\
358.699157413312	0.2554\\
378.210874529354	0.2548\\
397.569423928485	0.2635\\
417.335586887003	0.2353\\
437.24063726962	0.2491\\
456.746158216336	0.2685\\
475.471541454222	0.2619\\
495.114035136521	0.2363\\
515.952107669154	0.2418\\
535.907808138488	0.2526\\
555.515039381378	0.2456\\
575.671438628572	0.2316\\
597.274223137547	0.2252\\
618.513951682938	0.2069\\
640.845282600463	0.2153\\
662.847352539378	0.2171\\
684.823597322725	0.2105\\
707.302209113113	0.2045\\
729.775657737349	0.1965\\
753.632321138338	0.1886\\
778.714182256235	0.178\\
802.593652214711	0.1994\\
824.743444829917	0.2122\\
846.689240075515	0.22\\
868.174594709614	0.2192\\
889.140326137833	0.2163\\
910.826521410888	0.2275\\
931.252623786528	0.2479\\
950.778592311431	0.2499\\
970.203486804786	0.2635\\
989.393321073984	0.2682\\
1008.57800473826	0.2442\\
1028.55012483153	0.2404\\
1049.16520407573	0.2318\\
1069.68450480592	0.2362\\
1091.76325294207	0.2048\\
1114.38655243229	0.2037\\
1136.78937905318	0.2097\\
1158.65090688378	0.2125\\
1180.34913275655	0.2281\\
1200.426080748	0.2461\\
1220.26966725971	0.2374\\
1240.68323574728	0.2374\\
1261.0856361232	0.2278\\
1281.3357289068	0.2464\\
1301.32028159126	0.2452\\
1321.82265470383	0.2222\\
1343.53725985793	0.2285\\
1363.97227039357	0.2297\\
1384.5847419599	0.2331\\
1405.29524271432	0.2279\\
1425.8861201282	0.2353\\
1445.92482768604	0.237\\
1466.25981621982	0.2364\\
1486.50083699716	0.237\\
1506.91863139647	0.2374\\
1527.40087308012	0.2286\\
1548.41722071436	0.2314\\
1569.10998512108	0.2205\\
1590.40507111938	0.2223\\
1612.46345757477	0.2035\\
1634.02069279412	0.2275\\
1656.30589851918	0.2039\\
1678.59133733439	0.217\\
1700.33108607786	0.219\\
1723.04370928447	0.1896\\
1746.7895067621	0.1996\\
1768.90181298901	0.2236\\
1790.57196069696	0.227\\
1812.97700076021	0.2105\\
1834.89827671718	0.2231\\
1855.68255472959	0.2335\\
1876.45332657947	0.2271\\
1897.37705106334	0.2231\\
1918.0697202404	0.235\\
1938.66504722502	0.2478\\
1958.93645117302	0.2402\\
1980.45245231528	0.2064\\
2002.69817468199	0.2142\\
2024.38363084894	0.2304\\
2044.90526695773	0.2321\\
2064.32723024018	0.2593\\
2083.07717216645	0.2723\\
2100.87884960841	0.2821\\
2119.03340271521	0.2931\\
2137.38364462762	0.2768\\
2155.64363279569	0.2826\\
2173.67719280779	0.279\\
2191.29181875903	0.2886\\
2208.96937884522	0.2926\\
2226.51841476648	0.3136\\
2243.64360946896	0.3189\\
2260.26377702274	0.3209\\
2277.6190274916	0.2915\\
2295.6601860928	0.2829\\
2314.10921987431	0.2761\\
2332.36069598805	0.2805\\
2350.18457751312	0.311\\
2367.56083939442	0.293\\
2385.03264286863	0.3142\\
2402.21736040901	0.3152\\
2419.02921751877	0.3396\\
2435.21717362987	0.3723\\
2450.67554399914	0.3841\\
2466.23739632508	0.4097\\
2481.36837168584	0.4407\\
2496.35749667521	0.4875\\
2510.97688600468	0.5489\\
2526.69295181287	0.6382\\
2543.52847465648	0.6976\\
2562.16789615728	0.7502\\
2581.31464636675	0.7254\\
2599.96827658309	0.7338\\
2619.12834286869	0.7477\\
2638.77005825658	0.7409\\
2658.43924398078	0.7467\\
2677.52072110355	0.7509\\
2697.3186951968	0.7475\\
2716.55049086822	0.7543\\
2736.61348489718	0.7534\\
2756.88313093317	0.7588\\
2777.34718159707	0.7472\\
2796.86848178618	0.7296\\
2815.13820363758	0.7274\\
2834.27575293971	0.7413\\
2854.06114189287	0.7583\\
2873.72484981477	0.7421\\
2893.48722736699	0.7593\\
2913.7159195422	0.7629\\
2933.79478758524	0.7521\\
2953.33857774804	0.7424\\
2972.60472525056	0.752\\
2992.77688649746	0.7664\\
3012.97761049588	0.753\\
3033.21659336737	0.7718\\
3053.25261991038	0.7531\\
3073.31015184166	0.7485\\
3093.47478533371	0.7581\\
3113.05484558372	0.7361\\
3132.43578837821	0.7459\\
3152.22938417751	0.7423\\
3172.67458880219	0.7638\\
3193.02685833799	0.7538\\
3213.1357066327	0.754\\
3233.75916113505	0.7586\\
3254.03866867173	0.755\\
3273.78164300811	0.7529\\
3293.44833645521	0.7485\\
3313.81703034357	0.7509\\
3333.52258844034	0.7535\\
3353.60234263443	0.7579\\
3373.87980568965	0.7625\\
3393.9045865627	0.7558\\
3413.47200298758	0.7472\\
3433.26231848034	0.7454\\
3453.03457550541	0.7554\\
3472.7409371365	0.7514\\
3492.4105889763	0.756\\
3512.00926788841	0.7423\\
3531.63799029555	0.7529\\
3551.45724696338	0.7449\\
3570.75440031298	0.7401\\
3590.08496320331	0.7519\\
3609.53337066644	0.731\\
3628.82911792441	0.7528\\
3648.69416323868	0.7618\\
3669.06597198528	0.7626\\
3689.30225122306	0.7622\\
3709.34265422093	0.7454\\
3729.05496408349	0.7555\\
3748.95425222296	0.7529\\
3768.56003783909	0.7437\\
3788.07263957026	0.7489\\
3808.04068490182	0.7585\\
3828.05796991967	0.7664\\
3848.18614109625	0.7606\\
3868.13705034092	0.7522\\
3887.98017244671	0.7608\\
3908.41517248403	0.7768\\
3929.04970918301	0.7586\\
3949.29207504036	0.7609\\
3970.13348205611	0.7585\\
3990.38282736077	0.7433\\
4010.60199785126	0.7639\\
4030.39078788051	0.7553\\
4050.0661034239	0.7528\\
4070.24284052515	0.7684\\
4090.53801777588	0.753\\
4110.6021745439	0.7562\\
4130.74317753701	0.7638\\
4150.76935534439	0.7476\\
4170.0442760827	0.7451\\
4189.66798422434	0.7517\\
4208.89519746351	0.7437\\
4228.96194472473	0.7709\\
4249.63425813959	0.7647\\
4270.03884053411	0.754\\
4289.43868755816	0.7492\\
4309.30912857604	0.7452\\
4328.77245590538	0.7556\\
4347.80785771282	0.7418\\
4367.0733013341	0.7372\\
4386.25106370293	0.7495\\
4405.62839819208	0.7441\\
4424.91564087573	0.7629\\
4445.03289408939	0.7543\\
4464.25157217172	0.7403\\
4483.75149366463	0.7456\\
4503.19850196866	0.7454\\
4522.65329547419	0.7566\\
4542.88642845812	0.7526\\
4563.04169792774	0.75\\
4582.87063760958	0.7456\\
4602.49290315959	0.7465\\
4622.17741470064	0.7407\\
4641.98269304413	0.7557\\
4661.36398593203	0.7379\\
4680.89269746761	0.7501\\
4700.70069831121	0.7502\\
4719.99076787835	0.7492\\
4739.6086722708	0.7598\\
4759.58757298349	0.7442\\
4779.04027699308	0.7466\\
4798.8062597574	0.7476\\
4818.40573912912	0.7489\\
4838.03419581814	0.7505\\
4858.08079763822	0.7581\\
4878.18820970768	0.7553\\
4898.70462176578	0.7605\\
4918.59419484809	0.7503\\
4937.98955740479	0.7484\\
4958.01745685002	0.7458\\
4978.13499042529	0.753\\
4998.02672364309	0.7638\\
5018.18833084702	0.7552\\
5037.83700776416	0.7473\\
5057.05423231292	0.7365\\
5075.90692540963	0.7317\\
5094.68421403143	0.7335\\
5114.0080222529	0.7459\\
5133.80575910132	0.7451\\
5153.08892725044	0.7376\\
5172.25316643284	0.7486\\
5191.25816715522	0.7374\\
5210.53089288978	0.7452\\
5229.98224784686	0.7332\\
5249.2430181296	0.7395\\
5268.97803999684	0.7511\\
5288.38858654318	0.7407\\
5307.61310101137	0.7353\\
5326.46867791178	0.7451\\
5346.65391778248	0.7725\\
5366.60421584125	0.7538\\
5386.28532012719	0.7582\\
5406.06583575276	0.751\\
5425.22361441868	0.7502\\
5444.63043548705	0.7326\\
5463.7051314354	0.7455\\
5483.80918414064	0.7631\\
5504.22356739578	0.7529\\
5523.91117907859	0.7525\\
5543.59027224845	0.7305\\
5562.52431945238	0.7459\\
5581.67886013864	0.741\\
5600.78896719158	0.7388\\
5619.99536588219	0.7406\\
5638.86829647867	0.732\\
5658.00390414531	0.7464\\
5676.72837446134	0.7351\\
5696.10334697382	0.7409\\
5715.94249909191	0.7591\\
5735.62807338632	0.7551\\
5755.00482673116	0.7537\\
5774.39156131934	0.7367\\
5793.68005404976	0.7402\\
5812.86157242175	0.7408\\
5832.06920881574	0.742\\
5851.03302313419	0.7426\\
5870.80928844348	0.7424\\
5889.78926033017	0.7381\\
5908.62092355668	0.7213\\
5927.2891568935	0.7335\\
5945.42993941291	0.7253\\
5964.03508164191	0.7339\\
5983.30143978385	0.7423\\
6002.17345249126	0.7368\\
6021.11659560116	0.745\\
6041.11688727824	0.756\\
6061.09765742033	0.7492\\
6080.20485787578	0.7368\\
6099.70145554604	0.7632\\
6119.96222472445	0.7575\\
6139.61928315936	0.7607\\
6159.47464866507	0.7529\\
6179.53975911979	0.7443\\
6198.9919340338	0.7433\\
6218.72711716974	0.7524\\
6238.40128129978	0.761\\
6257.75288839772	0.7388\\
6277.16902231173	0.7522\\
6296.77264008026	0.7456\\
6316.28180533875	0.7514\\
6336.28913838596	0.7485\\
6355.54429507585	0.7389\\
6375.32844738429	0.7513\\
6395.23062914586	0.7555\\
6414.74330257682	0.7373\\
6433.94719695943	0.7346\\
6452.47357019326	0.7352\\
6471.06187712408	0.725\\
6490.04840742976	0.7316\\
6509.04055192673	0.749\\
6528.84274184537	0.7522\\
6548.10165967755	0.7471\\
6567.28586924214	0.7501\\
6587.10387701937	0.7589\\
6607.00392367303	0.7545\\
6627.00065115804	0.7525\\
6647.40748098186	0.764\\
6668.258886585	0.7768\\
6688.46240604317	0.7536\\
6708.0981464491	0.7548\\
6727.74476017671	0.7518\\
6747.82663766161	0.7444\\
6767.5514670609	0.7639\\
6787.79863086823	0.7575\\
6807.78281058649	0.7557\\
6827.37572672118	0.7539\\
6847.06745353879	0.7303\\
6865.79689961864	0.7384\\
6884.82918101631	0.7404\\
6904.38829414951	0.7518\\
6923.96202440359	0.7392\\
6943.0262275385	0.7476\\
6962.57780558091	0.749\\
6982.27675178377	0.7549\\
7002.04030419424	0.7721\\
7022.69174910746	0.7683\\
7042.84689475845	0.7479\\
7062.60242888984	0.7497\\
7081.98611046728	0.7396\\
7101.36734185305	0.7264\\
7120.08023164738	0.7192\\
7138.71883283005	0.75\\
7158.73516618455	0.765\\
7178.83441935959	0.7462\\
7198.56094757987	0.7471\\
7218.34322203533	0.7533\\
7237.65070563964	0.7349\\
7256.82267490878	0.7351\\
7275.73035468233	0.7425\\
7294.41184700272	0.729\\
7313.21904398155	0.7468\\
7332.56629262883	0.7418\\
7352.11262722701	0.7632\\
7372.35254849071	0.7614\\
7392.09259083432	0.7616\\
7412.26088053668	0.7555\\
7432.29712969604	0.7493\\
7451.75706393219	0.7425\\
7470.88330535762	0.7299\\
7490.24526409173	0.7543\\
7510.08282755076	0.7513\\
7529.2942116851	0.7418\\
7549.51695605886	0.7648\\
7569.98212792893	0.7674\\
7591.03943434328	0.7646\\
7610.81868735468	0.746\\
7629.88142666532	0.7445\\
7649.05932689683	0.7411\\
7668.21086384102	0.7347\\
7687.32095513486	0.7331\\
7706.84363889775	0.7529\\
7726.78344292073	0.7523\\
7746.79770472819	0.7478\\
7765.94567917182	0.7498\\
7785.31585018463	0.745\\
7804.76916775929	0.7454\\
7824.24022794456	0.7322\\
7843.28576670074	0.7469\\
7862.9049928428	0.7325\\
7881.60475219306	0.7291\\
7900.62936739838	0.7347\\
7919.68668125564	0.7425\\
7939.48708656082	0.7451\\
7958.8643310953	0.7458\\
7979.42328535797	0.7708\\
8000.00219131313	0.749\\
8004.62673145036	0.74676\\
8023.89594268766	0.7492\\
8043.62980959515	0.74976\\
8063.04841404258	0.74824\\
8082.35676258104	0.74808\\
8101.38118283553	0.73912\\
8120.79438219008	0.74368\\
8139.97311493641	0.7416\\
8159.05319374578	0.7376\\
8178.34617548171	0.75424\\
8198.24630538341	0.75672\\
8217.94618926641	0.74848\\
8237.4269491758	0.74864\\
8257.03017095148	0.7444\\
8276.11835528176	0.74456\\
8295.39025586935	0.75116\\
8315.05473343644	0.74372\\
8334.40538165836	0.74524\\
8353.95398870958	0.75828\\
8374.07228473961	0.75292\\
8393.8121750522	0.74644\\
8413.24447188633	0.7502\\
8432.66182213996	0.74732\\
8452.23340720934	0.7502\\
8471.97428673343	0.7526\\
8491.40694475421	0.75036\\
8510.89092361496	0.74644\\
8530.21715936022	0.74916\\
8549.38814719365	0.74624\\
8568.76772227541	0.74648\\
8588.29843585663	0.7472\\
8607.90562549018	0.75072\\
8627.45806401599	0.75176\\
8647.15165351505	0.75464\\
8667.41343623704	0.76576\\
8687.91833000753	0.75864\\
8707.84284437751	0.758\\
8727.30069033624	0.7416\\
8746.51768495636	0.74912\\
8765.83720585856	0.74416\\
8785.01825343967	0.7428\\
8804.47054996194	0.75512\\
8824.42346067892	0.763\\
8844.58092822555	0.75796\\
8864.9448659823	0.76084\\
8884.83062266219	0.75172\\
8904.66299403727	0.75508\\
8924.64317007794	0.75628\\
8944.27398042035	0.75092\\
8963.94042590749	0.75676\\
8983.88202487635	0.75916\\
9003.89207737127	0.76268\\
9023.89732505154	0.74668\\
9043.0412771024	0.73988\\
9062.7262114129	0.7562\\
9082.31323421225	0.75256\\
9101.72279847166	0.74728\\
9121.33127171847	0.7504\\
9140.88467783921	0.75464\\
9160.42765093647	0.74584\\
9180.40233268285	0.75824\\
9200.25984259423	0.75424\\
9219.40986108075	0.74768\\
9238.73479596258	0.74456\\
9257.80656717743	0.74096\\
9276.7608805995	0.75176\\
9296.22201830145	0.75176\\
9316.00100072392	0.7556\\
9335.92130508381	0.74824\\
9355.70646104553	0.75668\\
9375.80155214327	0.75476\\
9395.4216644637	0.75324\\
9415.12401850753	0.75484\\
9434.96344761269	0.7578\\
9454.81383370311	0.75084\\
9474.57645340248	0.75692\\
9494.26449752371	0.74644\\
9513.5261867397	0.73436\\
9532.87933487594	0.74748\\
9552.36781594267	0.74492\\
9571.84698941593	0.7534\\
9591.10581989429	0.74132\\
9610.73211972824	0.75608\\
9630.51458504936	0.74544\\
9649.81256504446	0.75112\\
9669.55795922306	0.74472\\
9688.96219403061	0.75648\\
9708.74913917439	0.75688\\
9728.12376555112	0.74608\\
9747.38624299853	0.74256\\
9766.12383041322	0.74264\\
9785.18228973161	0.74056\\
9804.46193402539	0.75248\\
9823.73964954357	0.74088\\
9842.93611156381	0.7372\\
9861.70838141835	0.74192\\
9881.26073407161	0.75724\\
9901.07098489294	0.7506\\
9920.66835097678	0.75412\\
9940.56911325662	0.75532\\
9960.0134214225	0.75132\\
9979.85984100092	0.75868\\
10000.000009334	0.75668\\
};
\end{axis}\end{tikzpicture}}\subfloat[$X_1(0)=0.3$, $n=25000$]{\begin{tikzpicture}
\definecolor{mycolor1}{rgb}{0.00000,0.44700,0.74100}%
\begin{axis}[
 axis lines=middle,
 x   axis line style={->},
y   axis line style={-},
    width=\l cm,
height=\h cm,
at={(0cm,0cm)},
xticklabel style={
        /pgf/number format/fixed,
        /pgf/number format/precision=5
},
scaled x ticks=false,
scale only axis,
xmin=0,
xtick={0,3000,6000,9000},
xmax=10500,
xlabel={$t$},
ymin=0,
ymax=1,
ylabel={$X_1(t)$},
axis background/.style={fill=white},
]

\addplot [color=red,solid,forget plot,thick]
  table[row sep=crcr]{%
0	0.3\\
17.9489062758526	0.29864\\
35.7906455000322	0.30008\\
53.3912403881705	0.31828\\
70.6425457888403	0.32972\\
87.4377667197198	0.3362\\
104.065908478883	0.3594\\
120.248883593332	0.3836\\
135.994089454845	0.4112\\
151.228148735372	0.4592\\
166.263325878313	0.51688\\
181.595384589045	0.59096\\
197.769686507717	0.6506\\
215.040779630383	0.6938\\
233.50153466092	0.72652\\
252.933212509834	0.74196\\
272.588595332258	0.74464\\
292.369381406759	0.74616\\
312.729533280188	0.7576\\
333.075248111591	0.74784\\
353.059786143193	0.74452\\
373.134277144381	0.74412\\
393.068211609484	0.74676\\
412.840759393894	0.747\\
433.363957040206	0.75628\\
453.726287711464	0.76192\\
473.78259596848	0.74336\\
494.202630848376	0.76488\\
515.109636483169	0.76312\\
535.66433701257	0.75364\\
555.846325442123	0.75396\\
575.968166679348	0.75636\\
596.396820932541	0.74612\\
616.65087924079	0.755\\
636.541600972163	0.74888\\
656.415691106864	0.74344\\
676.077950831107	0.74552\\
696.292459518866	0.7504\\
716.121324170553	0.7446\\
736.175040751524	0.75324\\
756.422791345812	0.75132\\
776.650458038034	0.75724\\
797.192466316022	0.75528\\
817.546020710317	0.75344\\
837.620676788057	0.75176\\
857.563509992965	0.74728\\
877.468267921586	0.74752\\
897.503521327271	0.7478\\
917.435490806153	0.75148\\
937.152175838665	0.73844\\
957.074236187006	0.74796\\
976.970029285175	0.75344\\
997.085982264914	0.74992\\
1017.22524544562	0.75344\\
1037.46078839607	0.7468\\
1057.36570550993	0.74584\\
1077.31943056093	0.7534\\
1097.4400819762	0.75132\\
1117.3198389199	0.7458\\
1137.06615330256	0.74484\\
1156.85981525359	0.7428\\
1176.80312547132	0.74232\\
1196.91048690033	0.75456\\
1217.00031540157	0.7484\\
1236.76904948141	0.74848\\
1257.14668893449	0.7446\\
1277.21714376129	0.75356\\
1297.38418778356	0.74348\\
1317.41256826832	0.75228\\
1337.43909792796	0.7516\\
1357.36519816987	0.75304\\
1377.57472199602	0.74344\\
1397.76670775115	0.75048\\
1418.24255095427	0.76092\\
1439.01438989393	0.76284\\
1459.70214831152	0.75884\\
1479.84294277196	0.74124\\
1499.32607991569	0.74028\\
1518.85975493746	0.7472\\
1538.77320760473	0.74696\\
1558.42898188418	0.7356\\
1577.93848748106	0.73784\\
1597.94428627949	0.75076\\
1618.29252234715	0.75252\\
1638.24267009577	0.74732\\
1658.29780415746	0.74548\\
1678.03444460308	0.7378\\
1697.58934917562	0.74088\\
1717.1862762901	0.744\\
1736.91014744922	0.74176\\
1756.86720263442	0.74144\\
1776.61040206478	0.75356\\
1796.55056999158	0.75148\\
1816.78764171346	0.75892\\
1837.17142244073	0.75028\\
1857.47312172174	0.752\\
1877.44763016063	0.75048\\
1897.70029291546	0.74976\\
1918.1479517124	0.75888\\
1938.70568905068	0.75536\\
1958.70115054	0.74772\\
1978.78428842521	0.7566\\
1999.1545441603	0.7462\\
2019.31513747586	0.76228\\
2039.53892332969	0.7508\\
2060.01494815782	0.75912\\
2080.14789102802	0.744\\
2100.54946086016	0.7576\\
2121.41752345892	0.76448\\
2142.18942186523	0.75452\\
2162.45683832655	0.75444\\
2182.64998265755	0.75148\\
2202.85600966434	0.7514\\
2223.11555240739	0.75136\\
2243.14806594136	0.74704\\
2263.05051504345	0.74712\\
2282.7659678312	0.74144\\
2302.45208126969	0.74472\\
2322.65824294383	0.75028\\
2342.3574505172	0.74004\\
2362.05570185766	0.73652\\
2381.80263996596	0.74604\\
2401.97136284719	0.75552\\
2422.60207480542	0.75832\\
2442.82342803343	0.75088\\
2463.02461241027	0.74304\\
2483.08078624897	0.75412\\
2503.42758334371	0.74772\\
2523.03510523573	0.74508\\
2543.33294722893	0.75548\\
2563.71244703925	0.75532\\
2584.09698369433	0.7528\\
2604.41031927729	0.7508\\
2624.76118715691	0.75912\\
2645.24882105121	0.7572\\
2665.65368267812	0.75316\\
2685.95151939337	0.75276\\
2706.06122319093	0.75644\\
2726.21737120515	0.74588\\
2745.85928191919	0.73844\\
2766.2031604964	0.75736\\
2786.46651880206	0.74856\\
2806.92886726576	0.75632\\
2827.30351814947	0.75912\\
2847.76482642439	0.75284\\
2867.7733974707	0.74476\\
2888.15054216148	0.75284\\
2908.73185713653	0.75164\\
2928.91379329374	0.74776\\
2949.03646400937	0.75408\\
2969.16333100209	0.74496\\
2989.1978313097	0.74232\\
3009.15304014213	0.75392\\
3029.17910273976	0.7474\\
3049.46574852642	0.75492\\
3070.11453308664	0.74756\\
3090.30473953782	0.74892\\
3110.27911742599	0.74944\\
3129.93164743357	0.74112\\
3149.93100608213	0.75208\\
3169.993340211	0.73912\\
3189.80120238248	0.74376\\
3209.8411578465	0.74524\\
3229.90686852994	0.75604\\
3250.33339880589	0.75884\\
3270.47214153698	0.74524\\
3290.69747431825	0.75488\\
3310.88539303376	0.7468\\
3330.9394268462	0.75424\\
3350.94547797385	0.74528\\
3370.83321472633	0.7524\\
3391.16406619299	0.75028\\
3411.02360054776	0.74828\\
3431.35052110414	0.76044\\
3452.02129577035	0.757\\
3472.41771124067	0.74896\\
3492.57603217267	0.75184\\
3512.8502356644	0.75416\\
3532.853644135	0.74944\\
3553.04360829834	0.7482\\
3573.54109055991	0.75428\\
3593.73115139636	0.74612\\
3613.53787893227	0.74844\\
3633.10367734238	0.73652\\
3652.63855818214	0.738\\
3672.63790313029	0.7548\\
3693.15627679697	0.75736\\
3713.55410513151	0.74968\\
3733.29421527703	0.73852\\
3753.31639653338	0.74268\\
3773.3069695201	0.75172\\
3793.26466724275	0.75308\\
3813.61084314561	0.7566\\
3834.02136093504	0.75312\\
3854.39810029181	0.76048\\
3874.87096267146	0.74848\\
3894.94506363883	0.75616\\
3915.40599787198	0.75292\\
3936.21292408877	0.7598\\
3956.79771608769	0.75724\\
3977.26073214406	0.7562\\
3997.83310402147	0.75192\\
4017.80839291726	0.75392\\
4037.95723594953	0.74608\\
4057.69912100615	0.7416\\
4077.67980898795	0.7492\\
4098.12151604713	0.7566\\
4118.41500871568	0.745\\
4138.70808929593	0.75612\\
4159.39106274691	0.761\\
4179.97144858619	0.75248\\
4199.97294776015	0.74808\\
4220.22850378128	0.74792\\
4240.19600611563	0.75048\\
4260.39066267014	0.74768\\
4280.28420580061	0.74636\\
4300.07695892508	0.75348\\
4319.88128615204	0.74532\\
4339.59114582678	0.73652\\
4359.40268016339	0.75592\\
4379.92161397754	0.7568\\
4400.35124538419	0.7544\\
4420.71088264686	0.74624\\
4440.73417752007	0.75304\\
4460.91997539564	0.75716\\
4481.26896910876	0.75004\\
4501.39918078002	0.75188\\
4521.48043498287	0.75268\\
4541.71862435223	0.75472\\
4561.9210425636	0.75184\\
4582.37397934345	0.7468\\
4602.11904855023	0.7472\\
4622.02124736457	0.74812\\
4642.31135666157	0.75556\\
4662.25754061476	0.74716\\
4682.31734723231	0.7474\\
4702.11093762526	0.74476\\
4722.16560166861	0.74784\\
4742.41226056907	0.75632\\
4762.76057733408	0.75704\\
4783.21332186756	0.76104\\
4803.8068335418	0.75572\\
4824.41805794283	0.76508\\
4844.8540539509	0.7498\\
4864.82342471755	0.74924\\
4884.71246251461	0.74692\\
4904.30918124267	0.74552\\
4924.04969809219	0.74352\\
4943.90254230159	0.75888\\
4964.33967991381	0.75472\\
4984.7349747251	0.75796\\
5004.80125498584	0.75444\\
5024.90366210924	0.75156\\
5045.05370503742	0.74812\\
5065.29770420306	0.76128\\
5086.01772332355	0.74768\\
5106.2492941376	0.74968\\
5126.34517219187	0.74792\\
5146.58000375119	0.75344\\
5167.17317289878	0.75268\\
5187.53411653319	0.75092\\
5207.63133422569	0.74508\\
5227.10444596154	0.7346\\
5247.08466315337	0.74848\\
5267.38716958029	0.75704\\
5288.3076787434	0.76384\\
5308.47631913815	0.74032\\
5328.10400461117	0.74584\\
5348.18319420307	0.75116\\
5368.38035985645	0.7578\\
5388.59492428638	0.7554\\
5408.83521679971	0.74812\\
5428.74387114245	0.74232\\
5448.48216067236	0.7404\\
5468.42294387253	0.76256\\
5489.26325796525	0.76384\\
5509.69072007847	0.74876\\
5529.80383927091	0.75012\\
5549.88772776262	0.7406\\
5569.46693610971	0.73972\\
5589.23542633795	0.73852\\
5608.95850616955	0.75024\\
5629.06899265671	0.74544\\
5648.82973977846	0.74456\\
5669.05464497845	0.75152\\
5689.12408337937	0.74916\\
5709.15334185491	0.75532\\
5729.49966324712	0.75308\\
5749.70464422528	0.75244\\
5770.24177418681	0.75708\\
5790.5812699725	0.74944\\
5811.17838498092	0.75656\\
5831.48210818468	0.74912\\
5851.69236359464	0.75608\\
5872.16558175527	0.75116\\
5892.07736024842	0.73812\\
5911.88169785845	0.7522\\
5932.31085757418	0.75868\\
5952.45008495733	0.74508\\
5972.39602042595	0.75304\\
5992.55603844226	0.74752\\
6012.72106298133	0.746\\
6032.8372751565	0.75584\\
6053.21775610406	0.7498\\
6073.04447610447	0.74932\\
6093.09379759646	0.75412\\
6113.32935257863	0.75092\\
6133.32940447073	0.74408\\
6153.19715563674	0.74848\\
6173.38436913105	0.752\\
6193.56955310269	0.7532\\
6214.24026532323	0.7488\\
6234.3748819778	0.73972\\
6254.246705462	0.74628\\
6274.37322366532	0.75228\\
6294.86931554462	0.76124\\
6315.29750886281	0.75376\\
6335.50815667593	0.74928\\
6355.63316258824	0.74976\\
6375.53616770331	0.74784\\
6395.6271433215	0.74768\\
6415.59904271907	0.75332\\
6435.81061366807	0.7582\\
6456.19016333366	0.75532\\
6476.49599883108	0.75012\\
6496.46944090433	0.74456\\
6516.02090874734	0.73808\\
6535.91917547132	0.75152\\
6556.23101930434	0.75192\\
6576.42943874694	0.74812\\
6596.43194359162	0.75164\\
6616.62987384446	0.7514\\
6636.60326066888	0.74308\\
6656.5771189345	0.74996\\
6676.74655953548	0.74848\\
6696.43021295857	0.74496\\
6716.42123379368	0.74872\\
6736.39182754088	0.74528\\
6756.28702978753	0.74292\\
6776.70478114458	0.76268\\
6797.1472346443	0.7522\\
6817.32524327737	0.75676\\
6838.23864950158	0.76372\\
6858.70794965539	0.7524\\
6878.78810534969	0.74192\\
6898.49803526937	0.7432\\
6918.33941424721	0.75104\\
6938.89563162799	0.76252\\
6959.75216774034	0.75884\\
6979.71307064135	0.74484\\
6999.75589909677	0.74692\\
7020.07068418898	0.7546\\
7040.70048891655	0.75344\\
7060.66555113216	0.74232\\
7080.60754726459	0.748\\
7100.734631811	0.74976\\
7120.62521391477	0.74516\\
7140.61745204796	0.75436\\
7160.91383410867	0.74884\\
7180.81527012643	0.74988\\
7200.98821146515	0.75416\\
7220.99882607407	0.74272\\
7240.66677937686	0.75232\\
7260.79015682494	0.74304\\
7280.7941661805	0.7416\\
7300.4456243422	0.7454\\
7320.21567912613	0.74428\\
7340.21465871293	0.7486\\
7360.31103004649	0.75468\\
7380.7133483862	0.75552\\
7400.84769249898	0.7516\\
7421.05573118446	0.75536\\
7441.305644568	0.75608\\
7461.66009459036	0.74648\\
7482.10857022912	0.753\\
7502.47969455794	0.75212\\
7522.94665901243	0.75788\\
7543.37118235	0.74924\\
7563.77651907842	0.76144\\
7584.28359474289	0.74656\\
7604.29520849893	0.7528\\
7624.40986693789	0.7508\\
7644.58591759318	0.74956\\
7664.73422514834	0.74724\\
7684.91589890588	0.75172\\
7705.0775316748	0.7478\\
7725.15840952782	0.75756\\
7745.5965330023	0.75384\\
7765.84654611361	0.75592\\
7786.13703581784	0.75096\\
7806.22198074297	0.75272\\
7826.20283783539	0.7398\\
7845.76830013206	0.73924\\
7865.47620090892	0.74372\\
7885.5612451158	0.75716\\
7906.1171573354	0.7606\\
7926.85621031532	0.75872\\
7947.01051618815	0.75304\\
7966.91996541989	0.74784\\
7987.09840200232	0.75096\\
8007.29684324589	0.75108\\
8027.12870638327	0.73988\\
8046.69723135826	0.7462\\
8066.78760251356	0.7474\\
8087.174485432	0.75756\\
8107.32047555876	0.74976\\
8127.18888473474	0.74216\\
8147.15467454089	0.74576\\
8167.24395574926	0.7492\\
8187.223002123	0.74692\\
8206.84802708299	0.73732\\
8226.836322249	0.7462\\
8246.70591591788	0.7482\\
8266.71197216137	0.74992\\
8287.02233274004	0.74856\\
8306.89864394008	0.75128\\
8327.01659152916	0.75056\\
8347.38843351267	0.7616\\
8367.83302303419	0.75404\\
8388.1452425311	0.74884\\
8408.27954503848	0.75268\\
8428.3850292819	0.74388\\
8448.37088147764	0.74384\\
8468.39799203243	0.74968\\
8488.29201952648	0.742\\
8508.27188865658	0.75184\\
8528.4404127241	0.75048\\
8548.88951864484	0.76068\\
8569.32136249474	0.75268\\
8589.55129961609	0.74556\\
8609.98101403017	0.75932\\
8630.12861384593	0.75656\\
8650.68419711624	0.76032\\
8671.14173803813	0.7492\\
8691.2294310656	0.7512\\
8711.29011085985	0.74764\\
8731.40756346258	0.74692\\
8751.30803320615	0.74532\\
8771.12212345375	0.7446\\
8790.96955796066	0.74452\\
8810.8663226682	0.7532\\
8831.12578343381	0.74512\\
8851.11231090733	0.75416\\
8871.41455560021	0.75408\\
8891.82495855195	0.75532\\
8912.1665638668	0.75044\\
8932.48118265294	0.75636\\
8953.2150653814	0.76012\\
8973.67032991008	0.74996\\
8994.17489414246	0.76328\\
9014.5941433715	0.75288\\
9034.6511193944	0.74584\\
9054.52418294151	0.75112\\
9074.32709262381	0.75204\\
9094.85916963128	0.7546\\
9115.40520944196	0.75332\\
9135.20157040422	0.73764\\
9154.87892708122	0.74924\\
9174.96547242593	0.74776\\
9195.06889919316	0.7484\\
9214.91403109272	0.74016\\
9234.67903034677	0.74784\\
9254.50092507206	0.7446\\
9274.23198412313	0.74116\\
9294.41969969432	0.75444\\
9314.70877833646	0.74492\\
9334.53979347631	0.7524\\
9354.69501406056	0.74128\\
9374.70226400497	0.74536\\
9394.48487944497	0.74472\\
9414.4851968964	0.74208\\
9434.58209680648	0.75204\\
9454.88197562809	0.74484\\
9474.65675313447	0.74548\\
9494.8775432177	0.75188\\
9515.04642162793	0.752\\
9535.35302393791	0.7532\\
9555.42934710361	0.75032\\
9575.6516473106	0.75224\\
9596.0691785666	0.74832\\
9616.20771846426	0.759\\
9636.42018821635	0.75004\\
9656.23662605675	0.74532\\
9676.80590708346	0.75852\\
9697.54742720639	0.7584\\
9717.45688383643	0.74168\\
9737.62021609891	0.74952\\
9757.50413016611	0.74896\\
9777.50453483642	0.75004\\
9797.9745160867	0.75148\\
9818.03081339883	0.74204\\
9837.94877274626	0.749\\
9858.05176174879	0.75172\\
9878.29823394313	0.7572\\
9898.58662621485	0.7512\\
9918.70631196026	0.74192\\
9938.86560439931	0.75928\\
9959.7242010103	0.76348\\
9980.03402555713	0.74652\\
10000.0001178724	0.74948\\
};
\end{axis}\end{tikzpicture}}
\caption{Trajectories of the stochastic imitation dynamics in Example \ref{ex:atan} for the game in Example \ref{ex:example} with $\lambda=K_{12}=K_{21}=1$, $n=2500$ and $n=25000$, respectively.}
\label{fig:simulations}
\end{figure}
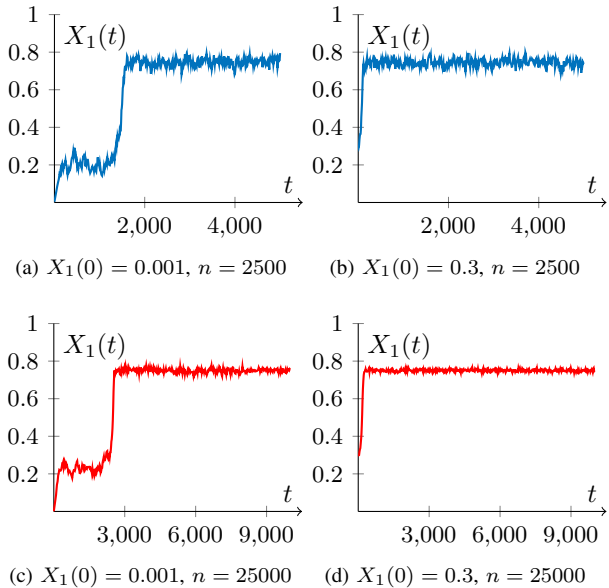
\end{example}\medskip

From this example, we may conjecture that the NE reached by the deterministic learning process can exhibit a different behavior, depending whether they are ESS or not: if the NE reached is so, then the process spends a very long time close to it. On the contrary, if it is not ESS, the process exits from it in a reasonable time, and converges close to another NE, until it reaches an ESS. The next section is devoted to formalize and prove this intuition, which is the main contribution of this paper.

\section{Long-run analysis}\label{sec:analysis}

In this section, we tackle the analysis of the stochastic imitation dynamics in terms of their long-run behavior. Specifically, we consider large-scale networks, i.e., with $n\to\infty$, and we show that the process spends most of its time close to $\mc S$, before being absorbed in a pure configuration. For the sake of simplicity, we make the following assumptions:
\begin{itemize}
\item[{\textit A1}:] the support of the initial condition is the whole $\mc A$;
\item[\textit A2:] the initial condition $X(0)$ is in the interior of its support;
\item[\textit A3:] the potential $\Phi$ admits a finite number of critical points;
\item[\textit A4:]  the pure configurations are not ESS.
\end{itemize}

\begin{remark}These assumptions are made to keep the presentation of the results and the readability of this section as clear as possible. However, they could be removed at the expense of having more complicated statements, notations, and/or weaker results. Specifically, $(A1)$ does not reduce the generality of our results: we can naturally reformulate every statement by considering the game restricted to the support of the initial condition. Assumptions $(A2)$ and $(A3)$ can be removed, obtaining slightly weaker results in terms of expected values, instead of in terms of probability. Finally, $(A4)$ is not too restrictive: when not verified, it means that some of the absorbing states of the system are ESSs and the absorbing event may occur in a short time (since it coincide with the convergence close to an ESS). In this case, our statements can be generalized by asserting that either the system  is absorbed into one of these pure configurations that are ESSs, or our results in the following hold.\end{remark}\medskip

At first, we recall that, according to Kurtz's Theorem \cite{Kurtz1970,Kurtz1971}, for any $T>0,\,\exists\, C_T>0$ such that  
\be\label{eq:kurtz}
\P\left[\text{sup}_{t\leq T}||X(t)-x(t)||\geq\eps\right]\leq 2(m-1)^2e^{-C_T n\eps^2},
\ee
where $x(t)$ is the solution of \eqref{eq:ode}, which converge to some $\bar x\in NE$, as extensively studied in \cite{cdc2017}. However, this result is not sufficient to understand the long-run behavior of the system, neither to guarantee convergence close to $\mc S$. In fact, i) Kurtz's Theorem has no validity for $T$ comparable with the population size $n$ (if $T$ is function of $n$, $C_T$ may depend on $n$ too, and the probability in \eqref{eq:kurtz} may not converge to $0$); and  ii) $x(t)$ may converge to some NE that is not an ESS (e.g., a saddle point of $\Phi$).

Therefore, to understand the long-run behavior of the system, we have to analyze the behavior of the stochastic process $X(t)$. Specifically, we first prove a pair of technical lemmas: the first one deals with the time needed by the process to decrease the potential, the second one with the time required to exit the neighborhood of unstable critical point of $\Phi$ (minima and saddle points). In order to improve the readability of the paper, we state here the two lemmas along with the new insight on the process they give. The detailed proofs are presented in the Appendix.\medskip

\begin{lemma}\label{lemma:decrease}
Let $X(t)$ be an imitation dynamics with transition rates \eqref{eq:rate} and a potential $\Phi$ that satisfies \eqref{eq:potential}. Let us focus on the stochastic process $\Phi(t)=\Phi(X(t))$ and let $X(t)$ such that $||X(t)-\bar x||_\infty\geq\delta>0$, for any $\bar x$ fixed point of \eqref{eq:ode} and $t\geq 0$. Then, $\forall\,\eps>0$ $\exists\,C_{\delta},K_{\delta}>0$ such that
\be
\P\left[\exists\, t<e^{C_{\delta}\eps n}:\Phi(t)<\Phi(0)-\eps\right]\leq e^{-K_{\delta}\eps n}.
\ee
\end{lemma}\medskip

Lemma \ref{lemma:decrease} allows us to tie the behavior of the stochastic process $X(t)$ and the one of the associated deterministic system of ODEs from \eqref{eq:ode}, far beyond the finite time ranges in which Kurtz's Theorem can be applied. In fact, from Lemma \ref{lemma:decrease} we deduce that  the asymptotically stable equilibria of  \eqref{eq:ode}, are also meta-stable equilibria of the stochastic imitation dynamics, since the process, when reached $\bar x\in \mc S$ (that is a local maximum of $\Phi$), needs an exponentially long time in $n$ to decrease the potential and move away from $\bar x$. However, the analysis of the stochastic imitation dynamics is not complete: we still have to study the behavior of the system close to critical points of $\Phi(x)$ that are not local maxima, in order to guarantee $X(t)$ to exit from their neighborhood in a time negligible with respect to the one spent close to $\mc S$. The following result serves this purpose.\medskip

\begin{lemma}\label{lemma:exit minima}
Let $X(t)$ be an imitation dynamics with transition rates \eqref{eq:rate} and a potential $\Phi$ that satisfies \eqref{eq:potential}. Let us focus on the stochastic process $\Phi(t)=\Phi(X(t))$ and let $X(t)$ such that $||X(t)-\bar x||\to 0$ as $n\to \infty$, where $\bar x$ is an unstable critical point of the potential. Then, $\exists\,\eps,K_\eps>0$, such that
\be\label{eq:fast exit}
\P\left[\exists\, t\leq K_\eps n\ln n:\Phi(t)\geq \Phi(\bar x)+\eps\right]\geq 1-\frac{1}{\ln N}.
\ee
\end{lemma}\medskip


Given an imitation dynamics with initial condition $X(0)$, we define the absorbing time as
\be
\tau:=\min\{t\in\R^+: X(t)=\delta^{(i)}, i\in \,A\},
\ee
and, given $\gamma>0$, we denote the fraction of time spent $\gamma$-close to the ESSs of the game as
\be
T_{\gamma}(t):=\frac1t\int_{0}^{t}\1_{\gamma}(X(s))ds,
\ee
where 
\be
\1_{\gamma}(X(s)):=\left\{\ba{ll}1&\text{if }\exists\,\bar x\in \mc S:||X(s)-\bar x||<\gamma\\0&\text{else.}\ea
\right.
\ee
All our results can be summarized in the following theorem, which is the main contribution of this paper.\medskip

\begin{theorem}\label{teo:main}
Let $X(t)$ be an imitation dynamics with transition rates \eqref{eq:rate} and a potential $\Phi$ that satisfies \eqref{eq:potential}. 
Then $\exists\,C_1,C_2,C_3>0$, such that, with probability converging to $1$ as $n$ grows, the following holds:
\begin{enumerate}[i)]
\item $\tau \geq e^{C_1n}$; and
\item $T_{\gamma}(e^{C_2 n})\geq 1-e^{-C_3n}$, $\forall\,\gamma>0$.
\end{enumerate}
\end{theorem}\medskip

\begin{proof}
Let $x(t)$ be the solution of \eqref{eq:ode} with initial condition $x(0)$. Results in \cite{cdc2017} guarantee $x(t)\to \bar x\in \mc N$. We distinguish two cases: 1) $\bar x\in \mc S$, or 2) $\bar x\notin \mc S$. 

\begin{enumerate}
\item For any $\delta'>0$, let $T_{\delta'}$ be the time such that $||x(t)-\bar x||<\delta'/2$. Then, applying \eqref{eq:kurtz} with $\eps=\delta'/2$, $||X(T_{\delta'})-\bar x||<\delta'$ holds with probability exponentially close to $1$. Since $\bar x$ is an isolated maximum of the Lipshitz-continuity potential \cite{Scutari2006}, it exists a neighborhood of $\bar x$ where $\Phi$ is strictly decreasing. Specifically, $\exists\,\delta$ such that, chosen $\delta'<\delta$, $\max_{x:||x-\bar x||_\infty=\delta}\Phi(x)\leq \Phi(\bar x)-\eps$ and $\min_{x:||x-\bar x||_\infty=\delta'}\Phi(x)\geq \Phi(\bar x)-\eps/2$, for some $\eps>0$. Hence, for $X(t)$ to exit a $\delta$ neighborhood of $\bar x$, the potential should decrease by at least $\eps/2$, while $||X(t)-\bar x||\geq \delta'/2$. Lemma \ref{lemma:decrease} bounds the probability that this event occur before $T_{\delta'}+e^{C_2n}$, for some $C_2>0$. Then, i) comes straightforward, being $\tau\geq e^{C_2n}$, and ii) is obtained $\forall\gamma>0$, by substituting $\eps$ instead of $\delta$, if $\gamma<\delta$, and by considering that the time spent $\gamma$-close to $\bar x$ is greater than the one spent $\delta$-close to it, otherwise. Finally, since $T_{\delta'}$ does not depend on $n$, $T_{\delta'}/e^{C_2n}\leq e^{-C_3n}$, for some $C_3>0$, as $n$ grows.

\item $\bar x$ is either a local minimum or a saddle point of $\Phi$. Similar to 1), after $T_{\delta'}$,  $||X(t)-\bar x||<\delta'/2$. Then, Lemma \ref{lemma:exit minima} guarantees that, after $\tilde T_{\bar x}=K_{\bar x'}n\ln n$, for some $K_{\bar x'}>0$, $\Phi(t)\geq \Phi(\bar x)+\eps$. Then, due to the  Lipshitz-continuity argument, $||x(t)-\bar x||\geq \delta$, so \eqref{eq:ode} yields convergence close to some $\bar x_1$. The core observation is that, since the potential cannot decrease along trajectories of \eqref{eq:ode}, $X(t)$ cannot visit the critical point of $\Phi$ twice. Being the number of critical points finite (A3), $X(t)$ spends at most a time $Kn\ln n$, for some $K\geq 0$ (given by the sum of all the times spent close to non-ESSs), before entering the basin of attraction of a local maximum. Then, follow 1).
\end{enumerate}
\end{proof}

Theorem \ref{teo:main} gives some interesting insights on the imitation dynamics, far beyond what can be deduced from the deterministic approximation. In fact, our result allows for performing an analysis of those phenomena that the deterministic imitation dynamics do not catch. Specifically, on the one hand, we give some bounds on the absorbing time of the process, guaranteeing that this event is not seen before an exponentially long time in the population size. On the other hand, we ensure that it spends most of the time before the absorbing event close to evolutionary stable NE of the (continuous) population game, that are the local maxima of the potential $\Phi$. Both these results hold true with probability converging to $1$ as the population size grows. We notice that these results are perfectly consistent with the conjectures deduced from the  simulations in Fig \ref{fig:simulations}.

\section{Numerical Simulations}\label{sec:simulations}

At this stage, it is natural to wonder what happens in situations where the players interact on a complex communication network, which is not a complete graph. The answer to this question is of course nontrivial: the presence of clustered communities or non-symmetric connections may influence the dynamics of the system, deviating it from the convergence close to $\mc S$, seen for the complete graph.

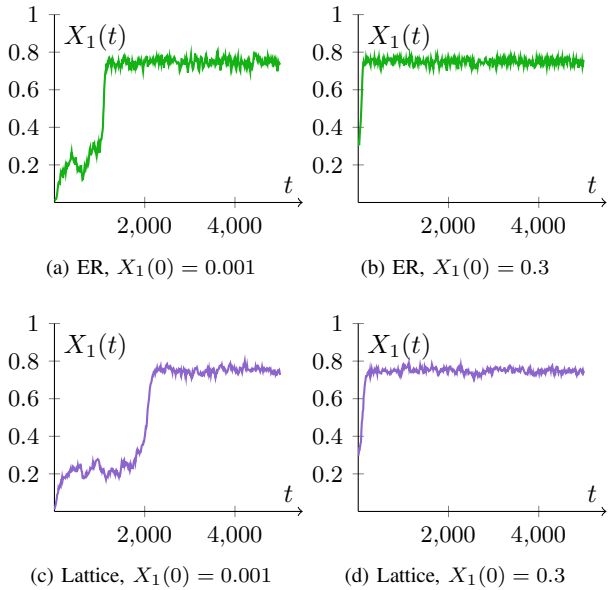
\begin{figure}
\centering
\subfloat[ER, $X_1(0)=0.001$]{\begin{tikzpicture}
\definecolor{mycolor1}{rgb}{0.08,0.7,0.08}%
\begin{axis}[
 axis lines=middle,
 x   axis line style={->},
y   axis line style={-},
    width=\l cm,
height=\h cm,
at={(0cm,0cm)},
scale only axis,
xmin=0,
xtick={0,2000,4000},
xmax=5500,
xlabel={$t$},
ymin=0,
ymax=1,
ylabel={$X_1(t)$},
axis background/.style={fill=white},
]

\addplot [color=mycolor1,solid,forget plot,thick]
  table[row sep=crcr]{%
0	0.01\\
10.0499015586447	0.0104\\
20.1176710296326	0.0154\\
30.1062699846674	0.019\\
40.1582540558374	0.0196\\
50.158853782658	0.0326\\
60.2068510516118	0.047\\
70.2112904241827	0.0696\\
80.2216466845434	0.0924\\
90.2175561299653	0.1058\\
100.230269634399	0.1094\\
110.303780136806	0.104\\
120.31063674713	0.1012\\
130.303062117274	0.142\\
140.363355464892	0.1574\\
150.228838845308	0.1474\\
160.223639153883	0.1456\\
170.254506414038	0.1622\\
180.233376006651	0.1698\\
190.217800743523	0.1298\\
200.177686572973	0.15\\
210.177404038942	0.1754\\
220.245222557647	0.1734\\
230.299193586438	0.1668\\
240.335121795474	0.1848\\
250.301341083375	0.1612\\
260.371046939274	0.1832\\
270.377813087769	0.199\\
280.435478648089	0.1918\\
290.459746965326	0.204\\
300.516488713182	0.2012\\
310.568174623007	0.1976\\
320.587256222427	0.2032\\
330.633045328692	0.2138\\
340.658360568154	0.2198\\
350.649939502218	0.2238\\
360.567402197836	0.2306\\
370.573356369486	0.2492\\
380.603143127737	0.2674\\
390.605007990837	0.2422\\
400.609534582705	0.245\\
410.671180079487	0.235\\
420.678971659777	0.2244\\
430.705782944521	0.2186\\
440.786233779902	0.2238\\
450.798192099736	0.2174\\
460.845000545943	0.2\\
470.834580860779	0.2404\\
480.847757533096	0.2184\\
490.792131440665	0.2086\\
500.890879544237	0.1992\\
510.979692507958	0.1714\\
521.013163100466	0.1772\\
531.030685325235	0.1782\\
541.054421488889	0.1832\\
551.037867991161	0.1654\\
560.988497430272	0.1644\\
570.967834732118	0.1722\\
580.999204538765	0.1768\\
591.058971437536	0.1988\\
601.119392693359	0.1644\\
611.188001708451	0.1472\\
621.195232673702	0.122\\
631.321210050011	0.1252\\
641.333232161406	0.141\\
651.41988193457	0.1536\\
661.448933203776	0.174\\
671.47656634853	0.1982\\
681.530915426553	0.1906\\
691.452421539104	0.181\\
701.441953373903	0.1974\\
711.456821701712	0.2048\\
721.47404732707	0.221\\
731.49739543951	0.206\\
741.508256193425	0.1966\\
751.51179809733	0.2048\\
761.554818039008	0.2304\\
771.602005085622	0.2562\\
781.698193518425	0.2804\\
791.729005959222	0.2902\\
801.738724697692	0.2588\\
811.738775973753	0.256\\
821.747498009296	0.2602\\
831.77998712834	0.2742\\
841.860127461615	0.2796\\
851.924450678961	0.289\\
861.929100826572	0.3132\\
871.904636645535	0.3122\\
881.975886849097	0.3162\\
891.986657213806	0.2854\\
901.973649149833	0.276\\
912.035414556338	0.2834\\
922.070967640405	0.267\\
932.159080139278	0.2874\\
942.201692006074	0.2572\\
952.213360655426	0.2756\\
962.27776887113	0.3012\\
972.250762435567	0.2906\\
982.282992703923	0.3134\\
992.318079513447	0.2922\\
1002.38515315058	0.2658\\
1012.39903990478	0.296\\
1022.38424004788	0.3338\\
1032.34387610579	0.3336\\
1042.35648284969	0.3368\\
1052.33090191419	0.3598\\
1062.38911828856	0.3746\\
1072.45051910527	0.3964\\
1082.42968523807	0.47\\
1092.41660902972	0.514\\
1102.42433874393	0.6022\\
1112.44667294619	0.6404\\
1122.50181263929	0.6812\\
1132.55117113402	0.6996\\
1142.57274201387	0.708\\
1152.60800258847	0.739\\
1162.64920887977	0.7182\\
1172.70404784171	0.735\\
1182.67834615262	0.725\\
1192.6885650099	0.735\\
1202.69831687217	0.7578\\
1212.74973508221	0.7518\\
1222.75753451994	0.765\\
1232.68332919557	0.7592\\
1242.69033026527	0.753\\
1252.69175028306	0.7372\\
1262.64514503473	0.7194\\
1272.59740221213	0.7512\\
1282.64159981637	0.7444\\
1292.57795906516	0.7598\\
1302.61422613241	0.7488\\
1312.58889696361	0.7454\\
1322.63437884871	0.7424\\
1332.67997145857	0.723\\
1342.67904901118	0.7248\\
1352.66551913616	0.7134\\
1362.75954235023	0.734\\
1372.76757929833	0.7354\\
1382.79631952239	0.7342\\
1392.82306195226	0.737\\
1402.83824535377	0.7418\\
1412.87085795251	0.7304\\
1422.8742858625	0.7526\\
1432.81556987566	0.7392\\
1442.83854224441	0.7406\\
1452.76797298774	0.7382\\
1462.8304925749	0.7168\\
1472.90253613892	0.7334\\
1482.96485993544	0.7326\\
1492.95093914824	0.7272\\
1502.9710420176	0.7614\\
1512.85485411248	0.7376\\
1522.92619748136	0.7652\\
1532.93882378508	0.7736\\
1542.92645633251	0.7668\\
1552.97419726971	0.761\\
1563.0665937142	0.7782\\
1573.10134149678	0.772\\
1583.01205048422	0.748\\
1593.06541526939	0.7364\\
1603.10577655094	0.7392\\
1613.13896652838	0.7514\\
1623.15241974689	0.7594\\
1633.15876396125	0.7348\\
1643.20305577613	0.7696\\
1653.23781650187	0.7192\\
1663.29317122314	0.725\\
1673.27477997363	0.7218\\
1683.31593750806	0.7078\\
1693.26929767563	0.7074\\
1703.32801520088	0.748\\
1713.28154844125	0.73\\
1723.38976019923	0.7532\\
1733.35233514526	0.74\\
1743.32885812358	0.7414\\
1753.35407600837	0.7288\\
1763.35769697554	0.7042\\
1773.31161514217	0.747\\
1783.37152139104	0.7456\\
1793.40461978723	0.7238\\
1803.48298995914	0.7402\\
1813.45119074971	0.7342\\
1823.41505237681	0.7198\\
1833.4973910506	0.7256\\
1843.47228152404	0.7506\\
1853.51023694423	0.7606\\
1863.54516885386	0.7506\\
1873.67304919982	0.7588\\
1883.62288136384	0.7454\\
1893.65326947213	0.75\\
1903.73961709805	0.74\\
1913.81400213019	0.7466\\
1923.86404570435	0.7582\\
1933.86377515742	0.7442\\
1943.85699492057	0.7414\\
1953.91343568109	0.7162\\
1963.9400164965	0.699\\
1973.94375279013	0.728\\
1983.94575056242	0.7468\\
1993.98262572754	0.741\\
2004.13963674616	0.745\\
2014.14746490827	0.7656\\
2024.16496420654	0.772\\
2034.15138396483	0.774\\
2044.18246578292	0.7658\\
2054.23456689906	0.7682\\
2064.12646656645	0.7268\\
2074.15659995577	0.7266\\
2084.20819286557	0.757\\
2094.18268035245	0.7362\\
2104.24019719061	0.7386\\
2114.31665482158	0.7432\\
2124.33514406523	0.7634\\
2134.36827725506	0.758\\
2144.3865037917	0.7476\\
2154.4408346406	0.7508\\
2164.45070242459	0.751\\
2174.44232672653	0.7594\\
2184.4917268957	0.7474\\
2194.42663783957	0.7592\\
2204.49200602859	0.7424\\
2214.50068751656	0.742\\
2224.57636609121	0.7358\\
2234.54553797538	0.7574\\
2244.54404732514	0.7656\\
2254.56588970446	0.7634\\
2264.60004769919	0.7778\\
2274.5639127202	0.761\\
2284.56739582459	0.7698\\
2294.5620493892	0.7656\\
2304.6954961365	0.7732\\
2314.72845217521	0.7392\\
2324.7210244729	0.7516\\
2334.70139997914	0.7604\\
2344.68645750428	0.7486\\
2354.666556989	0.7552\\
2364.68900794534	0.741\\
2374.69843939529	0.7866\\
2384.78873480162	0.7676\\
2394.78119621796	0.7554\\
2404.78700954725	0.732\\
2414.7613290483	0.76\\
2424.76859537692	0.7172\\
2434.78629534114	0.7502\\
2444.81610449075	0.7612\\
2454.85596053506	0.7572\\
2464.84822201971	0.7722\\
2474.85900746422	0.7742\\
2484.98203629722	0.7658\\
2494.96034910874	0.7732\\
2504.9236417385	0.7588\\
2514.99039419487	0.77\\
2524.9555155459	0.7604\\
2534.98983593004	0.7314\\
2545.02439051834	0.739\\
2554.98768204302	0.7744\\
2565.05209710804	0.7604\\
2575.08855943793	0.766\\
2585.11140453537	0.7454\\
2595.11211246492	0.7512\\
2605.13530071379	0.7502\\
2615.14937667648	0.7554\\
2625.06869787127	0.7498\\
2635.00198560749	0.7598\\
2645.06374561136	0.7686\\
2655.11139521387	0.774\\
2665.1677979447	0.769\\
2675.20488103637	0.7536\\
2685.25580495298	0.752\\
2695.27419717104	0.7498\\
2705.22176409151	0.7482\\
2715.2361498105	0.7394\\
2725.23357750322	0.7398\\
2735.2178472829	0.7426\\
2745.27047129145	0.7456\\
2755.24585554267	0.7432\\
2765.23833692585	0.7268\\
2775.19127245875	0.6986\\
2785.20170445215	0.7052\\
2795.21142631709	0.7172\\
2805.29197703815	0.7218\\
2815.37037407187	0.7626\\
2825.43333260719	0.7616\\
2835.452882993	0.746\\
2845.49195257398	0.754\\
2855.58841445812	0.7552\\
2865.63677648558	0.7566\\
2875.70401103647	0.756\\
2885.6912708478	0.732\\
2895.72215591296	0.763\\
2905.78528306748	0.7548\\
2915.79584239488	0.7458\\
2925.8862266379	0.7688\\
2935.91193633507	0.7534\\
2945.94697801237	0.7244\\
2956.03768372832	0.6912\\
2966.10046081739	0.7168\\
2976.24490176658	0.7326\\
2986.32014971815	0.7412\\
2996.31232980966	0.7426\\
3006.23947351911	0.7562\\
3016.27425662067	0.7678\\
3026.28456127812	0.7854\\
3036.32352636075	0.7486\\
3046.3926805992	0.7472\\
3056.44915131417	0.719\\
3066.38573095871	0.7084\\
3076.38634999223	0.7142\\
3086.39758100238	0.7294\\
3096.42585785462	0.7434\\
3106.43240316601	0.7602\\
3116.51834847475	0.7924\\
3126.57323799976	0.769\\
3136.57746088317	0.7542\\
3146.54161712721	0.7546\\
3156.53642381726	0.7642\\
3166.5895835233	0.7506\\
3176.60580933705	0.769\\
3186.62672987035	0.715\\
3196.61207360025	0.7038\\
3206.63580903193	0.6988\\
3216.68169055408	0.722\\
3226.68755637201	0.7494\\
3236.64943652195	0.7686\\
3246.64552318593	0.7752\\
3256.62243837814	0.7724\\
3266.64412745354	0.7698\\
3276.63924858338	0.7362\\
3286.59388013237	0.713\\
3296.60926448947	0.7184\\
3306.7043096063	0.72\\
3316.75082443699	0.7298\\
3326.75125802538	0.7478\\
3336.72785820079	0.753\\
3346.78930237899	0.7704\\
3356.81376035049	0.759\\
3366.84966212459	0.7764\\
3376.81853546467	0.7616\\
3386.90877174882	0.7436\\
3396.9305726078	0.7384\\
3406.94844833985	0.7272\\
3416.92904185956	0.7226\\
3426.93950609642	0.7266\\
3436.92916343528	0.736\\
3446.93031388078	0.762\\
3456.93160359981	0.7344\\
3466.97086807614	0.7314\\
3476.96225710172	0.755\\
3486.99422031484	0.7624\\
3497.04501538557	0.7576\\
3507.07666308382	0.755\\
3517.01500891996	0.742\\
3527.06632460396	0.7484\\
3536.99138059612	0.755\\
3546.96386499451	0.7508\\
3557.08011429288	0.7518\\
3567.10226330739	0.7288\\
3577.12767983329	0.7342\\
3587.1510368531	0.7662\\
3597.16783769466	0.786\\
3607.1584604373	0.8018\\
3617.25116020505	0.808\\
3627.27481714314	0.7754\\
3637.32652682788	0.7328\\
3647.35516196204	0.7564\\
3657.36044046114	0.7624\\
3667.37458702512	0.7598\\
3677.33853089502	0.703\\
3687.36424331588	0.7184\\
3697.38762040895	0.733\\
3707.44870386935	0.7492\\
3717.44231935302	0.737\\
3727.47432735378	0.7392\\
3737.51844982021	0.7684\\
3747.57506371393	0.7778\\
3757.52741775661	0.7672\\
3767.49327095036	0.765\\
3777.4779909284	0.7622\\
3787.53960122904	0.7794\\
3797.48784214858	0.7536\\
3807.46343821575	0.7398\\
3817.48100190897	0.7502\\
3827.50376662695	0.744\\
3837.50324263744	0.7252\\
3847.56862816302	0.7766\\
3857.59226975024	0.7588\\
3867.67233388646	0.7726\\
3877.70199453442	0.7848\\
3887.70018163363	0.7782\\
3897.68582970889	0.7488\\
3907.74641265617	0.721\\
3917.75816156076	0.7396\\
3927.77564018785	0.7182\\
3937.75819753546	0.738\\
3947.81466453016	0.745\\
3957.8454062229	0.7342\\
3967.8636557074	0.739\\
3977.90006474456	0.7388\\
3987.94798281601	0.7672\\
3997.94565528739	0.7728\\
4007.95023586381	0.7286\\
4017.99496445873	0.718\\
4028.04851403976	0.7402\\
4038.14226444461	0.723\\
4048.16852776184	0.7418\\
4058.25125882977	0.7644\\
4068.32154389578	0.7782\\
4078.36970279884	0.742\\
4088.33756919693	0.7416\\
4098.37098077714	0.7286\\
4108.45041671407	0.7078\\
4118.45285396359	0.7006\\
4128.45582721485	0.744\\
4138.50708771588	0.7624\\
4148.58213186447	0.7504\\
4158.62143813391	0.7434\\
4168.56955317076	0.7464\\
4178.54797275468	0.7416\\
4188.61185494908	0.7342\\
4198.6277344536	0.7352\\
4208.64181311721	0.7256\\
4218.60066319366	0.7554\\
4228.55087953658	0.7528\\
4238.59172823213	0.7288\\
4248.60087611898	0.7196\\
4258.68970840554	0.7544\\
4268.77128869036	0.7344\\
4278.80035384331	0.7576\\
4288.86979064444	0.763\\
4298.85552402261	0.7298\\
4308.83748614875	0.7364\\
4318.78584637381	0.7474\\
4328.75650080588	0.7138\\
4338.73148298647	0.698\\
4348.87147290668	0.7396\\
4358.92389562996	0.7378\\
4368.96944331237	0.7164\\
4378.98294003447	0.7124\\
4388.9851966313	0.7252\\
4398.91520803031	0.7378\\
4408.93147690303	0.7608\\
4418.92910090291	0.7838\\
4428.93956100362	0.7772\\
4438.90808140422	0.7874\\
4448.79608873696	0.7806\\
4458.80455613893	0.7798\\
4468.81471321924	0.7886\\
4478.91202271332	0.7832\\
4488.87801560524	0.7636\\
4498.90007768637	0.7552\\
4508.85202339245	0.7668\\
4518.87497626275	0.7752\\
4528.85928929275	0.7604\\
4538.84031770433	0.7736\\
4548.97156329404	0.775\\
4558.95586908994	0.767\\
4568.94728738415	0.7468\\
4578.97451121783	0.7514\\
4589.07066533334	0.7334\\
4599.03742857893	0.7404\\
4609.05000137744	0.7496\\
4619.04089668582	0.7324\\
4629.08082951673	0.734\\
4639.00329368918	0.7426\\
4648.96200070818	0.7432\\
4658.93227255459	0.7444\\
4668.95051142736	0.7758\\
4678.87496662458	0.7358\\
4688.8721497434	0.7348\\
4698.8786273813	0.733\\
4708.8819497724	0.7454\\
4718.89997955093	0.7378\\
4728.92124951297	0.7804\\
4738.99500692076	0.7696\\
4748.99606135959	0.7362\\
4759.02382712347	0.7572\\
4769.06959459414	0.7474\\
4779.06525274322	0.7502\\
4789.08951320957	0.7518\\
4799.05078466112	0.7652\\
4809.11205129269	0.7532\\
4819.14270588833	0.7588\\
4829.18808619831	0.7658\\
4839.14576701554	0.727\\
4849.24672720882	0.7256\\
4859.26174568158	0.7314\\
4869.27640078757	0.7442\\
4879.33574282268	0.762\\
4889.35801707232	0.7546\\
4899.43509529592	0.7626\\
4909.55845221264	0.742\\
4919.52740370069	0.7484\\
4929.60235459759	0.734\\
4939.63763229805	0.6984\\
4949.77330870175	0.7206\\
4959.86357276865	0.7472\\
4969.86962219907	0.7378\\
4979.91321020463	0.742\\
4990.00263683998	0.7382\\
5000.00004418879	0.742\\
};
\end{axis}\end{tikzpicture}}\subfloat[ER, $X_1(0)=0.3$]{\begin{tikzpicture}
\definecolor{mycolor1}{rgb}{0.08,0.7,0.08}%
\begin{axis}[
 axis lines=middle,
 x   axis line style={->},
y   axis line style={-},
    width=\l cm,
height=\h cm,
at={(0cm,0cm)},
scale only axis,
xmin=0,
xtick={0,2000,4000},
xmax=5500,
xlabel={$t$},
ymin=0,
ymax=1,
ylabel={$X_1(t)$},
axis background/.style={fill=white},
]

\addplot [color=mycolor1,solid,forget plot,thick]
  table[row sep=crcr]{%
0	0.3\\
9.98384269874069	0.3348\\
20.0680372549653	0.3056\\
30.1316958867476	0.343\\
40.1798252601232	0.385\\
50.2063276818369	0.413\\
60.2178231189461	0.451\\
70.2718725942829	0.5138\\
80.3272762380685	0.5748\\
90.3631390736912	0.6554\\
100.449438878419	0.7064\\
110.46778514075	0.7376\\
120.484724932875	0.75\\
130.511748336716	0.7358\\
140.544411920304	0.7298\\
150.600458215753	0.7306\\
160.615673532723	0.735\\
170.73200319736	0.7632\\
180.752859084917	0.745\\
190.752306281713	0.7476\\
200.790390879663	0.7752\\
210.760707628751	0.773\\
220.77771702925	0.7516\\
230.753077649517	0.7556\\
240.775037311385	0.7572\\
250.768154230242	0.749\\
260.844015439338	0.74\\
270.819588756353	0.7592\\
280.861809633436	0.7412\\
290.831979951298	0.764\\
300.847086727242	0.763\\
310.890306822971	0.7642\\
320.990536747812	0.7652\\
331.042553872487	0.759\\
341.142252693011	0.7566\\
351.188097985009	0.7446\\
361.124833729452	0.7702\\
371.146233983578	0.7522\\
381.174180735	0.7546\\
391.197368308081	0.738\\
401.230021078205	0.742\\
411.284835272015	0.741\\
421.356150328537	0.7594\\
431.30527542941	0.765\\
441.312432258568	0.7392\\
451.328661569722	0.7492\\
461.375968918566	0.7696\\
471.450356448723	0.7478\\
481.441556301407	0.7498\\
491.440582498062	0.7134\\
501.434084985034	0.705\\
511.531191698151	0.721\\
521.488785975967	0.725\\
531.485017147763	0.7388\\
541.499410463739	0.7526\\
551.487862027014	0.7388\\
561.537301512983	0.7568\\
571.603409061923	0.7728\\
581.561679398795	0.7648\\
591.63463094276	0.7508\\
601.661401532622	0.7542\\
611.73930361904	0.7384\\
621.76378262531	0.736\\
631.831691058866	0.7438\\
641.905475285426	0.7536\\
651.89657679784	0.772\\
661.859932553989	0.7604\\
671.882955312677	0.763\\
681.869879693774	0.7574\\
691.90977226783	0.7586\\
701.941563635139	0.7566\\
711.959094164341	0.7666\\
721.965889858012	0.7646\\
731.98334905817	0.75\\
741.985142054256	0.758\\
752.052799280224	0.7486\\
762.180765930437	0.7542\\
772.259108592987	0.7258\\
782.321011486747	0.748\\
792.344188169374	0.7782\\
802.414419400622	0.7742\\
812.417539690953	0.7578\\
822.514974236546	0.767\\
832.574699591197	0.7552\\
842.606641240893	0.7648\\
852.609212782387	0.7676\\
862.59125582194	0.7774\\
872.698020047764	0.7672\\
882.702977825113	0.771\\
892.640903312848	0.7616\\
902.62346840924	0.7498\\
912.686077693481	0.7668\\
922.754898408152	0.7478\\
932.824137324076	0.7676\\
942.850221452918	0.7772\\
952.810683969279	0.7588\\
962.765509457239	0.7474\\
972.810133070429	0.7462\\
982.807647442318	0.7446\\
992.785644471112	0.7588\\
1002.83745118416	0.7628\\
1012.8419403368	0.728\\
1022.92881988579	0.7528\\
1033.00378391695	0.7518\\
1043.04575765866	0.7414\\
1053.03457988198	0.7396\\
1063.09049536261	0.7614\\
1073.15383318704	0.7438\\
1083.2409824631	0.7468\\
1093.27222414216	0.7624\\
1103.25822847988	0.7626\\
1113.30514913013	0.7656\\
1123.33410873181	0.7334\\
1133.44109683708	0.7294\\
1143.38196416999	0.7792\\
1153.32818877101	0.768\\
1163.36750980126	0.7548\\
1173.36480922406	0.7612\\
1183.39776814482	0.7796\\
1193.35494997802	0.7694\\
1203.39925021773	0.7454\\
1213.44915139008	0.731\\
1223.49706631022	0.7194\\
1233.5607010414	0.7226\\
1243.61640162412	0.7326\\
1253.57642536203	0.7274\\
1263.6689414844	0.7338\\
1273.67020784293	0.7608\\
1283.72236592592	0.7544\\
1293.75569740306	0.7626\\
1303.80505871596	0.7666\\
1313.81904900884	0.7484\\
1323.81541086045	0.7662\\
1333.81203068899	0.7624\\
1343.81520677426	0.7424\\
1353.86269673866	0.7612\\
1363.77968829555	0.7512\\
1373.81340071107	0.7564\\
1383.82394560399	0.746\\
1393.87535627843	0.7382\\
1403.80698799897	0.7424\\
1413.74311411685	0.7474\\
1423.81494744058	0.7466\\
1433.89524683649	0.7456\\
1443.91985584727	0.7568\\
1453.96426166948	0.7404\\
1464.03866037607	0.7418\\
1474.04745712981	0.7348\\
1484.05113859114	0.7332\\
1494.02907256307	0.7452\\
1504.06495222075	0.73\\
1514.04958030795	0.7524\\
1524.11801316358	0.7632\\
1534.17404512585	0.7582\\
1544.23290811815	0.7652\\
1554.25265555124	0.7514\\
1564.21458027085	0.7418\\
1574.21730133891	0.7242\\
1584.31483784537	0.7082\\
1594.39450616305	0.7204\\
1604.46304379079	0.7392\\
1614.42720634072	0.7668\\
1624.50297996883	0.7548\\
1634.40235461293	0.756\\
1644.41840357415	0.7246\\
1654.41575815377	0.7224\\
1664.42824772487	0.7102\\
1674.45116228826	0.74\\
1684.52107257404	0.7496\\
1694.53274524424	0.745\\
1704.5323829908	0.746\\
1714.54243801021	0.7504\\
1724.57706606062	0.7776\\
1734.59735834096	0.7656\\
1744.6057527007	0.767\\
1754.67977594507	0.7814\\
1764.71899067349	0.7588\\
1774.8137502766	0.7384\\
1784.78676068478	0.7376\\
1794.7877462585	0.7152\\
1804.80022891402	0.7292\\
1814.78062886708	0.7462\\
1824.83615347202	0.7504\\
1834.90064725018	0.7406\\
1844.94648932324	0.7512\\
1854.92813720729	0.7626\\
1864.94452073664	0.7408\\
1874.94791020473	0.7514\\
1884.87547779852	0.754\\
1894.89931570154	0.7516\\
1904.89023187676	0.7584\\
1914.87779937338	0.7608\\
1924.84101098075	0.7422\\
1934.91936096003	0.7326\\
1944.94266647688	0.7554\\
1954.96807157523	0.7428\\
1965.03434433667	0.7004\\
1975.00929002246	0.7244\\
1985.01363502261	0.7612\\
1995.0287410767	0.7778\\
2005.0593629255	0.7682\\
2015.12262367953	0.7494\\
2025.07782700737	0.7426\\
2035.03834108819	0.729\\
2045.06093694113	0.7456\\
2055.05704422804	0.7574\\
2065.04927777849	0.7674\\
2075.11256930904	0.763\\
2085.13480992528	0.749\\
2095.14289747826	0.7556\\
2105.10656580639	0.764\\
2115.14412132263	0.7438\\
2125.10449275719	0.7384\\
2135.00822243861	0.74\\
2144.9714054899	0.7756\\
2154.90118157684	0.77\\
2164.89482334018	0.7614\\
2174.9308492773	0.7306\\
2184.9917809962	0.7406\\
2195.0214358881	0.7524\\
2205.06431847319	0.7294\\
2215.08774619292	0.7096\\
2225.14307595379	0.7548\\
2235.15578446893	0.7698\\
2245.25169111568	0.7524\\
2255.24514020978	0.7474\\
2265.28657342761	0.7412\\
2275.31758200375	0.7374\\
2285.27424590183	0.7268\\
2295.3138968672	0.7464\\
2305.32682314053	0.7646\\
2315.34393579596	0.7566\\
2325.38102387233	0.7544\\
2335.40096710858	0.7624\\
2345.43867246186	0.767\\
2355.41023699844	0.7512\\
2365.38339945399	0.7394\\
2375.37389091863	0.753\\
2385.3769935034	0.7568\\
2395.3966933941	0.749\\
2405.38461689822	0.7754\\
2415.37424495055	0.7792\\
2425.35286986902	0.7768\\
2435.30820016276	0.7596\\
2445.31906808378	0.7622\\
2455.32346066958	0.7642\\
2465.38353971802	0.7486\\
2475.3462034564	0.7576\\
2485.37354173161	0.73\\
2495.33139179641	0.7316\\
2505.35987702439	0.7186\\
2515.37752372462	0.7532\\
2525.42458492168	0.7636\\
2535.466171708	0.7362\\
2545.49332943708	0.7508\\
2555.5109987128	0.7578\\
2565.5139763156	0.7234\\
2575.57003105819	0.7282\\
2585.69108445317	0.7188\\
2595.75933806547	0.7164\\
2605.81681320543	0.7236\\
2615.88933077691	0.7204\\
2625.96159966511	0.7398\\
2635.99398669792	0.74\\
2645.99983684292	0.7222\\
2656.11403716895	0.7448\\
2666.16327840382	0.7562\\
2676.22794696357	0.7516\\
2686.32318198356	0.7508\\
2696.31027902428	0.7564\\
2706.33347803378	0.75\\
2716.33579810322	0.7396\\
2726.36258111193	0.7452\\
2736.30568272688	0.747\\
2746.26704074347	0.742\\
2756.29551171381	0.7422\\
2766.27257217232	0.7426\\
2776.27292830042	0.7356\\
2786.25466704609	0.7398\\
2796.25240679681	0.7462\\
2806.26383827116	0.7588\\
2816.22932476625	0.7674\\
2826.28349848259	0.7704\\
2836.40866736985	0.755\\
2846.46439924456	0.745\\
2856.44544189926	0.7466\\
2866.49178972121	0.7398\\
2876.50448112904	0.7442\\
2886.55678831126	0.7382\\
2896.56324599096	0.7378\\
2906.68740999062	0.7288\\
2916.72869069295	0.767\\
2926.64642713062	0.745\\
2936.68384608687	0.753\\
2946.64496244597	0.7576\\
2956.62418650171	0.7588\\
2966.67032895035	0.7624\\
2976.69913475469	0.722\\
2986.74547794722	0.7234\\
2996.8256583038	0.7488\\
3006.89286986703	0.7476\\
3016.89522151317	0.7572\\
3026.90355393264	0.7538\\
3036.91076106346	0.7448\\
3046.91241392332	0.7286\\
3056.96226035831	0.7468\\
3066.95412621656	0.7232\\
3077.00257427343	0.7426\\
3087.04451666318	0.7744\\
3097.13386334802	0.7618\\
3107.20086980255	0.7404\\
3117.1803311783	0.7754\\
3127.20630190271	0.7462\\
3137.16288078303	0.7522\\
3147.13799152727	0.7466\\
3157.20926606743	0.7618\\
3167.24640621461	0.746\\
3177.23990249003	0.7448\\
3187.29509876611	0.7314\\
3197.33736066691	0.7404\\
3207.37127259371	0.7736\\
3217.38823096884	0.7788\\
3227.40553236007	0.747\\
3237.3594382114	0.7396\\
3247.41062782973	0.7472\\
3257.42053377751	0.7666\\
3267.4255759493	0.7564\\
3277.34979637146	0.7554\\
3287.31534235281	0.7388\\
3297.3141343279	0.7598\\
3307.3080144488	0.7636\\
3317.26445166189	0.7438\\
3327.28893457433	0.7628\\
3337.33745214317	0.7518\\
3347.32515868629	0.7624\\
3357.31285514827	0.7524\\
3367.25363864863	0.7694\\
3377.33009634552	0.741\\
3387.34809546111	0.7576\\
3397.24643571582	0.7726\\
3407.34577539405	0.759\\
3417.35133607957	0.7708\\
3427.37920320478	0.771\\
3437.37043834708	0.7556\\
3447.41984634062	0.773\\
3457.44724175053	0.7808\\
3467.45955303824	0.7612\\
3477.43840473463	0.7516\\
3487.36621225605	0.7592\\
3497.44259021104	0.7442\\
3507.49522575235	0.747\\
3517.56580086489	0.7432\\
3527.61919523352	0.7386\\
3537.55268755677	0.7412\\
3547.56668198839	0.7296\\
3557.51929352819	0.7514\\
3567.52021759621	0.7328\\
3577.53630462016	0.7358\\
3587.58616002766	0.749\\
3597.62299878524	0.7748\\
3607.58933641669	0.7626\\
3617.64869188999	0.7516\\
3627.67999247809	0.7582\\
3637.7121168784	0.746\\
3647.74287208863	0.7484\\
3657.69434594687	0.7452\\
3667.67849482592	0.7242\\
3677.79542568256	0.7386\\
3687.8554483053	0.772\\
3697.78175793236	0.7666\\
3707.78986221477	0.746\\
3717.77502233004	0.7572\\
3727.74456059145	0.7404\\
3737.7310717366	0.738\\
3747.7672769357	0.7232\\
3757.78064684	0.7338\\
3767.80160093834	0.7466\\
3777.83960281768	0.7594\\
3787.89230082789	0.7604\\
3797.92485698874	0.7524\\
3807.99792771664	0.7396\\
3817.9831406038	0.736\\
3828.08666577849	0.744\\
3838.10051344142	0.7324\\
3848.10202352679	0.7334\\
3858.20849649117	0.7306\\
3868.20185371622	0.7554\\
3878.22893237235	0.744\\
3888.2990008117	0.7416\\
3898.30262709443	0.7336\\
3908.27499100833	0.7178\\
3918.35869821675	0.7204\\
3928.34626687884	0.726\\
3938.36103435974	0.7308\\
3948.48110445353	0.7628\\
3958.49587706218	0.744\\
3968.48583311546	0.7438\\
3978.52309423543	0.7594\\
3988.6173676768	0.743\\
3998.50250823603	0.746\\
4008.52425518182	0.743\\
4018.56753095817	0.7712\\
4028.5649453388	0.7626\\
4038.62853069547	0.7456\\
4048.66526577266	0.7588\\
4058.68574079725	0.7358\\
4068.70320354342	0.7432\\
4078.71575846784	0.7558\\
4088.75062748641	0.7392\\
4098.81370243484	0.7268\\
4108.74268482668	0.7312\\
4118.76310170545	0.7442\\
4128.71759333411	0.7544\\
4138.69900646577	0.763\\
4148.74964057849	0.7554\\
4158.75773580776	0.7656\\
4168.64806064831	0.74\\
4178.67294917166	0.72\\
4188.7095659171	0.7418\\
4198.70297540583	0.7652\\
4208.7937097375	0.7508\\
4218.72277120928	0.7598\\
4228.75063723768	0.749\\
4238.77091444638	0.7264\\
4248.74457163353	0.7348\\
4258.81525672618	0.7378\\
4268.78297106174	0.7808\\
4278.73184371269	0.7586\\
4288.67772359125	0.7646\\
4298.74362669179	0.7478\\
4308.77761514605	0.7626\\
4318.75093576397	0.7486\\
4328.75133459606	0.7556\\
4338.78828073898	0.7472\\
4348.75732903333	0.7304\\
4358.81798029724	0.7448\\
4368.8267127108	0.7378\\
4378.81112911321	0.7336\\
4388.85657947823	0.7618\\
4398.93557087435	0.7688\\
4408.98313503888	0.7432\\
4418.89913872833	0.746\\
4428.83977774152	0.7634\\
4438.81439364299	0.7312\\
4448.81359566629	0.7414\\
4458.76610708138	0.758\\
4468.81526921343	0.7646\\
4478.8662771915	0.7692\\
4488.9293089319	0.7654\\
4498.96716447098	0.7468\\
4509.03700221975	0.7544\\
4519.10056211567	0.7698\\
4529.16452856363	0.7336\\
4539.16880444542	0.7358\\
4549.15761082573	0.7284\\
4559.23141047328	0.7608\\
4569.1836718318	0.7652\\
4579.17503701035	0.7486\\
4589.17834953538	0.7498\\
4599.23868233994	0.7754\\
4609.24787574101	0.7476\\
4619.26996589717	0.7416\\
4629.33515651146	0.7354\\
4639.37457250115	0.7722\\
4649.40367126966	0.7684\\
4659.39807213749	0.758\\
4669.38983503781	0.7562\\
4679.38956356012	0.77\\
4689.46676330775	0.7604\\
4699.48311024615	0.7374\\
4709.482019397	0.7234\\
4719.5153161862	0.7328\\
4729.55667215846	0.7306\\
4739.55202467902	0.7374\\
4749.5702469811	0.7298\\
4759.56938115207	0.7436\\
4769.57913359012	0.7672\\
4779.55841854727	0.7518\\
4789.55474919813	0.732\\
4799.62621774506	0.7452\\
4809.6414328587	0.7476\\
4819.66954320334	0.7526\\
4829.68362068537	0.7654\\
4839.68271206259	0.7442\\
4849.71644982563	0.7442\\
4859.71321108264	0.7412\\
4869.67661619776	0.7602\\
4879.69299995886	0.717\\
4889.72043137388	0.7444\\
4899.76752993421	0.7384\\
4909.8802754664	0.7486\\
4919.84163072339	0.7568\\
4929.91898250489	0.7508\\
4939.92885055423	0.7304\\
4949.92821145958	0.7288\\
4959.92991859304	0.7488\\
4969.98059076404	0.7224\\
4979.97773474437	0.7366\\
4989.98099325846	0.7548\\
5000.00027348846	0.7592\\
};
\end{axis}\end{tikzpicture}}\\
\subfloat[Lattice, $X_1(0)=0.001$]{\begin{tikzpicture}
\definecolor{mycolor1}{rgb}{0.55,0.4,0.8}%
\begin{axis}[
 axis lines=middle,
 x   axis line style={->},
y   axis line style={-},
    width=\l cm,
height=\h cm,
at={(0cm,0cm)},
scale only axis,
xmin=0,
xtick={0,2000,4000},
xmax=5500,
xlabel={$t$},
ymin=0,
ymax=1,
ylabel={$X_1(t)$},
axis background/.style={fill=white},
]

\addplot [color=mycolor1,solid,forget plot,thick]
  table[row sep=crcr]{%
0	0.00991866693116445\\
10.0175570549527	0.0216226939099385\\
20.0007960841541	0.0287641341003769\\
29.9288911161732	0.0382860543542948\\
39.9445891983268	0.0527673080737949\\
49.9843717579637	0.0660583217615553\\
60.006325750602	0.0886728823646102\\
70.1298005268305	0.0934338424915691\\
80.1388694198442	0.100773656020631\\
90.1597962776877	0.126562190041658\\
100.211446383573	0.121404483237453\\
110.203455292023	0.124975203332672\\
120.323723920607	0.1362824836342\\
130.325803502134	0.145010910533624\\
140.33114262282	0.139258083713549\\
150.46178789948	0.166435231104939\\
160.393278946414	0.166831977782186\\
170.349067976961	0.182701844872049\\
180.353231515279	0.167427097798056\\
190.429947225471	0.183296964887919\\
200.457551115159	0.181709978178933\\
210.483426891942	0.181313231501686\\
220.468111584549	0.179329498115453\\
230.51486771016	0.179527871454077\\
240.549469010079	0.17734576472922\\
250.613208577831	0.185677444951399\\
260.62472149928	0.192223765125967\\
270.634614022142	0.198174965284666\\
280.695010433312	0.199365205316405\\
290.742972089814	0.185875818290022\\
300.712108106575	0.190438405078357\\
310.772374833426	0.200357072009522\\
320.801444705719	0.195199365205316\\
330.872740281805	0.210672485617933\\
340.892262817486	0.213648085697282\\
350.912925243698	0.219996032533228\\
360.947357996893	0.228724459432652\\
370.981181789309	0.212061098988296\\
381.061221413079	0.213846459035906\\
391.058491672126	0.220789525887721\\
401.044019168614	0.22495536599881\\
411.084969451093	0.23943661971831\\
421.052016374377	0.236857766316207\\
431.010030187335	0.233287046220988\\
440.975421723308	0.218409045824241\\
450.974827898973	0.217417179131125\\
460.969010915302	0.222178139258084\\
471.002950884604	0.218210672485618\\
481.073632252333	0.22614560603055\\
491.090503129911	0.215235072406269\\
501.066957958871	0.214441579051775\\
511.084774254392	0.219797659194604\\
521.202876967605	0.243800833168022\\
531.247375474403	0.232096806189248\\
541.269892741258	0.231700059512002\\
551.323002173578	0.233485419559611\\
561.402956573763	0.238444753025193\\
571.352711427166	0.249156913310851\\
581.366455732074	0.249156913310851\\
591.413382184988	0.227732592739536\\
601.461221571569	0.215235072406269\\
611.488319473585	0.211465978972426\\
621.448902097223	0.179924618131323\\
631.520963974061	0.17853600476096\\
641.555236458893	0.182106724856179\\
651.542909516187	0.176552271374727\\
661.571322033342	0.176750644713351\\
671.602261772428	0.178932751438207\\
681.593800212536	0.183098591549296\\
691.593563084882	0.198968458639159\\
701.559620555909	0.207300138861337\\
711.564962338401	0.211267605633803\\
721.560503855624	0.190240031739734\\
731.587188249079	0.194405871850823\\
741.543609085544	0.201348938702638\\
751.555989484801	0.199960325332275\\
761.552088148354	0.204919658797858\\
771.485325650241	0.209482245586193\\
781.486716844018	0.217615552469748\\
791.487813684928	0.229517952787145\\
801.542026997801	0.225352112676056\\
811.515647895992	0.22614560603055\\
821.512553346355	0.228922832771276\\
831.530785885482	0.229319579448522\\
841.516673402733	0.238048006347947\\
851.480911736893	0.258480460226146\\
861.508903757647	0.240825233088673\\
871.603264644212	0.250347153342591\\
881.620819628033	0.238048006347947\\
891.624257777546	0.245586193215632\\
901.614785405387	0.250743900019837\\
911.63666020585	0.264034913707598\\
921.755591370498	0.269390993850427\\
931.788528581224	0.272564967268399\\
941.867416741863	0.273755207300139\\
951.876397987966	0.275342194009125\\
961.854354928519	0.278317794088475\\
971.843665272451	0.275540567347748\\
981.911659895782	0.270581233882166\\
991.920337894468	0.261456060305495\\
1001.93733218483	0.242015473120413\\
1011.93783745172	0.226740726046419\\
1021.97081710713	0.220789525887721\\
1031.99029086751	0.230113072803015\\
1042.12362804135	0.2231700059512\\
1052.19383226406	0.212656219004166\\
1062.24363629302	0.222971632612577\\
1072.26661754682	0.223566752628447\\
1082.29067516272	0.214044832374529\\
1092.33541430246	0.219599285855981\\
1102.33024032025	0.209482245586193\\
1112.42315987687	0.197183098591549\\
1122.4243057551	0.185479071612775\\
1132.40422334408	0.179527871454077\\
1142.34156663447	0.170204324538782\\
1152.30444923035	0.183296964887919\\
1162.34598248108	0.184685578258282\\
1172.35578723303	0.189049791707994\\
1182.33327459214	0.185082324935529\\
1192.31798427254	0.201547312041262\\
1202.34297559927	0.213052965681412\\
1212.38068360933	0.213449712358659\\
1222.46137350574	0.207696885538584\\
1232.4514167669	0.199166831977782\\
1242.49082563837	0.213846459035906\\
1252.51101504763	0.215036699067645\\
1262.52903691042	0.228129339416782\\
1272.51979499931	0.228526086094029\\
1282.52654981169	0.220392779210474\\
1292.48477543782	0.215631819083515\\
1302.54586512146	0.219202539178734\\
1312.58860726716	0.21047411227931\\
1322.68280373339	0.204721285459234\\
1332.73220080669	0.213449712358659\\
1342.72664940742	0.214441579051775\\
1352.74030427225	0.196984725252926\\
1362.80850998772	0.206308272168221\\
1372.88250676754	0.219400912517358\\
1382.87880625734	0.215433445744892\\
1392.94214423966	0.196984725252926\\
1402.99650942421	0.19599285855981\\
1413.07113061276	0.190041658401111\\
1423.06758068779	0.195794485221186\\
1432.9984396808	0.184288831581036\\
1442.96964534676	0.193810751834953\\
1453.02173602698	0.191231898432851\\
1462.94750962228	0.205713152152351\\
1472.8885430513	0.224161872644317\\
1482.88441135417	0.245586193215632\\
1492.94882023428	0.266812140448324\\
1503.08182882864	0.260067446935132\\
1513.10426128253	0.233287046220988\\
1523.1607382867	0.262447926998611\\
1533.24989292841	0.268200753818687\\
1543.2653335786	0.273358460622892\\
1553.24375600809	0.246379686570125\\
1563.25199600125	0.256695100178536\\
1573.22335598923	0.250743900019837\\
1583.22116061305	0.257290220194406\\
1593.24394599202	0.253521126760563\\
1603.22711323386	0.258877206903392\\
1613.21605802558	0.257091846855783\\
1623.18026179953	0.239833366395556\\
1633.24128284153	0.231700059512002\\
1643.30189152594	0.241023606427296\\
1653.31075117156	0.247173179924618\\
1663.26036733627	0.235270779607221\\
1673.30444972975	0.218012299146995\\
1683.29055041643	0.227335846062289\\
1693.32330644974	0.248363419956358\\
1703.35492802426	0.237849633009324\\
1713.35718067035	0.233882166236858\\
1723.41959680832	0.233287046220988\\
1733.47631627539	0.23824637968657\\
1743.47663099335	0.235072406268597\\
1753.491711715	0.259273953580639\\
1763.58426503068	0.26839912715731\\
1773.57397445918	0.275738940686372\\
1783.62670769872	0.280499900813331\\
1793.63633730449	0.263043047014481\\
1803.62488695532	0.283078754215433\\
1813.61126520598	0.296568141241817\\
1823.63539558965	0.321959928585598\\
1833.68900485781	0.32870462209879\\
1843.73393888786	0.317397341797262\\
1853.74204481155	0.321761555246975\\
1863.75609446914	0.314024995040667\\
1873.77684485046	0.314620115056536\\
1883.75194546727	0.308073794881968\\
1893.78808675024	0.320174568537988\\
1903.80076866696	0.321166435231105\\
1913.8150645185	0.319182701844872\\
1923.9045112088	0.334655822257489\\
1933.96905864428	0.348740329299742\\
1944.10883619237	0.353898036103948\\
1954.17093862482	0.355485022812934\\
1964.22384043592	0.355286649474311\\
1974.2631313478	0.37274350327316\\
1984.25065535819	0.372941876611783\\
1994.2645138123	0.387026383654037\\
2004.27516908255	0.403888117437016\\
2014.35687847498	0.424915691331085\\
2024.41249376534	0.448522118627256\\
2034.37577566327	0.446538385241024\\
2044.3646052915	0.459631025590161\\
2054.44019412353	0.501686173378298\\
2064.36786293607	0.540170601071216\\
2074.37140522169	0.563181908351518\\
2084.49418553307	0.579051775441381\\
2094.47300291166	0.59769886927197\\
2104.48971384413	0.625272763340607\\
2114.54754110997	0.654830390795477\\
2124.59872390282	0.669708391192224\\
2134.66434170044	0.681610791509621\\
2144.67725023516	0.68458639158897\\
2154.65735919212	0.697282285260861\\
2164.69870910372	0.712953779012101\\
2174.62639676157	0.723070819281889\\
2184.63911494962	0.721682205911526\\
2194.67769040313	0.739139059710375\\
2204.69131529883	0.744098393175957\\
2214.70099460392	0.74608212656219\\
2224.76092961386	0.750644713350526\\
2234.70775320416	0.734378099583416\\
2244.72772597364	0.744098393175957\\
2254.70683584408	0.746875619916683\\
2264.72574055972	0.753223566752628\\
2274.79631931605	0.764332473715533\\
2284.78759491578	0.751239833366396\\
2294.83728210149	0.755405673477485\\
2304.79772492197	0.752628446736759\\
2314.75239696235	0.755207300138861\\
2324.72211027641	0.7623487403293\\
2334.77441627128	0.755207300138861\\
2344.82609100895	0.759174766911327\\
2354.87321863475	0.769688553858361\\
2364.9276881504	0.7478674866098\\
2374.9497231578	0.750049593334656\\
2384.96075287279	0.744495139853204\\
2394.88123205324	0.741321166435231\\
2404.90279830994	0.754810553461615\\
2414.85587344622	0.751834953382265\\
2424.90773802993	0.755604046816108\\
2434.9439622158	0.749652846657409\\
2445.04267580674	0.739734179726245\\
2455.13365989105	0.743900019837334\\
2465.12351926186	0.771077167228724\\
2475.08528466629	0.77742511406467\\
2485.08047291457	0.781789327514382\\
2495.19436193107	0.784764927593731\\
2505.21921041437	0.770878793890101\\
2515.21699428284	0.765324340408649\\
2525.18588710867	0.761951993652053\\
2535.24864082785	0.766712953779012\\
2545.24149625913	0.771473913905971\\
2555.31938561369	0.768101567149375\\
2565.30519221316	0.76175362031343\\
2575.37092624449	0.768299940487998\\
2585.42846916991	0.761555246974807\\
2595.43827855021	0.757587780202341\\
2605.49446161635	0.751834953382265\\
2615.49121818192	0.739535806387622\\
2625.46934816575	0.759571513588574\\
2635.46583242903	0.7623487403293\\
2645.46450485978	0.754612180122991\\
2655.50915822126	0.7478674866098\\
2665.52295085639	0.745883753223567\\
2675.60203814027	0.743701646498711\\
2685.57622641294	0.729617139456457\\
2695.53083735932	0.745288633207697\\
2705.58694346197	0.752033326720889\\
2715.58826519447	0.74727236659393\\
2725.6709819974	0.738147193017259\\
2735.71116216641	0.739139059710375\\
2745.80110250179	0.744495139853204\\
2755.85183234643	0.758579646895457\\
2765.86019483843	0.756595913509224\\
2775.86430476632	0.757587780202341\\
2785.90660172122	0.756595913509224\\
2795.87060699266	0.738345566355882\\
2805.93134367723	0.750049593334656\\
2815.9099821391	0.730013886133704\\
2825.98620007171	0.745288633207697\\
2836.02791119136	0.741717913112478\\
2846.09605317686	0.747073993255306\\
2856.0801182534	0.746478873239437\\
2866.18691964505	0.75996826026582\\
2876.1915393713	0.754810553461615\\
2886.26133549406	0.748264233287046\\
2896.20739901954	0.727633406070224\\
2906.24649496833	0.737948819678635\\
2916.30521088832	0.717913112477683\\
2926.3238956944	0.744693513191827\\
2936.33376550208	0.750843086689149\\
2946.43236279937	0.750644713350526\\
2956.52610229198	0.740130926403491\\
2966.51555263907	0.743900019837334\\
2976.56870870528	0.749454473318786\\
2986.6164959419	0.741916286451101\\
2996.58612259806	0.743106526482841\\
3006.59374291347	0.725054552668121\\
3016.57128015979	0.724459432652252\\
3026.61813654004	0.723070819281889\\
3036.64540038149	0.711565165641738\\
3046.60822218948	0.723269192620512\\
3056.65317872728	0.738147193017259\\
3066.72242860917	0.724062685975005\\
3076.76323312451	0.731799246181313\\
3086.80940980778	0.742709779805594\\
3096.8728444632	0.741717913112478\\
3106.9035159452	0.757786153540964\\
3116.88376836364	0.761951993652053\\
3126.91155757316	0.760365006943067\\
3136.90689153934	0.758381273556834\\
3147.01974003533	0.75937314024995\\
3157.03171798682	0.765125967070026\\
3167.05265525461	0.756000793493354\\
3177.09108632478	0.740527673080738\\
3187.14924987565	0.732394366197183\\
3197.04966475085	0.730609006149573\\
3207.04831713406	0.727831779408847\\
3217.07974084512	0.728030152747471\\
3227.12247772414	0.732592739535806\\
3237.1192343676	0.739337433048998\\
3247.06486692322	0.733187859551676\\
3257.08701319725	0.722475699266019\\
3267.15290649055	0.722078952588772\\
3277.09473743992	0.736361832969649\\
3287.21239307329	0.739337433048998\\
3297.18632805863	0.726244792699861\\
3307.24101753907	0.728030152747471\\
3317.30332197633	0.725252926006745\\
3327.36158977109	0.737155326324142\\
3337.36066545071	0.744098393175957\\
3347.33641082333	0.750843086689149\\
3357.42483552492	0.748859353302916\\
3367.41396968069	0.744693513191827\\
3377.36894905367	0.747470739932553\\
3387.36699740798	0.757191033525094\\
3397.34517001249	0.742313033128348\\
3407.30807840825	0.751834953382265\\
3417.30013031757	0.736560206308272\\
3427.32090236537	0.756000793493354\\
3437.35862500718	0.737552073001389\\
3447.4173755212	0.748065859948423\\
3457.41597781205	0.752430073398135\\
3467.45343955102	0.726839912715731\\
3477.45694515898	0.742511406466971\\
3487.50807681149	0.743304899821464\\
3497.53920592968	0.749057726641539\\
3507.50327718309	0.745090259869074\\
3517.5079512354	0.758381273556834\\
3527.52623937513	0.76115850029756\\
3537.5236347327	0.752826820075382\\
3547.48005406681	0.736163459631026\\
3557.50312497541	0.725451299345368\\
3567.50192828604	0.711565165641738\\
3577.44841239057	0.725252926006745\\
3587.43727401536	0.734973219599286\\
3597.5280276605	0.746875619916683\\
3607.52879717183	0.742908153144217\\
3617.51928313787	0.751834953382265\\
3627.57723109309	0.756199166831978\\
3637.59570331113	0.759174766911327\\
3647.58747582216	0.759769886927197\\
3657.58237061593	0.772862527276334\\
3667.57013385673	0.774846260662567\\
3677.59238989184	0.774251140646697\\
3687.62614455324	0.7768299940488\\
3697.62425512422	0.772862527276334\\
3707.62547050682	0.780400714144019\\
3717.71537599252	0.766316207101766\\
3727.68256519729	0.760960126958937\\
3737.72009817673	0.763935727038286\\
3747.77904824748	0.751239833366396\\
3757.85817523713	0.753818686768498\\
3767.90992811795	0.761356873636183\\
3777.90102817916	0.756000793493354\\
3787.88093975015	0.745288633207697\\
3797.86459518206	0.738345566355882\\
3807.89251207498	0.746875619916683\\
3817.91194114098	0.744098393175957\\
3827.91146763875	0.742313033128348\\
3837.8569956222	0.748859353302916\\
3847.87086035811	0.756000793493354\\
3857.83814995412	0.765125967070026\\
3867.88572129358	0.77623487403293\\
3877.93705544553	0.779805594128149\\
3887.92401761363	0.778615354096409\\
3898.06312315464	0.786947034318588\\
3908.08692085943	0.777028367387423\\
3918.08100354808	0.756199166831978\\
3928.11882520385	0.758381273556834\\
3937.98187083476	0.751239833366396\\
3948.01148742062	0.766316207101766\\
3958.04171300259	0.76115850029756\\
3968.1105098057	0.760166633604444\\
3978.10437243336	0.756794286847848\\
3988.13087282403	0.766712953779012\\
3998.21399750189	0.757786153540964\\
4008.2072906973	0.749256099980163\\
4018.21202103845	0.740924419757985\\
4028.19452642775	0.74608212656219\\
4038.2153983344	0.755802420154731\\
4048.26384633709	0.748065859948423\\
4058.26293668724	0.735171592937909\\
4068.26793180821	0.73041063281095\\
4078.27606220211	0.736758579646896\\
4088.27079421541	0.738345566355882\\
4098.24842390553	0.750843086689149\\
4108.24843286036	0.765919460424519\\
4118.34321113257	0.765324340408649\\
4128.44375915345	0.757191033525094\\
4138.54458742376	0.758579646895457\\
4148.50447925133	0.747669113271176\\
4158.56387964493	0.739139059710375\\
4168.6021185384	0.753421940091252\\
4178.58950587771	0.7623487403293\\
4188.56998066321	0.779012100773656\\
4198.59554979761	0.775838127355683\\
4208.66467144299	0.783574687561992\\
4218.7510457108	0.798254314620115\\
4228.71928876895	0.778416980757786\\
4238.75288868259	0.76175362031343\\
4248.82619645445	0.768101567149375\\
4258.78313755753	0.759571513588574\\
4268.90001866699	0.775838127355683\\
4278.92227332811	0.756794286847848\\
4288.9135761634	0.755405673477485\\
4298.91653994103	0.754215433445745\\
4308.97940930361	0.753025193414005\\
4318.97660809068	0.752231700059512\\
4328.97701667185	0.752628446736759\\
4338.96247514629	0.747669113271176\\
4348.9386727249	0.752033326720889\\
4358.98251955232	0.755207300138861\\
4368.97032296012	0.758579646895457\\
4378.96169021011	0.745883753223567\\
4389.02124026764	0.74667724657806\\
4399.06793849257	0.758381273556834\\
4409.09432243011	0.759769886927197\\
4419.09092683217	0.759174766911327\\
4429.10449213299	0.757984526879587\\
4439.1168003177	0.755604046816108\\
4449.10160983918	0.753025193414005\\
4459.09668831531	0.771473913905971\\
4469.14853363869	0.76115850029756\\
4479.16902185925	0.770482047212855\\
4489.20562591424	0.769886927196985\\
4499.19547986027	0.773259273953581\\
4509.1372217894	0.763935727038286\\
4519.18605444505	0.769291807181115\\
4529.15837469743	0.765919460424519\\
4539.06800478292	0.757191033525094\\
4549.11556653223	0.751041460027772\\
4559.04985020871	0.744495139853204\\
4569.04743974918	0.753421940091252\\
4579.07816280369	0.757191033525094\\
4589.12438056224	0.756595913509224\\
4599.11950318205	0.769291807181115\\
4609.15018138733	0.754017060107122\\
4619.15694100747	0.74489188653045\\
4629.20177922603	0.756992660186471\\
4639.28315423396	0.76175362031343\\
4649.30341496021	0.753025193414005\\
4659.27720232576	0.758976393572704\\
4669.30071463583	0.757389406863717\\
4679.31158110929	0.753223566752628\\
4689.31982814463	0.749057726641539\\
4699.33313816214	0.755604046816108\\
4709.31174463237	0.76294386034517\\
4719.30409657406	0.761555246974807\\
4729.39410966759	0.75937314024995\\
4739.39847000389	0.750843086689149\\
4749.45816200927	0.756992660186471\\
4759.51058830481	0.760365006943067\\
4769.53306294789	0.750446340011902\\
4779.55736197337	0.751041460027772\\
4789.58011013817	0.755802420154731\\
4799.64369845234	0.739337433048998\\
4809.63275679425	0.746280499900813\\
4819.59452001232	0.747470739932553\\
4829.62471490572	0.734179726244793\\
4839.66955343728	0.73160087284269\\
4849.76718914763	0.73041063281095\\
4859.76173650028	0.735766712953779\\
4869.81053824451	0.751239833366396\\
4879.8653500331	0.750843086689149\\
4889.91525904924	0.743900019837334\\
4899.9418093668	0.728030152747471\\
4909.91900914507	0.733584606228923\\
4919.93799818351	0.727831779408847\\
4929.97731029787	0.728030152747471\\
4939.99815880362	0.743304899821464\\
4950.01926138325	0.74846260662567\\
4960.02249736153	0.744693513191827\\
4969.98994975306	0.748859353302916\\
4980.05171437588	0.752231700059512\\
4990.01299974159	0.737948819678635\\
5000.00029267898	0.726046419361238\\
};
\end{axis}\end{tikzpicture}}\subfloat[Lattice, $X_1(0)=0.3$]{\begin{tikzpicture}
\definecolor{mycolor1}{rgb}{0.55,0.4,0.8}%
\begin{axis}[
 axis lines=middle,
 x   axis line style={->},
y   axis line style={-},
    width=\l cm,
height=\h cm,
at={(0cm,0cm)},
scale only axis,
xmin=0,
xtick={0,2000,4000},
xmax=5500,
xlabel={$t$},
ymin=0,
ymax=1,
ylabel={$X_1(t)$},
axis background/.style={fill=white},
]

\addplot [color=mycolor1,solid,forget plot,thick]
  table[row sep=crcr]{%
0	0.297560007934934\\
10.0915573719789	0.304304701448125\\
20.1485342029323	0.332275342194009\\
30.239658515109	0.330291608807776\\
40.3133853167187	0.343979369172783\\
50.2850388791781	0.360245982939893\\
60.3218942236025	0.369172783177941\\
70.3191642192038	0.427891291410434\\
80.3321930847158	0.47649275937314\\
90.4287463906386	0.490180519738147\\
100.465743243088	0.532037294187661\\
110.416223714457	0.540765721087086\\
120.471934255623	0.61218012299147\\
130.509926812253	0.62785161674271\\
140.488640579715	0.662170204324539\\
150.45880356922	0.682801031541361\\
160.585480943775	0.698075778615354\\
170.674011741972	0.71771473913906\\
180.726512850161	0.716326125768697\\
190.73620917178	0.720095219202539\\
200.693160562492	0.727038286054354\\
210.7125992247	0.737948819678635\\
220.723657261838	0.736956952985519\\
230.795683433536	0.742908153144217\\
240.765450521366	0.73100575282682\\
250.819724198217	0.725054552668121\\
260.759585087848	0.745090259869074\\
270.77338974485	0.717317992461813\\
280.839854486652	0.728823646101964\\
290.861737343605	0.735766712953779\\
300.896843549817	0.756595913509224\\
310.911356725186	0.744693513191827\\
320.925326335692	0.749057726641539\\
330.935315045283	0.746478873239437\\
340.969657633869	0.734973219599286\\
351.066803756137	0.74429676651458\\
361.095126200288	0.755604046816108\\
371.142539236247	0.750049593334656\\
381.15068407551	0.745288633207697\\
391.099167295241	0.753620313429875\\
401.177530130868	0.762150366990676\\
411.214403521751	0.747470739932553\\
421.194286641221	0.733584606228923\\
431.165014025257	0.75937314024995\\
441.187775452222	0.764530847054156\\
451.228539286563	0.749256099980163\\
461.189092973558	0.746280499900813\\
471.185035937763	0.740527673080738\\
481.189134900985	0.739139059710375\\
491.165831502761	0.735369966276532\\
501.160190635802	0.734774846260663\\
511.165434721573	0.733981352906169\\
521.131252590179	0.74489188653045\\
531.163078923559	0.751239833366396\\
541.173268400402	0.756992660186471\\
551.216423863648	0.753223566752628\\
561.272187128037	0.7478674866098\\
571.247561135704	0.748065859948423\\
581.287977788384	0.753421940091252\\
591.218885521756	0.752430073398135\\
601.202179830733	0.756992660186471\\
611.234046007652	0.745685379884943\\
621.208476087657	0.745288633207697\\
631.277339299977	0.742709779805594\\
641.299959314521	0.745685379884943\\
651.306519972348	0.735171592937909\\
661.341546218503	0.739337433048998\\
671.324136961789	0.739932553064868\\
681.391465367602	0.742313033128348\\
691.442145325949	0.742511406466971\\
701.485272089171	0.733782979567546\\
711.505537780729	0.738543939694505\\
721.482703823301	0.7478674866098\\
731.555435970397	0.736758579646896\\
741.563477677414	0.741519539773854\\
751.599423776222	0.741717913112478\\
761.607434061535	0.74608212656219\\
771.605403959892	0.748264233287046\\
781.649208700471	0.754017060107122\\
791.614045825862	0.756992660186471\\
801.563927197697	0.748660979964293\\
811.533170751211	0.749851219996033\\
821.649222749469	0.753620313429875\\
831.696257722218	0.754215433445745\\
841.74786533605	0.749454473318786\\
851.777405751008	0.741122793096608\\
861.790953400533	0.743106526482841\\
871.80176258186	0.743900019837334\\
881.777676014899	0.755207300138861\\
891.784054503555	0.739734179726245\\
901.787080891536	0.743304899821464\\
911.927514138118	0.724261059313628\\
921.959737975771	0.719103352509423\\
931.949264152435	0.742313033128348\\
941.950465244086	0.743701646498711\\
951.985616290421	0.748065859948423\\
961.969870534132	0.730807379488197\\
972.040537045711	0.739734179726245\\
982.027711234572	0.769291807181115\\
992.088337161603	0.779408847450903\\
1002.06834623877	0.784368180916485\\
1012.1581097587	0.780599087482642\\
1022.20604209418	0.760761753620313\\
1032.24416040741	0.765721087085896\\
1042.29259863783	0.771473913905971\\
1052.30091363164	0.76353898036104\\
1062.29359018532	0.762745487006546\\
1072.31920378219	0.766712953779012\\
1082.30031634406	0.769093433842492\\
1092.29287341385	0.769688553858361\\
1102.33636208838	0.77742511406467\\
1112.31528124364	0.770482047212855\\
1122.34961488343	0.784169807577862\\
1132.37378607448	0.756992660186471\\
1142.308225218	0.766117833763142\\
1152.33104487061	0.751438206705019\\
1162.37324102893	0.744098393175957\\
1172.43947396371	0.740329299742115\\
1182.47684467495	0.741321166435231\\
1192.46333939236	0.741916286451101\\
1202.57147142723	0.74846260662567\\
1212.54483329605	0.74429676651458\\
1222.60866434131	0.752826820075382\\
1232.58600280018	0.760761753620313\\
1242.65464625778	0.758182900218211\\
1252.60841173237	0.743900019837334\\
1262.53885778181	0.745288633207697\\
1272.55221084122	0.746478873239437\\
1282.57569317664	0.745090259869074\\
1292.56951132414	0.755604046816108\\
1302.58950974879	0.750644713350526\\
1312.56665776637	0.739139059710375\\
1322.57515433513	0.736758579646896\\
1332.55275093477	0.735965086292402\\
1342.62051391911	0.735171592937909\\
1352.64169476673	0.751041460027772\\
1362.59234375997	0.758381273556834\\
1372.58117403334	0.776036500694307\\
1382.67105709322	0.7623487403293\\
1392.75709486771	0.760365006943067\\
1402.75638803891	0.75937314024995\\
1412.81691599919	0.766316207101766\\
1422.78126291763	0.751834953382265\\
1432.76215125641	0.748859353302916\\
1442.81089652988	0.735568339615156\\
1452.82807592003	0.736361832969649\\
1462.79670305501	0.740726046419361\\
1472.84632893486	0.770085300535608\\
1482.89859222741	0.762150366990676\\
1492.90119488414	0.756595913509224\\
1502.93459812285	0.746280499900813\\
1512.9623923678	0.764134100376909\\
1522.99811992029	0.761356873636183\\
1533.07284884761	0.762150366990676\\
1543.07875865833	0.75996826026582\\
1553.0427843252	0.751834953382265\\
1562.9967206084	0.746478873239437\\
1573.02296473952	0.74846260662567\\
1583.05541746574	0.749454473318786\\
1593.08211753954	0.750843086689149\\
1603.05846859392	0.756595913509224\\
1613.07488617514	0.76175362031343\\
1623.08595192156	0.7623487403293\\
1633.15980015876	0.782979567546122\\
1643.20022795348	0.774647887323944\\
1653.25822916903	0.766712953779012\\
1663.25612861635	0.753818686768498\\
1673.29330081921	0.76056338028169\\
1683.18725422395	0.75937314024995\\
1693.11370140937	0.76115850029756\\
1703.16569420174	0.762745487006546\\
1713.2509814364	0.765919460424519\\
1723.22827434428	0.757786153540964\\
1733.24651399624	0.762547113667923\\
1743.27280700752	0.754810553461615\\
1753.30571908183	0.740527673080738\\
1763.23330758629	0.730013886133704\\
1773.20040670627	0.733782979567546\\
1783.22513205335	0.728426899424717\\
1793.20213877843	0.734774846260663\\
1803.15597222886	0.74608212656219\\
1813.21783280336	0.7478674866098\\
1823.21199948783	0.745288633207697\\
1833.1761746405	0.756794286847848\\
1843.25819816224	0.756199166831978\\
1853.30384379183	0.758381273556834\\
1863.30968142667	0.7478674866098\\
1873.27590969529	0.74727236659393\\
1883.27587769964	0.743701646498711\\
1893.29576985444	0.750644713350526\\
1903.30538626025	0.767109700456259\\
1913.3761590768	0.76115850029756\\
1923.32919313247	0.752826820075382\\
1933.33908736978	0.753025193414005\\
1943.38981198455	0.755802420154731\\
1953.47769821795	0.764134100376909\\
1963.50933946066	0.754612180122991\\
1973.53641493664	0.741916286451101\\
1983.53718066694	0.749851219996033\\
1993.53638531764	0.739734179726245\\
2003.47678169057	0.750843086689149\\
2013.48557817291	0.764332473715533\\
2023.49814047942	0.770283673874231\\
2033.47773754559	0.772069033921841\\
2043.42066644449	0.771473913905971\\
2053.4192531716	0.757786153540964\\
2063.49018899687	0.753025193414005\\
2073.46332583315	0.750446340011902\\
2083.52509269818	0.755207300138861\\
2093.55531168336	0.749256099980163\\
2103.62473486684	0.756000793493354\\
2113.68202956666	0.754612180122991\\
2123.65106303081	0.749851219996033\\
2133.62677634878	0.743701646498711\\
2143.56763800942	0.733981352906169\\
2153.61256905775	0.740130926403491\\
2163.62797268525	0.736560206308272\\
2173.66203834658	0.739734179726245\\
2183.66871295941	0.734774846260663\\
2193.64589214656	0.755405673477485\\
2203.7138071931	0.764332473715533\\
2213.79262736147	0.767704820472129\\
2223.75078255285	0.754612180122991\\
2233.7887971486	0.752826820075382\\
2243.78270036583	0.759769886927197\\
2253.87862030293	0.768895060503868\\
2263.94674891734	0.758182900218211\\
2274.0188214727	0.777028367387423\\
2284.07704164557	0.759174766911327\\
2294.0330120921	0.751438206705019\\
2304.02467748227	0.75937314024995\\
2313.99337213102	0.747073993255306\\
2324.02635016166	0.745685379884943\\
2334.10056646774	0.730212259472327\\
2344.07753558472	0.722674072604642\\
2354.12229293314	0.737155326324142\\
2364.09093857225	0.740329299742115\\
2374.07093328945	0.74489188653045\\
2384.0491227771	0.737155326324142\\
2394.06255044849	0.748660979964293\\
2404.10241616699	0.754612180122991\\
2414.08628709126	0.754612180122991\\
2424.08523857675	0.751438206705019\\
2434.15015728839	0.740329299742115\\
2444.15549335149	0.747470739932553\\
2454.11307300727	0.734378099583416\\
2464.13249341574	0.736956952985519\\
2474.15302638407	0.739337433048998\\
2484.17668031974	0.737155326324142\\
2494.21062720152	0.735369966276532\\
2504.22681207196	0.738147193017259\\
2514.25375796308	0.739337433048998\\
2524.22440485605	0.727038286054354\\
2534.20699837847	0.729022019440587\\
2544.19402777986	0.720690339218409\\
2554.22216684943	0.710771672287245\\
2564.22539713864	0.723269192620512\\
2574.27795630109	0.712755405673478\\
2584.28673902274	0.728228526086094\\
2594.31548114156	0.72981551279508\\
2604.29346422786	0.742709779805594\\
2614.25009284651	0.734973219599286\\
2624.20326818256	0.740924419757985\\
2634.28219805465	0.742908153144217\\
2644.36126937883	0.733187859551676\\
2654.42915482412	0.719103352509423\\
2664.46785617654	0.724856179329498\\
2674.50373761602	0.730013886133704\\
2684.58863185834	0.723864312636382\\
2694.51934421293	0.73279111287443\\
2704.6143162388	0.729022019440587\\
2714.65399717109	0.730212259472327\\
2724.69743372367	0.719698472525293\\
2734.7537642272	0.731799246181313\\
2744.8749120217	0.737353699662765\\
2754.88046638367	0.736361832969649\\
2764.88394470865	0.748660979964293\\
2774.88403721845	0.740130926403491\\
2784.87591371465	0.74489188653045\\
2794.99549517207	0.738147193017259\\
2805.03089194925	0.741321166435231\\
2815.04001159605	0.735568339615156\\
2825.0971188512	0.742511406466971\\
2835.01987159921	0.746478873239437\\
2844.99064092217	0.746875619916683\\
2854.97608940649	0.756595913509224\\
2865.02612513034	0.766316207101766\\
2875.09595945278	0.7623487403293\\
2885.11467376547	0.74608212656219\\
2895.21728113451	0.741321166435231\\
2905.20558584887	0.746280499900813\\
2915.21453194906	0.756397540170601\\
2925.30481870195	0.76175362031343\\
2935.30606105281	0.750843086689149\\
2945.27749397887	0.743106526482841\\
2955.28506632342	0.73279111287443\\
2965.28604438133	0.745685379884943\\
2975.30113001836	0.738345566355882\\
2985.27804610881	0.755405673477485\\
2995.35914538569	0.759571513588574\\
3005.40442419835	0.76115850029756\\
3015.44341395241	0.757587780202341\\
3025.48255854379	0.762150366990676\\
3035.51885273028	0.760166633604444\\
3045.56746016539	0.768696687165245\\
3055.56691563102	0.766712953779012\\
3065.70615764283	0.764530847054156\\
3075.68708374619	0.765721087085896\\
3085.75186208125	0.755604046816108\\
3095.74184980216	0.755604046816108\\
3105.72222388242	0.751041460027772\\
3115.76357100432	0.758182900218211\\
3125.83054057997	0.76115850029756\\
3135.90834127089	0.764729220392779\\
3145.94625035739	0.766712953779012\\
3155.98304541359	0.755405673477485\\
3166.02653205573	0.749652846657409\\
3176.02831647235	0.74846260662567\\
3186.01484455498	0.751239833366396\\
3196.03659249424	0.748264233287046\\
3206.11689152938	0.749454473318786\\
3216.08744946295	0.751438206705019\\
3226.07262685078	0.754413806784368\\
3236.08728924184	0.762745487006546\\
3246.14536513738	0.762150366990676\\
3256.16035470889	0.745685379884943\\
3266.26273545327	0.740130926403491\\
3276.2901488415	0.743503273160087\\
3286.26455345797	0.750049593334656\\
3296.30108705967	0.749057726641539\\
3306.33202674025	0.752033326720889\\
3316.37146465592	0.758976393572704\\
3326.36274893721	0.753025193414005\\
3336.40514208235	0.76056338028169\\
3346.42724224308	0.773656020630827\\
3356.44436973504	0.776036500694307\\
3366.50364240816	0.768299940487998\\
3376.5638114423	0.771077167228724\\
3386.57871374171	0.767109700456259\\
3396.64176491495	0.763340607022416\\
3406.67868588906	0.74429676651458\\
3416.73078993211	0.737948819678635\\
3426.74799663611	0.739337433048998\\
3436.79990113537	0.754810553461615\\
3446.88898646014	0.745883753223567\\
3456.88841349035	0.745685379884943\\
3466.89222924702	0.741321166435231\\
3476.82285193109	0.74548700654632\\
3486.78652592077	0.759769886927197\\
3496.81661572026	0.758182900218211\\
3506.85806294628	0.758778020234081\\
3516.92826375444	0.754612180122991\\
3526.97519233446	0.756199166831978\\
3537.04841057975	0.751239833366396\\
3547.0778580785	0.745090259869074\\
3557.08235960172	0.74429676651458\\
3567.11398269176	0.754215433445745\\
3577.18905346758	0.763142233683793\\
3587.15289439451	0.753421940091252\\
3597.14751189587	0.745685379884943\\
3607.15461593032	0.752033326720889\\
3617.1492071607	0.758182900218211\\
3627.12950175707	0.7623487403293\\
3637.19241398963	0.768299940487998\\
3647.27862313593	0.765919460424519\\
3657.2733144991	0.75937314024995\\
3667.3138729571	0.753421940091252\\
3677.31014640298	0.74489188653045\\
3687.31633968169	0.739139059710375\\
3697.39018368278	0.735568339615156\\
3707.44835502577	0.732592739535806\\
3717.52419800706	0.736163459631026\\
3727.57858028671	0.736163459631026\\
3737.6511314027	0.73100575282682\\
3747.67137637857	0.72922039277921\\
3757.68675130394	0.746875619916683\\
3767.71974526018	0.754612180122991\\
3777.70002743555	0.744693513191827\\
3787.72515133428	0.749454473318786\\
3797.70120336268	0.757786153540964\\
3807.73237236978	0.745883753223567\\
3817.69613173481	0.758579646895457\\
3827.69017151221	0.74429676651458\\
3837.76531032801	0.744693513191827\\
3847.77034474749	0.730807379488197\\
3857.77337939949	0.730807379488197\\
3867.67474553989	0.727435032731601\\
3877.68003636601	0.739535806387622\\
3887.72402674569	0.727038286054354\\
3897.7805156629	0.727633406070224\\
3907.86633945622	0.724261059313628\\
3917.81541830299	0.733187859551676\\
3927.83981852284	0.739932553064868\\
3937.83641388292	0.739932553064868\\
3947.90829304459	0.729418766117834\\
3957.95654705395	0.734378099583416\\
3968.01747505631	0.742313033128348\\
3978.07475667347	0.748660979964293\\
3988.16280872521	0.754810553461615\\
3998.23752633953	0.748660979964293\\
4008.2710706636	0.734973219599286\\
4018.23201522108	0.749256099980163\\
4028.26060460285	0.752628446736759\\
4038.32914546502	0.754017060107122\\
4048.41714432525	0.739337433048998\\
4058.39691184271	0.729022019440587\\
4068.45303770238	0.734179726244793\\
4078.50893462101	0.753025193414005\\
4088.55203728504	0.749057726641539\\
4098.55299088594	0.742908153144217\\
4108.47521612907	0.738543939694505\\
4118.51646911569	0.737155326324142\\
4128.57265425235	0.74667724657806\\
4138.68375660458	0.739139059710375\\
4148.72390931273	0.73279111287443\\
4158.65136695405	0.742709779805594\\
4168.72495016133	0.758182900218211\\
4178.81342910268	0.758976393572704\\
4188.85934261413	0.751438206705019\\
4198.79356008172	0.754810553461615\\
4208.81416096629	0.751041460027772\\
4218.83644472754	0.754413806784368\\
4228.87123996065	0.764530847054156\\
4238.8823729612	0.759174766911327\\
4248.94172039473	0.761356873636183\\
4258.94487190211	0.758182900218211\\
4268.93672720309	0.746875619916683\\
4278.98688243474	0.76353898036104\\
4289.01158372057	0.771275540567348\\
4298.95346343465	0.775441380678437\\
4308.95150733668	0.755802420154731\\
4318.9471536197	0.74727236659393\\
4328.91278762902	0.737948819678635\\
4338.90069901091	0.729617139456457\\
4348.8464993978	0.746280499900813\\
4358.77320765482	0.741122793096608\\
4368.70638720737	0.735766712953779\\
4378.71794118332	0.752628446736759\\
4388.73024509448	0.768101567149375\\
4398.77061852623	0.769886927196985\\
4408.77915988473	0.766514580440389\\
4418.80097430116	0.766712953779012\\
4428.7852076587	0.748264233287046\\
4438.85081088692	0.737552073001389\\
4449.01654034907	0.746875619916683\\
4459.03686900274	0.74489188653045\\
4468.97366963435	0.744098393175957\\
4478.97825631651	0.74548700654632\\
4489.06246331053	0.746875619916683\\
4499.02822066898	0.73041063281095\\
4509.01338027607	0.741122793096608\\
4519.02661611455	0.736163459631026\\
4529.00489394812	0.742709779805594\\
4539.03169428677	0.734973219599286\\
4549.13077263389	0.741122793096608\\
4559.15878890355	0.745883753223567\\
4569.15135964656	0.755405673477485\\
4579.20861489147	0.745090259869074\\
4589.21390839947	0.744495139853204\\
4599.29375991123	0.736956952985519\\
4609.28322688841	0.740527673080738\\
4619.29414516837	0.74608212656219\\
4629.34704742482	0.739932553064868\\
4639.30489984226	0.741519539773854\\
4649.43285779814	0.740924419757985\\
4659.38393301449	0.760166633604444\\
4669.39448713907	0.76175362031343\\
4679.39117623684	0.752826820075382\\
4689.40842923978	0.753025193414005\\
4699.45624796816	0.739734179726245\\
4709.48710321036	0.741122793096608\\
4719.56126902123	0.746478873239437\\
4729.58795618984	0.740726046419361\\
4739.62647275266	0.743701646498711\\
4749.61084634687	0.750446340011902\\
4759.60830733248	0.742313033128348\\
4769.69391985079	0.743304899821464\\
4779.67163165942	0.755207300138861\\
4789.70717892393	0.756595913509224\\
4799.73095384156	0.752628446736759\\
4809.73540527058	0.75937314024995\\
4819.73950765678	0.749057726641539\\
4829.72500772617	0.731799246181313\\
4839.72980900805	0.725252926006745\\
4849.77174083496	0.737552073001389\\
4859.77573072298	0.73219599285856\\
4869.75509775386	0.72922039277921\\
4879.72314169948	0.730013886133704\\
4889.76626395585	0.738345566355882\\
4899.7272804421	0.749851219996033\\
4909.72698356306	0.754215433445745\\
4919.74181087249	0.756595913509224\\
4929.78860631917	0.749652846657409\\
4939.82525688107	0.746875619916683\\
4949.85792251997	0.742511406466971\\
4959.95242000332	0.748065859948423\\
4969.9313527852	0.739932553064868\\
4979.95244384079	0.750247966673279\\
4989.96124407341	0.739535806387622\\
5000.00022695224	0.74608212656219\\
};
\end{axis}\end{tikzpicture}}
\caption{Sample paths of the stochastic imitation dynamics in Example \ref{ex:atan}  for the game in Example \ref{ex:example} with $\lambda=K_{12}=K_{21}=1$ on an ER with $p=0.02$ and $n=5000$, and on a square lattice with $n=5041$.}
\label{fig:er}
\end{figure}

However, in this section, we show that there are at least some relevant cases in which the behavior predicted for a complete graph in Theorem \ref{teo:main} seems to coincide with the one on non-complete topologies. In Fig. \ref{fig:er}, we present some simulations of the imitation dynamics proposed in Example \ref{ex:atan} with the potential in Example \ref{ex:example} on some relevant network of interactions. Specifically, in (a) and (b) we consider an Erd\H{o}s-R\'enyi (ER) random graph, where each couple of individuals are connected with probability $p$ (independently on the others); in (c) and (d), instead, we simulate the system on a regular square lattice. The behavior of this system in both cases is very similar to the one predicted for the complete graph (as one can see by comparing Figs. \ref{fig:simulations} and \ref{fig:er}), suggesting the possibility to extend our analytical results to more general networks of interactions.

\section{Conclusion}\label{sec:conclusion}

In this paper, we extended our comprehension of imitation dynamics, gaining new insight with respect to those that can be obtained by analyzing their deterministic counterparts. In particular, we studied the different behavior of NE, depending on their evolutionary stability. Our main results, formalized in Theorem \ref{teo:main}, concern the long-run behavior of the system under the imitation dynamics. Specifically, we show that under some reasonable assumptions, the process spends almost all the time close to the ESSs of the game, for a time exponentially long time in the population size.

In future research, first we seek to extend our results to more general cases. On the one hand, we would like to generalize the formulation of Theorem \ref{teo:main}, removing the assumptions made in this work. On the other hand, inspired by the promising preliminary numerical results in Section \ref{sec:simulations}, we are  extending our analytical finding to the case in which players interact on a non-complete network of connections. We also plan to study cases in which the learning process interacts with the dynamics of a physical system \cite{Como2013}.

\appendix
\setcounter{lemma}{0}
\begin{lemma}
Let us consider $X(t)$ with transition rates \eqref{eq:rate} and a potential $\Phi$ that satisfies \eqref{eq:potential}. Let us focus on the stochastic process $\Phi(t)=\Phi(X(t))$ and let $X(t)$ such that $||X(t)-\bar x||_\infty\geq\delta>0$, for any $\bar x$ fixed point of \eqref{eq:ode} and $t\geq 0$. Then, $\forall\,\eps>0$ $\exists\,C_{\delta},K_{\delta}>0$ such that
\be\label{eq:formula1}
\P\left[\exists\, t<e^{C_{\delta}\eps n}:\Phi(t)<\Phi(0)-\eps\right]\leq e^{-K_{\delta}\eps n}.
\ee
\end{lemma}\medskip

\begin{proof}
Without any loss in generality, we re-scale $\lambda=1$. Let us focus on the stochastic process $\Phi(t)$ and the effect of each transition of the process $X(t)$ on it. At first, we notice that, when $X(t)=x$, the process $\Phi(t)$ increases or decreases with rates that can be explicitly written as functions of $x$. Notably, called $q^{\pm}(x)$ the increasing and the decreasing rate of $\Phi(t)$ when $X(t)=x$, respectively, we have
\be
q^-(x)=n\sum_{i\in\mc A}\sum_{j|r_j<r_i}x_ix_jf_{ij}(x),
\ee
and
\be
q^+(x)=n\sum_{i\in\mc A}\sum_{j|r_j>r_i}x_ix_jf_{ij}(x).
\ee
Due to \eqref{eq:sign}, it is straightforward that $q^+(x)\geq q^-(x)$, and that the strict inequality holds for all the states $x$ that are not fixed points of \eqref{eq:ode}. Moreover, Lipschitz-continuity of $f_{ij}$s guarantees that $\exists\,\delta'>0$ such that, $\forall\,x\in\mc X$ with $||x-\bar x||_\infty>\delta$, it holds \be\label{eq:drift}q^+(x)\geq(1+\delta')q^-(x).\ee 

Named $C$ the maximum possible reward, i.e., $C=\max_{x,i} r_i(x)$. Then, as a straightforward consequence, for each jump of the Markov process $X(t)$ in the direction of the decreasing potential, the potential decreases by no more than $C/n$, according to\eqref{eq:potential}. We recall that each change of potential is associated with a player that updates its action from some $i\in\mc A$ to some $j\in\mc A$ with $j\neq i$, and all the possible pairs of distinct actions $(i,j)$ are exactly $m(m-1)/2$. 

Hence, in order for the process $\Phi(t)$ to go beyond $\Phi(0)-\eps$, at least $\eps n/C$ steps  in the direction of the decreasing potential have to be present. As a consequence, there must be at least a pair of actions $(i,j)$ such that the number of steps in the direction of the decreasing potential due to action swapping from $i$ to $j$ (or vice versa) is greater than the one in the opposite direction by at least $2\eps n/Cm(m-1)$.

The rest of the proof is focused on bounding the probability that this event before an exponentially large time is passed. Fixed a pair of actions $(i,j)$, we name $E_{ij}(T)$ the event: \textit {the number of steps in the direction of the decreasing potential, consequence of action flipping from $i$ to $j$ (or vice versa) is greater than the one in the direction of the increasing potential at least by a quantity $2\eps n/Cm(m-1)$ within a time-horizon of duration $T$}. The goal of the main part of this proof consists in demonstrating that $\exists\,C_\eps>0$ such that \be\label{exp}
\P\left[E_{ij}(n^{-1}\exp\{C_{\delta'}\eps n\})\right]\leq10\exp\{-C_{\delta'}\eps n\},
\ee
and this proof is performed tailoring the argument used in \cite{Fagnani2017} to the specific features of this model.

From now on, to improve the readability, we call jump an event in which a player flips his/her action from $i$ to $j$ or vice versa. Let $\Lambda(t)$ be the number of jumps  the process $X(t)$ does in the time interval $[0,t]$.  $\Lambda(t)$ is a time-varying Poisson process, naturally dominated by an homogeneous Poisson process with rate $\mu=n$ (that is the rate of the Poisson process governing the player's activation, not regarding on the action they play).

Let us denote by $\Phi^+_k$ the potential of the process after its $k$-th jump and let us define \be A(t):=\{k=1,\dots ,\Lambda(t): \Phi^+_{k-1}\in (\Phi(0)-\eps, \Phi(0)]\}.\ee
Let $\xi_k$ be the Bernoulli random variable that assumes value $1$ if the $k$-th jump  increases the potential.  It holds
\be\label{sojour-ber}\mathbb{P}(\xi_k=1)=\frac{q^+(x)}{q^+(x)+q^-(x)}\geq\frac{1+\delta'}{2+\delta'}=:p.\ee
Now, we bound the number of jumps occurring during a fixed time range $T$, with
\be\label{eq:bound jump}\ba{lll}\mathbb P[\Lambda(T)> K\mu T]&\leq&\ds\sum\limits_{k=\lceil K\mu T+1\rceil}^{+\infty}e^{-\mu T}\frac{(\mu T)^k}{k!}\\[10pt]&\leq &\ds\frac{(\mu T)^{\lceil K\mu T\rceil}}{\lceil K\mu  T\rceil!}\leq \left(\frac{e}{K}\right)^{\lceil K\mu T\rceil}_.\ea\ee
For any $K>e$, \eqref{eq:bound jump} guarantees an exponential decay of the probability of having more than $K\mu T$ jumps during a time range T.

We now estimate $\P[E_{ij}(T)]$, by conditioning on the number of jumps in the time range $T$, and splitting the summation into two parts, where each one can be bounded using different techniques. To simplify the notation we define \be
\P[E^L]:=\P[E_{ij}(T)\,|\,\Lambda(T)=L],
\ee
and we estimate as follows:
\be\label{sojour-estim1}\ba{lll}\P[E_{ij}(T)]&=&\ds\sum_{L\in\N} \mathbb P\left[E^L\right]\cdot\mathbb P[\Lambda(T)=L]\\[10pt]
\quad&\leq&\displaystyle\sum_{L=1}^{3\mu T}\mathbb P\left[E^L\right]\cdot\mathbb P[\Lambda(T)=L]\\[8pt]&&\ds+\mathbb P[\Lambda(T)>3\mu T].\ea\ee
The second term of the right hand side of \eqref{sojour-estim1} can be bounded using \eqref{eq:bound jump}. Therefore, we focus on estimating the first one: the probability of having a sequence of $l\geq 2\eps n/Cm(m-1)$ consecutive jumps while the potential is in $(\Phi(0)-\eps,\Phi(0)]$, for which the number of overall jumps in the direction of decreasing potential exceeds the ones in the other direction at least by  $2\eps n/Cm(m-1)$. Using the union bound, we have 
\begin{equation}\label{sojour-estim2}\begin{array}{l}\mathbb P\left[E^L\right]\leq
\displaystyle\sum\limits_{k=1}^L\sum\limits_{l=\lceil\eps n\rceil}^L\mathbb P[E_l],\end{array}\end{equation}
where the event $E_l$ is defined as
\be E_l=\bigcap_{i=0}^{l-1}\left\{k+i\in A(t)\right\}\,\bigcap\left\{\sum_{i=0}^{l-1}\xi_{k+i}\leq \frac{l}{2}-\frac{\eps n}{Cm(m-1)}\right\}\ee
Using Chernoff bound, we estimate
\be\label{eq:prob el}\ba{lll}
\ds\mathbb P\left[E_l\right]&\leq&\ds\exp\left\{-lp\left(\frac{2p-1}{2p}\right)^2\right\}\\&\leq&\ds\exp\left\{-\frac{(2p-1)^2l}{8p}\right\}\\&\leq&\ds\exp\left\{-\frac{\delta^2 l}{8(1+\delta)(2+\delta)}\right\}.
\ea\ee
Combining (\ref{sojour-estim2}) and (\ref{eq:prob el}), we bound
\begin{equation}\label{sojour-estim3}\begin{array}{lll}\mathbb P\left[E^L\right]&\leq&
\displaystyle\sum\limits_{k=1}^L\sum\limits_{l=\lceil\eps n\rceil}^L\mathbb P[E_l]
\\&\leq& \displaystyle\sum\limits_{k=1}^L\sum\limits_{l=\lceil\eps n\rceil}^L\exp\left\{-\frac{\delta^2 l}{8(1+\delta)(2+\delta)}\right\}\\&\leq& \ds L^2\exp\left\{-\eps\frac{\delta^2}{8(1+\delta)(2+\delta)}n\right\}.\end{array}\end{equation}
Being the bound in \eqref{sojour-estim3} monotonically increasing in $L$, we can conclude
\be\label{sojour-estim4}\ba{l}\displaystyle\sum_{L=1}^{3\mu T}\mathbb P\left[E^L\right]\mathbb P[\Lambda(T)=L]\\[12pt]\qquad\leq \displaystyle\mathbb P\left[E^{3\mu T}\right]\ds\sum_{L=1}^{3\mu T}\mathbb P[\Lambda(T)=L] \\[12pt]\qquad\leq \displaystyle(3\mu T)^2\exp\left\{-\eps\frac{\delta^2}{8(1+\delta)(2+\delta)}n\right\}.\ea\ee
We bound the second term of \eqref{sojour-estim1} using \eqref{eq:bound jump} with $K=3$, being the smallest integer number greater than $e$, obtaining the following exponential decay:
\be\label{sojour-estim5}\ds\mathbb P[E_{ij}(T)]\leq (3\mu T)^2\exp\left\{\frac{\delta^2\eps n}{8(1+\delta)(2+\delta)}\right\}
+\left(\frac{e}{3}\right)^{\lceil 3\mu T\rceil}.\ee
Fix now $\epsilon >0$ and put \be\label{eq:choice T}T=\frac1n\exp\left\{\frac{\delta^2}{24(1+\delta)(2+\delta)}\eps n\right\}.\ee Using the fact that $(e/3)^x<x^{-2}$ for all $x>0$, we finally prove the validity of \eqref{exp} with $C_{\delta'}$ equal to the expression multiplying $\eps n$ in the argument of \eqref{eq:choice T}. 

Finally, using the union bound on all the $m(m-1)/2$ pairs of actions, we estimate
\be\label{exp2}\ba{l}
\displaystyle\P\left[\exists\, t\leq n^{-1}\exp\{C_{\delta'}\eps n\}:\Phi(t)\leq\Phi(0)-\eps\right]\\[6pt]\qquad \leq\displaystyle\sum_{(i,j)}\P\left[E_{ij}(n^{-1}\exp\{C_{\delta'}\eps n\})\right]\\[6pt]
\qquad\leq\displaystyle \sum_{(i,j)}10\exp\{-C_{\delta'}\eps n\}\\[6pt]
\qquad\leq\displaystyle 5m(m-1)\exp\{-C_{\delta'}\eps n\}.
\ea\ee
Since $\delta'$ is defined in \eqref{eq:drift}, depending on $\delta$, $C_{\delta'}$ is ultimately function of $\delta$. We conclude the proof by noticing that \eqref{exp2} can be re-formulated as \eqref{eq:formula1}, for any $C_\delta,K_\delta>0$ such that
\be
C_\delta \leq C_\delta' -\frac{\ln n}{\eps n},\quad K_\delta \leq C_\delta' -\frac{\ln (5m(m-1))}{\eps n},\quad 
\ee
which has solutions if $n$ is sufficiently large.
\end{proof}\medskip

\begin{lemma}
Let us consider $X(t)$ with transition rates \eqref{eq:rate} and a potential $\Phi$ that satisfies \eqref{eq:potential}. Let us focus on the stochastic process $\Phi(t)=\Phi(X(t))$ and let $X(t)$ such that $||X(t)-\bar x||\to 0$ as $n\to \infty$, where $\bar x$ is an unstable critical point of the potential. Then, $\exists\,\eps>0$ such that, for $K_\eps>0$ it holds
\be\label{eq:fast exit}
\P\left[\exists\, t\leq K_\eps n\ln n:\Phi(t)\geq \Phi(\bar x)+\eps\right]\geq 1-\frac{1}{\ln N}.
\ee
\end{lemma}\medskip

\begin{proof}
Without any loss in generality, we re-scale $\lambda=1$. The proof is straightforward when $\bar x$ is a local minimum of $\Phi$. We adopt the same formalism used in the proof of Lemma \ref{lemma:decrease}. We notice that $q^+(x)\geq q^-(x)$, but \eqref{eq:drift} is not satisfied as far as $x$ is close to $\bar x$. Being $\bar x$ a local minimum of $\Phi(x)$, the Lipshitz-continuity argument guarantees that $\exists\,\delta>0$ such that $\text{min}_{x:||x-\bar x||=\delta}\Phi(x)=\Phi(\bar x)+\eps$, for some $\eps>0$. We notice that $\delta n$  jumps in the direction of the increasing potential guarantee to reach $\Phi(\bar x)+\eps$. The probability of having a jump in the direction of the increasing potential is greater than or equal to $1/2$, which immediately yields the process $\Phi(t)$ to stochastically dominate a (fair) voter model $P(t)$ with the same jumps and copying probability $1/2$. The expected time needed for $P(t)$ to reach $\Phi(\bar x)+\eps$ is $K_\eps n$, for some $K_\eps>0$ only depending on $\delta$ and, ultimately, on $\eps$ \cite{Aldous2013}. Finally, Markov inequality yields the proof.

If $\bar x$ is a saddle point, the presence of a stable manifold poses some technical issues to be taken into account. Lemma \ref{lemma:decrease}, however, guarantees that the process can not exit from the neighborhood of $\bar x$ along a stable manifold (along which the potential decreases), before an exponentially long time in $n$. Then, since all transitions can occur with non-null probability, at each activation of a node the system has a non-zero probability to deviate from a stable manifold, so after $n$ transitions the probability of observing at least a deviation is exponentially close to $1$. Considering that, with probability converging exponentially close to $1$ we have at least $n$ transitions in a time-horizon of duration $\ln n$, we conclude that in a time-window of duration $\ln n$ the process deviates from a stable manifold, with probability converging exponentially to $1$. Finally, the result proved above for the exit time from a minimum can be directly applied when the process is not on a stable manifold of the saddle point. 
\end{proof}

\end{document}